\documentclass[preprint]{aastex}        
\usepackage{graphicx,times,natbib,lscape}

\def\arcsec{$^{\prime\prime}$}

\def\spose#1{\hbox to 0pt{#1\hss}}
\def\lta{\mathrel{\spose{\lower 3pt\hbox{$\mathchar"218$}}
     \raise 2.0pt\hbox{$\mathchar"13C$}}}

\shorttitle{UV-Optically Selected Galaxies at z$\sim$3}
\shortauthors{Guo et al.}

\begin{document}

\title{Rest-frame UV--Optically Selected Galaxies at $2.3\lesssim z \lesssim 3.5$: Searching for Dusty Star-forming and Passively Evolving Galaxies}
\author{Yicheng Guo$^{1,2}$, Mauro Giavalisco$^{1}$, Paolo Cassata$^{1}$, 
Henry C. Ferguson$^{3}$, Christina C. Williams$^{1}$, Mark Dickinson$^{4}$, 
Anton Koekemoer$^{3}$, Norman A. Grogin$^{3}$, Ranga-Ram Chary$^{5}$,
Hugo Messias$^{6}$, Elena Tundo$^{7}$, Lihwai Lin$^{8}$, 
Seong-Kook Lee$^{9}$, Sara Salimbeni$^{1}$, Adriano Fontana$^{10}$, 
Andrea Grazian$^{10}$, Dale Kocevski$^{11}$, Kyoung-Soo Lee$^{12}$, 
Edward Villanueva$^{13}$, and Arjen van der Wel$^{14}$
}
\affil{$^1$ Astronomy Department, University of Massachusetts,
710 N. Pleasant Street, Amherst, MA 01003, USA}
\affil{$^2$ email: {\texttt yicheng@astro.umass.edu}}
\affil{$^3$ Space Telescope Science Institute, 3700 San Martin Drive,
Baltimore, MD, 21218, USA}
\affil{$^4$ NOAO-Tucson, 950 North Cherry Avenue, Tucson, AZ 85719, USA}
\affil{$^5$ Spitzer Science Center, California Institute of Technology, MS 220-6, Pasadena, CA 91125, USA}
\affil{$^6$ Centro de Astronomia e Astrofísica da Universidade de Lisboa, Observatório Astronómico de Lisboa, Tapada da Ajuda, 1349-018 Lisboa, Portugal }
\affil{$^7$ INAF - Osservatorio Astronomico di Trieste, Via Tiepolo 11, I-34131 Trieste, Italy}
\affil{$^8$ Institute of Astronomy \& Astrophysics, Academia Sinica, Taipei 106, Taiwan}
\affil{$^9$ School of Physics, Korea Institute for Advanced Study, Hoegiro 87, Dongdaemun-Gu, Seoul 130-722, Republic of Korea}
\affil{$^{10}$ INAF - Osservatorio Astronomico di Roma, Via Frascati 33, I–00040, Monteporzio, Italy}
\affil{$^{11}$ UCO/Lick Observatory, University of California, Santa Cruz, CA 95064, USA}
\affil{$^{12}$ Yale Center for Astronomy and Astrophysics, Department of Physics, Yale University, New Haven, CT 06520, USA}
\affil{$^{13}$ Carnegie Observatories, 813 Santa Barbara Street, Pasadena, CA 91101-1292, USA}
\affil{$^{14}$ Max-Planck Institut für Astronomie, Königstuhl 17, D-69117 Heidelberg, Germany}



\begin{abstract} 
A new set of color selection criteria (VJL) analogous with the BzK method is
designed to select both star-forming galaxies (SFGs) and passively evolving
galaxies (PEGs) at $2.3\lesssim z \lesssim 3.5$ by using rest-frame UV--optical
(V-J vs. J-L) colors. The criteria are thoroughly tested with theoretical
stellar population synthesis models and real galaxies with spectroscopic
redshifts to evaluate their efficiency and contamination. We apply the
well-tested VJL criteria to the HST/WFC3 Early Release Science field and study
the physical properties of selected galaxies. The redshift distribution of
selected SFGs peaks at z$\sim$2.7, slightly lower than that of Lyman break
galaxies at z$\sim$3. Comparing the observed mid-infrared fluxes of selected
galaxies with the prediction of pure stellar emission, we find that our VJL
method is effective at selecting massive dusty SFGs that are missed by the
Lyman break technique. About half of the star formation in massive (${\rm
M_{star} > 10^{10}M_\odot}$) galaxies at $2.3\lesssim z \lesssim 3.5$ is
contributed by dusty (extinction E(B-V)$>$0.4) SFGs, which however, only
account for $\sim$20\% of the number density of massive SFGs. We also use the
mid-infrared fluxes to clean our PEG sample, and find that galaxy size can be
used as a secondary criterion to effectively eliminate the contamination of
dusty SFGs. The redshift distribution of the cleaned PEG sample peaks at
z$\sim$2.5. We find six PEG candidates at z$>$3 and discuss possible methods to
distinguish them from dusty contamination. We conclude that at least part of
our candidates are real PEGs at z$\sim$3, implying that this type of galaxies
began to form their stars at z$\gtrsim$5. We measure the integrated stellar
mass density (ISMD) of PEGs at z$\sim$2.5 and set constraints on it at z$>$3.  We find
that the ISMD grows by at least about factor of 10
in 1 Gyr at 3$<$z$<$5 and by another factor of 10 in next 3.5 Gyr (1$<$z$<3$).
\end{abstract}

\keywords{Cosmology: observations --- Galaxies: evolution --- Galaxies: formation --- Galaxies: fundamental parameters --- Galaxies: general --- Galaxies: high-redshift --- Infrared: galaxies}

\section{Introduction}
\label{intro}

Understanding galaxy formation and evolution remains one of the most
outstanding questions in astronomy. According to the standard paradigm, galaxies
are initially formed in the center of small cold dark matter halos, gradually
assembled with time through hierarchical processes, and eventually evolved into
populations with various color, size, morphology, etc., as observed in our
local universe \citep{white78}.  However, the physics behind this scenario is
still poorly understood. Theoretical models
\citep[e.g.,][]{benson03,bower06,croton06,delucia06} require complex ingredients (e.g.,
using feedback to quench star formation in galaxies) in addition to simple gas
falling and cooling to reproduce even basic observations of nearby galaxies,
such as luminosity function \citep[e.g.,][]{blanton01,norberg02,blanton03a} and
color bimodality \citep[e.g.,][]{kauffmann03,bell04,blanton05}.  Since these
ingredients are predicted (or designed) to begin to work since the universe is
young, it is essential to test them through observational studies on the
physical properties of high-redshift galaxies.

During the 13.7 billion years of cosmic time, the era of $1<z<4$ is of
particular interest, in terms of star formation, stellar mass content, and
galaxy morphology.  First, although the increase of cosmic star formation rate
density (SFRD) with redshift is well studied out to $z \sim 1$
\citep[e.g.,][]{hopkinsa04,hopkinsa06}, the question of whether the SFRD has a broad
peak during $2<z<4$ is still far from being finally solved \citep[e.g.,
][]{hopkinsa04, hopkinsa06,pg08,chary10}. Furthermore, if such a peak exists,
what is the mechanism that turns off the bulk of star formation in the
universe?  Second, being related to the evolution of the SFRD, the assembly
history of massive (${\rm >10^{11} M_{\odot}}$) galaxies is still in question.
A large number of massive galaxies is found at z$\sim$2
\citep[e.g.,][]{daddi04a,fontana04,glazebrook04,saracco05}, but only a few of them are
found at z$>$3.5 \citep{mobasher05,dunlop07,rodighiero07,wiklind08,mancini09}.
This dearth of massive galaxies at high redshift raises the question of when
and how these giants were largely assembled in the universe.  Last but not the
least, the morphology of galaxies also undergoes a transition at z$\sim$3.
Although being studied in detail in the local universe and even traced back to
z$\sim$1.5 \citep{vandenbergh00}, the Hubble sequence of galaxy morphology is
not believed to be in place at z$\sim$3
\citep{giavalisco96,conselice04,ravindranath06} because a large fraction of
galaxies in that epoch has irregular shapes (chain-like, clumpy, multiple
cores, etc.).  Therefore, the origin and emergence of the Hubble sequence
remains an open question.  To answer all the above questions requires
observational studies on the physical properties (e.g., star formation rate
[SFR], stellar mass, and morphology) of galaxies at z$\sim$3.

High-redshift galaxies can be effectively selected from deep sky surveys
through their broad-band colors. Star-forming galaxies (SFGs) at z$\sim$3 and above
are prevalently selected with the dropout method by locating the position of
the Lyman break from their rest-frame UV colors
\citep[e.g.,][]{giavalisco02,steidel03,giavalisco04}.  This technique has
been proved to be very successful because galaxies selected in this way, namely,
Lyman break galaxies \citep[LBGs, see][for a review]{giavalisco02}, are
spectroscopically confirmed as SFGs at high redshift
\citep{steidel96a,steidel96b,steidel99, steidel03} with little contamination.
Recently, this technique has been extended to select galaxies at 1.4$<$z$<$2.5
\citep[BX/BM galaxies,][]{adelberger04,steidel04}.  However, the Lyman break
technique misses one interesting galaxy population, namely dusty SFGs.
How much this population contributes to the cosmic SFRD and number
density of galaxies at z$\sim$3 is still controversial.  Studies using far-IR
or sub-millimeter (sub-mm) emission from cold dust show that some dusty galaxies, for
example sub-mm galaxies \citep[e.g.,][]{blain02,chapman03,chapman05, swinbank06},  have
SFRs up to $\sim {\rm 1000 M_{\odot}/yr}$.  The high SFRs of dusty galaxies
imply that the contribution of this population to the cosmic SFRD at z$\sim$3
may not be ignored.  To avoid underestimating the SFRD due to the exclusion of
this population, a new color selection method is required to select
SFGs independent of dust reddening.

Besides dusty SFGs, passively evolving galaxies (PEGs) at high redshift are also missed by the Lyman break
technique.  Although PEGs contribute little to the SFRD, they are directly
related to the ceasing of star formation in galaxies and to the history of stellar
mass assembly in the universe.  To search for this population, several color
selection criteria have been proposed. Among them, the most commonly used two
are the Extremely Red Objects
\citep[EROs;][]{thompson99,daddi00,roche02,roche03,mccarthy04} and Distant Red
Galaxies \citep[DRGs;][]{franx03,vandokkum03,vandokkum04,
vandokkum06,papovich06}.  EROs are selected with very red optical to near-IR
color, typically $(R-K)_{Vega}>5$, while DRGs have a red near-IR color with
$(J-K)_{Vega}>2.3$. Both methods use the red color as an indicator of the large
amount of old stars in galaxies.  However, due to the strong degeneracy between
age and dust reddening, the red color of a galaxy could be caused by either
old stars or high dust extinctions. As a result, samples selected by both
methods contain both massive PEGs and dusty SFGs with similar fractions, as
showed by spectroscopic observations
\citep{cimatti02,cimatti03,fs04,yanl04}.  To exclude the
contamination of SFGs, a more efficient way of selecting PEGs at z$\sim$3 is
needed.

A selection method that satisfies the above requirements already exists for
galaxies at z$\sim$2, as proposed by \citet{daddi04bzk}.  This method uses the
B-, z-, and K-band photometry to select both SFGs and PEGs at z$\sim$2.
Samples selected through the BzK method are now widely used to investigate
several aspects of galaxies at z$\sim$2, from physical properties
\citep[e.g.,][]{daddi04bzk,daddi05,reddy05,daddi07a, daddi07b,blanc08},
abundance \citep{kong06,lane07,blanc08}, stellar mass function
\citep{grazian07}, to clustering \citep{kong06,blanc08}. 

In this work, we try to design an analogous method that selects and classifies
simultaneously both SFGs (with different dust extinctions) and PEGs at
2.3$<$z$<$3.5.  For this purpose, we extend the successful BzK method from
z$\sim$2 to z$\sim$3 by replacing the selection bands with the V-, J-, and IRAC
3.6$\mu$m band (hereafter L-band), according to the relative shift of galaxy
spectra between the two redshifts.  Our selection method (hereafter VJL) uses
the same rest-frame colors as the BzK method so that galaxies selected by both
methods have same spectral types.  However, due to the different depth and
sensitivities of the bands used in each method, the VJL selected sample may
have different incompleteness and contamination from the BzK selected sample. 

Nowadays, photometric redshift (photo-z) can be fairly accurately measured with
relative error of only a few percent \citep[e.g.,][]{ilbert09,dahlen10} and is 
hence widely used to select galaxies within a certain redshift range.  However,
the bias of photo-z selection is not explicit. It is common to characterize
photo-z errors with a redshift probability distribution function (PDF).  The accuracy
of the distribution function strongly depends on the assumed mix of galaxy
templates in the spectral energy distribution (SED) library.  Unfortunately,
our knowledge on the true SED types is limited and the commonly used SED
libraries are often not good representatives of real galaxies.  Let alone the
mystery of dust extinction curve, initial mass function (IMF), metallicity
of high-z galaxies, one major uncertainty of fitting high-z galaxies is the
unknown star formation history (SFH). The commonly used exponentially declined
SFH ($\tau$-model) may be a suitable approximation for low-z galaxies, but
is not a realistic model for high-z galaxies. Recently postulated hypotheses on SFH
of high-z galaxies include exponentially increasing \citep{maraston10} or
roughly linearly increasing \citep{joshualee10} SFH. Using an unrealistic SFH
would eventually result in a mis-interpretation of the bias of photo-z
selection.

On the other side, the bias of color selection can be fairly explicitly
determined. One easy way to do so is applying the color criterion to simulated
galaxies that have a certain range of redshift, SFHs and extinctions and
calculating the success and failure rate of the selection. Thus, one can
robustly measure the expected redshift distribution as well as the
incompleteness of the selection as a function of several variables, such as
magnitude, size, and color of galaxies. Moreover, color selection is easier to
reproduce. Unlike photo-z selections, results of which may vary from people to
people, depending on the used SED-fitting codes or SED libraries, color
selection results are robust and make the comparison of different works easy
for the whole community.  The success of color selection method has been proved
by the prevalence of Lyman break technique \citep[see the review
of][]{giavalisco02}.

In this paper, we apply our VJL selection method to the HST/WFC3 Early Release
Science \citep[ERS,][]{windhorst11ers} observations in the south field of the
Great Observatories Origins Deep Survey \citep[GOODS,][]{giavalisco04} South
field (GOODS-S).  Serving as an ideal test field of our selection method, ERS
brings three advantages for us to calibrate and optimize our method. First, its
deep ($\sim$27 AB mag) J-band allows us to select galaxies that are faint in
their rest-frame optical bands. These galaxies could be dusty SFGs and the
ability to detect and correctly classify them is a key of our method. Second,
embedded within GOODS South field, ERS is augmented by several existing data
sets, from X-ray, optical, to mid-infrared band and sub-mm band. The
multi-wavelength data enable us to accurately understand the nature of our
selected galaxies. Third, ERS has similar depth on J- and H-bands as the
upcoming CANDELS observation \citep{candelsoverview,candelshst} so that our
method calibrated in ERS can be easily adapted to apply to CANDELS data.

Throughout the paper, we adopt a flat ${\rm \Lambda CDM}$ cosmology with
$\Omega_m=0.3$, $\Omega_{\Lambda}=0.7$ and use the Hubble constant in terms of
$h\equiv H_0/100 {\rm km~s^{-1}~Mpc^{-1}} = 0.70$.  All magnitudes in the paper
are in AB scale \citep{oke74} unless otherwise noted.

\section{The Data}
\label{data}

\subsection{Images}
\label{data:image}

The ERS observation \citep{windhorst11ers} covers 40--50 ${\rm arcmin^2}$ of
the GOODS South field in 10 bands. The data used in this work are its near-IR
observations, i.e., F098M (Ys), F125W (J), and F160W (H) images.  The 50\%
completeness limit for 5-$\sigma$ detections for typical compact objects
(circular aperture with radius of 0.4\arcsec\ ) is 27.2, 27.55, and 27.25 for
Ys, J, and H. We re-processed the images and drizzled them to a 0.06\arcsec\
per pixel scale and registered to the GOODS WCS.

The GOODS south field has been observed with various telescopes and instrument
combinations, from X-ray to sub-mm and radio. Relevant to our analysis here are
imagings of Very Large Telescope (VLT)/VIMOS ultra-deep U-band
\citep[]{nonino09}, {\it HST}/ACS BViz \citep{giavalisco04}, VLT/ISAAC JHKs
\citep{retzlaff10}, Spitzer/IRAC 3.6, 4.5, 5.7, 8.0 $\mu$m (M. Dickinson et al.
in preparation), and Spitzer/MIPS 24 $\mu$m. Table \ref{tb:band} summarizes the
sensitivity (limiting magnitude of S/N=5 for a point source) and resolution,
namely, the full width at half-maximum (FWHM) of point-spread function (PSF),
of each band used in this study.

In our work, we also try to select an LBG sample at z$\sim$3 through the U-band
dropout technique as a reference sample to compare to our VJL sample. The
VLT/VIMOS U-band used in GOODS-S is blueward to the traditional U-band and
would bias selected galaxies toward higher redshift. The other U-band in
GOODS-S, namely the CTIO U-band, is a traditional U-band, but the depth of its
imaging is about 1.5 mag shallower than that of VIMOS U-band. In order to
select a relatively complete U-band dropout sample at z$\sim$3, we turn to use
the multi-wavelength catalog of the GOODS North field (GOODS-N), where the KPNO
U-band imaging satisfies the requirements of being both traditional U-band and
deep to valid our z$\sim$3 LBG selection. Besides having the KPNO U-band and
HST/ACS and Spitzer IRAC observations, GOODS-N also has the ground-based NIR
images observed through CFHT WIRCAM J- and K-band \citep[][]{linlihwai11, wangweihao10}. We list their sensitivity and
resolution, together with those of KPNO U-band, in Table \ref{tb:band}.

\begin{table}[h]
\caption{Sensitivity and Resolution of GOODS Filters \label{tb:band}}
\begin{tabular}{ccc}
\hline\hline
Filter & Sensitivity &  Resolution \\
       & (Limiting Magnitude of  & (FWHM of  \\
       & S/N=5 for Point Source) &  PSF) \\
\hline
VIMOS U & 28.00 & 0.8\arcsec\ \\
CTIO U & 25.8 & $\sim$1.5\arcsec\ \\
KPNO U & 27.1 & 1.15\arcsec\ \\
ACS F435W (B) & 28.7 & 0.08\arcsec\ \\
ACS F606W (V) & 28.8 & 0.08\arcsec\ \\
ACS F775W (i) & 28.3 & 0.08\arcsec\ \\
ACS F850LP (z) & 28.1 & 0.09\arcsec\ \\
WFC3/IR F098M (Ys) & 27.2 & 0.12\arcsec\ \\
WFC3/IR F125W (J) & 27.55 & 0.13\arcsec\ \\
WFC3/IR F160W (H) & 27.25 & 0.15\arcsec\ \\
ISAAC J & 25.0 & $\sim$0.5\arcsec\ \\
ISAAC H & 24.5 & $\sim$0.5\arcsec\ \\
ISAAC Ks & 24.4 & $\sim$0.5\arcsec\ \\
CFHT/WIRCAM J & 24.6 & $\sim$0.8\arcsec\ \\
CFHT/WIRCAM K & 24.2 & $\sim$0.8\arcsec\ \\
IRAC 3.6 $\mu$m (ch1 or L) & 26.1 & 1.7\arcsec\ \\
IRAC 4.5 $\mu$m (ch2) & 25.5 & 1.7\arcsec\ \\
IRAC 5.8 $\mu$m (ch3) & 23.5 & 1.7\arcsec\ \\
IRAC 8.0 $\mu$m (ch4) & 23.4 & 1.9\arcsec\ \\
MIPS 24 $\mu$m & 20.4 & 6\arcsec \\
\hline
\end{tabular}
\end{table}

\subsection{Catalogs}
\label{data:cat}

To robustly measure the photometry of objects in all above bands with mixed
resolutions, we use a software package with object template-fitting method
\citep[TFIT;][]{laidler07}.  For each object, TFIT uses the spatial position
and morphology of the object in a high-resolution image to construct a
template.  This template is then fit to the images of the object in all
other low-resolution bands. During the fitting, the fluxes of the object in
low-resolution bands are left as free parameters. The best-fit fluxes are
considered as the fluxes of the object in low-resolution bands. These
procedures can be simultaneously done for several objects which are close
enough to each other in the sky so that the deblending effect of these objects
on the flux measurement would be minimized. Experiments on both simulated and
real images show that TFIT is able to measure accurate isophotal photometry of
objects to the limiting sensitivity of the image \citep{laidler07}.

Catalogs of different fields (ERS, GOODS-S and GOODS-N) are generated based on
different detection bands.  In the ERS catalog, we use WFC3/IR H-band as the
detection band as well as TFIT high-resolution template. ACS BViz and WFC3 YJH
isophotal photometry is measured in dual-image mode by SExtractor based on
H-band detection. U-band, ISAAC Ks-band, and IRAC 4 channels' photometry is
measured through TFIT. For both GOODS fields, the ACS z-band is chosen as the
detection band. ACS isophotal photometry is measured in dual-mode by
SExtractor, while other bands' photometry is measured by TFIT with z-band
template (N. A. Grogin et al. in preparation). All SExtractor isophotal and
ground-based TFIT (isophotal) fluxes are converted to total fluxes by
multiplying an aperture correction factor, which is the ratio of SExtractor
FLUX\_AUTO and FLUX\_ISO of the detection band of each filed (H-band for ERS
and z-band for both GOODS fields).

In addition to the above bands, the GOODS fields are also observed by the
Spitzer MIPS 24$\mu$m channel. Fluxes of sources in MIPS images are measured by
fitting PSF to prior positions of objects detected in the Spitzer IRAC
3.6$\mu$m image. During the fitting, the positions of MIPS sources are allowed
to wander by less than 0.6\arcsec\  from the IRAC prior position. After the
first pass of fitting and subtraction of fitted MIPS sources, a second pass of
fitting is run for MIPS sources that do not have IRAC counterparts. Sources
detected and fit in both passes are combined together into the final MIPS
catalogs. Details about the catalog of 24$\mu$m photometry can be referred from
\citet{magnelli11}. To combine the TFIT and MIPS catalogs, we match sources in
the two catalogs with positions, allowing a maximum matching distance of
1.0\arcsec\ .

\section{Color Selection Criteria}
\label{vjl}

\begin{figure*}[htbp]
\center{\includegraphics[scale=0.45, angle=0]{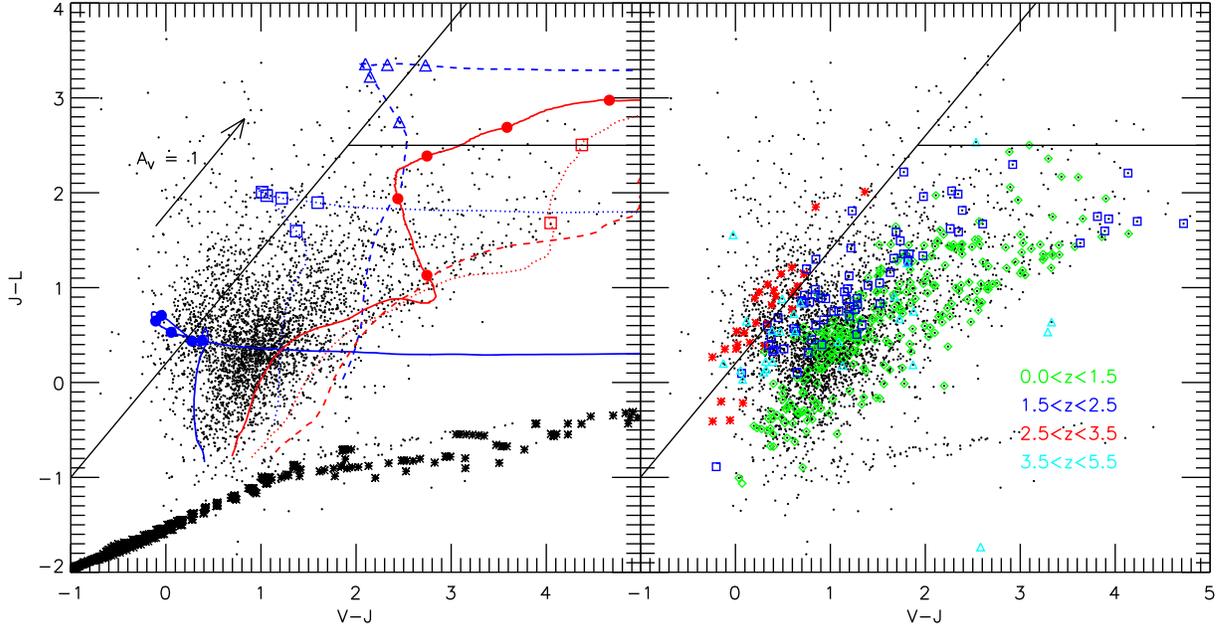}}
\caption[]{{\it Left}: tracks of galaxy models placed at different redshifts in
the (J-L) vs. (V-J) two-color diagram. Symbols in each track are for redshift
z=2, 2.5, 3, 3.5 and 4. Blue tracks are models with a constant star formation
(CSF) rate, age of 0.5 Gyr and various dust reddenings (solid with filled
points: E(B-V)=0.0; dotted with squares: E(B-V)=0.3; dashed with triangles:
E(B-V)=0.6).  Red lines show the tracks of dust-free SSP models with ages of 
0.5 (solid with filled points), 1 (dotted with
squares), and 2 Gyr (dashed with triangles). Black stars show the locus of
stars of \citet{lejeune97}.  {\it Right}: ERS galaxies with spec-z in the (J-L)
vs. (V-J) two-color diagram. Galaxies at different redshift ranges are color
coded as labels show.  The two solid black lines in each panel show our
designed selection windows (upper left for sVJL and upper right for pVJL). In
each panel, overplotted small dots show all ERS galaxies with S/N$>=$5 in J-
and L-bands. \\
\label{fig:model}}
\vspace{-0.2cm}
\end{figure*}

In order to simultaneously select both SFGs and PEGs at z$\sim$3, we extend the
BzK method at z$\sim$2 \citep{daddi04bzk} to z$\sim$3 by replacing the B-z and
z-K colors in the BzK criteria with the V-J and J-L color, as the rest-frame
wavelengths observed by the BzK bands for a galaxy at z$\sim$2 are redshifted
to the observation windows of the VJL bands at z$\sim$3. Because ratios of
central wavelengths of the VJL bands to the BzK bands are not a constant, we
also adjust the coefficient in the original BzK criteria so that the dust
reddening vector is parallel to our selection window, which would ideally make
our selection criteria independent of dust reddening. We determine the
intersection terms of each selection equation through the distributions of
galaxies with different redshifts in the J-L versus V-J color diagram (the right
panel of Fig. \ref{fig:model}). Since the slope of the star-forming VJL
criterion (Eq. \ref{eq:svjl}) is fixed based on the dust reddening vector, we
shift the criterion line (the diagonal line in Fig. \ref{fig:model}) to get the
term (+0.2) in Eq. \ref{eq:svjl} that optimally separates galaxies at
1.5$<$z$<$2.5 (blue squares in the right panel of Fig. \ref{fig:model}) from
those at 2.5$<$z$<$3.5 (red stars in the same panel). For the intersection term
(2.5) in Eq. \ref{eq:pvjl}, since we do not have passive galaxies at
2.5$<$z$<$3.5 that have been spectroscopically observed in our sample to help
calibrate the selection window, we choose to use this term to exclude
low-redshift interlopes as much as possible and meanwhile to keep the single
stellar population (SSP) model of galaxies with age of 1 Gyr (the red dotted line
with squares in the left panel of Fig. \ref{fig:model} at z=2.5 within the
selection window. Thus, our VJL criteria are
\begin{equation}
  J-L\geq1.2\times(V-J)+0.2 
\label{eq:svjl}
\end{equation}
for selecting SFGs and
\begin{equation}
  J-L\geq2.5 \bigwedge J-L<1.2\times(V-J)+0.2
\label{eq:pvjl}
\end{equation}
for selecting PEGs, where $\bigwedge$ means the logical {\it and}. Our method,
similar to the BzK method, uses the strength and slope of the Balmer break,
which is between the J- and L-bands for galaxies around z$\sim$3, to select SFGs
and distinguish them from PEGs. For simplicity, in this paper, we call galaxies
selected or selection window defined by Equation \ref{eq:svjl} as sVJLs, while
those by Equation \ref{eq:pvjl} as pVJLs.

An extension of the Bzk method to higher redshift has been already
proposed by \citet{daddi04bzk}, who suggested to use R-, J-, and L-band colors
to select galaxies at z$>$2.5. \citet{daddi04bzk} tested the validity of the
selection criterion in their K20 sample. They ended up with few detections at
z$>$2.5, as the K20 sample does not cover the redshift range z$>$2.5. They also
claimed that using GOODS ACS+ISAAC+{\it Spitzer} data set would be deep enough
in all of the RJL bands to detect galaxies at z$>$2.5. The RJL method is quite
similar to out VJL, however, we use all space-based bands in our selection to
ensure a deep sensitivity. Another color selection aiming toward selecting
galaxies at 1.5$<$z$<$3.5 by using rest-frame UV/optical colors has been
proposed by \citet{cameron11}. They use {\it HST} Y-H versus V-z colors to
identify and characterize 1.5$<$z$<$3.5 galaxies in the HUDF and ERS field.
While their criteria have the advantage of having similar resolutions in all
bands that are used for selection, our criteria cover a much longer wavelength
baseline. And our reddest band (the L-band), a close proxy of stellar mass in
the interested redshift range, enables our selected to be easily compared with
a stellar mass selected sample.

We test the validity of our VJL selection criteria in two ways. First, we study
the evolutionary track of stellar population synthesis models along redshift in
the (J-L) versus (V-J) two-color diagram.  Second, we study the distribution of
real galaxies from the ERS field with spectroscopic redshift (spec-z) in the
two-color diagram. 

The left panel of Figure \ref{fig:model} shows tracks of shifting galaxy models
along redshift (from z=0 to z=7) in the (J-L) versus (V-J) two-color diagram.
Symbols in each track stand for models at (starting from the lowest one) z=2,
2.5, 3, 3.5, and 4. Galaxy models are retrieved from an updated version (CB09)
of the stellar population synthesis library of \citet[][,BC03]{bc03} with the
Salpeter IMF \citep{salpeter55} and solar metallicity. The Calzetti law
\citep{calzetti97,calzetti00} and the recipe of \citet{madau95} are applied to
each model to account for the dust reddening and the opacity of intergalactic
medium (IGM) in the universe.  Our selection windows corresponding to Equation
\ref{eq:svjl} (the upper left region) and \ref{eq:pvjl} (the upper right
region) are outlined by black lines.

Blue tracks stand for models with a constant star formation (CSF) rate, age of
0.5 Gyr and various dust reddenings (solid: E(B-V)=0.0; dotted: E(B-V)=0.3;
dashed: E(B-V)=0.6). These tracks show two facts: (1) all CSF models enter our
sVJL selection window in the redshift range of $2.3 \leq z \leq 3.5$ and (2)
the reddening vector (the black arrow in the panel) is almost parallel to our
sVJL selection window (the diagonal black line).  We also test our criteria
with models with older ages (2 Gyr) as well as models with an exponentially
declining SFH ($\tau$-model, where $\tau$, the
characteristic time scale of star formation, is fixed to 1.0 Gyr). Both types
of models have similar tracks as that of the CSF model with t=0.5 Gyr. These
results demonstrate that our sVJL criterion can select SFGs with various SFH,
age, and SFR independent of dust reddening.

However, the CSF model with E(B-V)=0.6 enters our pVJL selection window twice,
at $1.5 \lesssim z \lesssim 2.5$ and $z \geq 4.0$.  Models with different
SFH but same age and dust reddening also enter the pVJL selection window at
similar redshifts.  The behavior of these models suggests that our pVJL
galaxies may be contaminated by highly obscured SFGs from both
lower ($z\leq 2.0$) and higher ($z\geq 4.0$) redshift, regardless of their
exact SFH.

Red tracks show the evolutionary track of dust-free SSP models with 
ages of 0.5, 1, and 2 Gyr. All three SSP models enter our
pVJL selecting window, but at different redshifts: $\sim$3.5, $\sim$3.0, and
$\sim$2.5 for models with age of 0.5 Gyr (blue), 1.0 Gyr (green), and 2.0 Gyr
(red).  Overall, our pVJL criteria are able to select PEGs around z=2.5 and
above.

The right panel of Figure \ref{fig:model} shows our second test, that is the
position of galaxies with different spec-zs in the (J-L) versus (V-J) diagram.
This test with real galaxies supplements the first one in two ways: (1) it
helps in understanding the effect of photometry uncertainty and (2) it shows
how our method works for galaxies with unknown and perhaps more complex SFH.
Galaxies with spec-z in the ERS are divided into different redshift ranges and
shown by colors and labels.

In this panel, the edge of our sVJL selection window effectively separates
galaxies at $2.3 \leq z \leq 3.5$ from others, satisfying our expectation.
However, several galaxies with lower redshift and a few with higher ones also
enter our sVJL window. We suspect that photometry uncertainty is the main
reason that scatters them into our sVJL window, although we cannot rule out the
effect of a complex SFH. Few galaxies with (J-L) color redder than 2.5 are
found in our spec-z sample. The lack of red galaxies is caused by the fact that
spectroscopic observations are biased against dusty SFGs and PEGs because of
their faint and featureless rest-frame optical spectra. It is also possible
that red galaxies are really rare in the high-z universe. We note that a few
galaxies from lower redshift ($z<2.5$) and higher redshift ($z>4.5$) enter the
pVJL selection window. The existence of these types of contamination is
consistent with our above analysis with theoretical models (see the {\it left}
panel). We will discuss how to eliminate contamination in both selection
methods later. 

Active galactic nucleus (AGN) sources could also contaminate our VJL selected sample. As shown in
\citet{civano11}, about 30\% of AGN at z$\gtrsim$3 show a typical optical
spectrum of an SFG, but have X-ray luminosity $> 10^{44} erg/s$, a typical value
of quasars. In order to evaluate their contamination, we study the redshift
tracks of AGN templates of \citet{polletta07} in the (J-L) versus (V-J) plot.
Templates of type 2 QSO (QSO2), type 1 QSO with the lowest optical--to--IR ratio
(BQSO1), and type 1 QSO with the highest optical--to--IR ratio (TQSO1) are all
within our sVJL selection window at z=0. However, QSO2 leaves the window
quickly before z=0.5, and BQSO1 also leaves the window around z=1. Given the
small cosmic volume that our surveys observe at z$<$1, we argue that these two
types of AGN would not severely contaminate our sample. The only template that
stays in our sVJL selection window up to z$>$3.5 is TQSO1. We also examine the
track of AGN + star-burst template of I19254 of \citet{polletta07}. The
template enters our sVJL selection window at z$>$2 and evolves to redder (J-L)
direction within the window as redshift increases. It suggests that our red
(J-L$>$2.5) dusty SFG sample could be contaminated by Seyfert 2 galaxy. We will
discuss the possible contamination in detail later in Sec. \ref{svjl:dusty}. We
also note that no QSO template enters our pVJL selection window, which
indicates that our selected PEG sample is in principle immune from AGN
contamination.

As a summary, using both theoretical models (CSF, $\tau$-model, and SSP) and
spectroscopically observed galaxies, we show that our sVJL selection window
(defined by Equation \ref{eq:svjl}) can select SFGs with various levels of 
SFR independent of dust reddening at $2.3 \leq z \leq 3.5$.  And our
pVJL selection window (defined by Equation \ref{eq:pvjl}) can select PEGs
around z=2.5 and above, although such a selected sample may be contaminated by
dusty SFGs at $z\leq 2.0$ and $z\geq 4.0$. Also, no template at $z\leq 1.5$ or
galaxies with spec-z$<$1.5 enters either of our selection windows, suggesting
that our criteria are effective at excluding low-redshift galaxies.

\section{Star-forming VJL Galaxies}
\label{svjl}

We apply our sVJL criterion, defined by Eq. \ref{eq:svjl}, to the
multi-wavelength catalog of the ERS field, which is based on the WFC3
H-band detection, as discussed in \S\ref{data:cat}, to select SFGs at
z$\sim$3. To ensure an accurate measure of galaxy colors, we require all
selected galaxies to have S/N$>$10 in J- and L-bands. We also construct a
samples with S/N$>$20 in the two bands. Comparison between the two samples
would show us how photometric uncertainty affects our selection results. For
V-band photometry, if S/N$<$1, we use the 1$\sigma$ photometric uncertainty as
the upper limit of flux. The two samples contain 354 and 146 galaxies,
respectively.

We note that the BzK color criterion of \citet{daddi04bzk} was constructed
to be applied to K-selected samples. Similarly, one would expect the VJL
criterion to be applied to L-band limited samples. Without an L-band detection,
thus with an upper limit (at best) on the J-L color, no VJL galaxy can be
unambiguously selected. Moreover, using L-band also ensures the closest proxy
for mass selection of the sample. In this paper, we choose to apply the 
sigal-to-noise ratio (S/N) cut
on both bands instead of on only the L-band so that we could have accurate J-L
color. This is not contradictory with selecting an L-band limited sample.
Instead, it asks for more strict constraint on the J-L color to exclude
interlopes. This is well fitting the purpose of this paper to demonstrate the
validity of the selection criterion. We also acknowledge that the use of S/N
cut on two bands would bring a more complicated selection effect on the
completeness of sample, because now the completeness is not only dependent on
the proxy of mass, but also on the color. However, we will argue later (in
\S\ref{svjl:dustcontribution} and \S\ref{pvjl:massdensity}) that the induced
selection effect would not significantly change the quantitative results of
comparing our VJL samples with other samples.

\subsection{Deriving Physical Properties of Galaxies}
\label{svjl:prop}

\begin{figure}[htbp]
\center{\includegraphics[scale=0.6, angle=0]{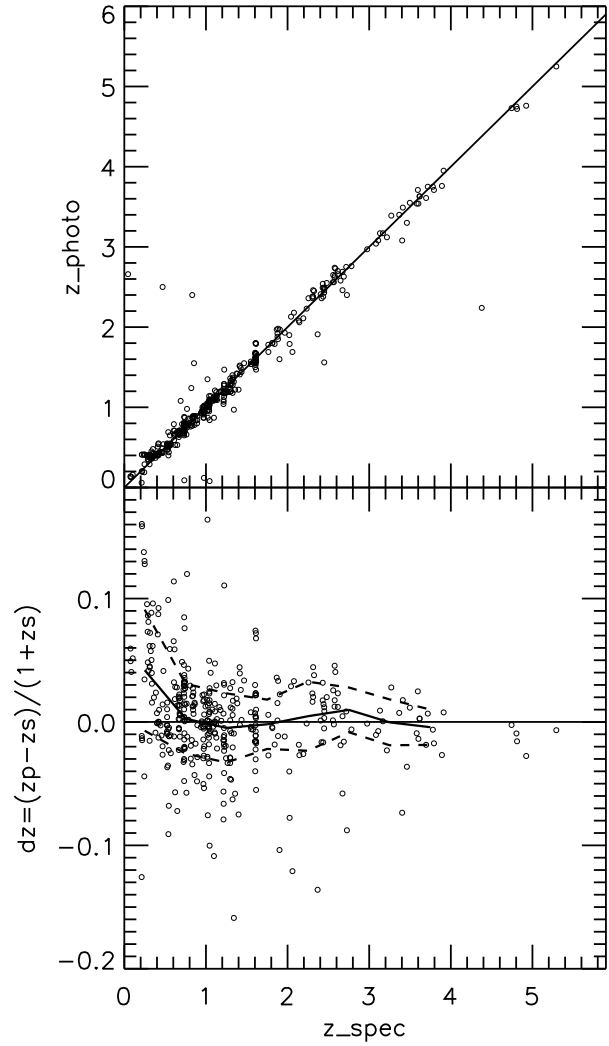}}
\caption[]{Accuracy of our photo-z measurement. {\it Top}: the comparison of
photo-z and spec-z. The solid line shows the one-to-one correspondence. {\it
Bottom}: the relative error as a function of redshift. For clarity, the 
bottom panel zooms into the range of -0.2$<$dz$<$0.2. The solid curve shows 
the mean of the relative error, while two dashed curves cover the 1$\sigma$
confidence level. \\
\label{fig:zcomp}}
\vspace{-0.2cm}
\end{figure}

We derive photo-zs and physical properties of selected galaxies by fitting 
their SEDs to stellar population synthesis
models. Models used to measure photo-zs are extracted from the library of
PEGASE 2.0 \citep{pegase}. Instead of using the redshift with the least
$\chi^2$, we integrate the probability distribution function of redshift (zPDF)
and derive the likelihood-weighted average redshift. When the zPDF has two or
more peaks, we only integrate the main peak that has the largest power. The
accuracy of our photo-z measurement is shown in Figure \ref{fig:zcomp}, where
we compare our photo-zs and spec-zs of galaxies that are spectroscopically
observed in the ERS field. The {\it top} panel shows a very good agreement between
photo-zs and spec-zs. The relative error (defined as ${\rm
(z_{phot}-z_{spec})/(1+z_{spec})}$) has an almost zero mean (0.0005) and a very
small deviation (0.037 after 3$\sigma$ clipping). And the fraction of outliers,
defined as ${\rm |\Delta z|/(1+z) > 0.15}$, is about 3.4\%. The {\it bottom}
panel shows the mean and standard deviation (after 3-sigma clipping) of
relative errors in each redshift bin with a bin size 0.5. The
means of the relative errors have no significant offset from zero at all
redshift bins, especially for the range of $2<z<4$, which is of the most
interest in this study. The high accuracy of our photo-z measurement enables us
to statistically study the physical properties of our selected galaxies without
spectroscopic redshifts.

The physical properties (stellar mass, specific star formation rate [SSFR], and
dust reddening) of galaxies are measured through SED-fitting models retrieved
from the library of CB09 with the Salpeter IMF \citep{salpeter55}. The lower
and upper cuts on mass in the IMF are ${\rm 0.1 M_{\odot}}$ and ${\rm 100
M_{\odot}}$, respectively.  The models consist of grid points in a parameter
space spanned by redshift, dust extinction E(B-V), SFH (characterized by $\tau$
and age), and metallicity. The available values of each parameter are shown in
Table \ref{tb:param}. We apply the Calzetti law \citep{calzetti97,calzetti00}
and the recipe of \citet{madau95} to the models to account for dust extinction
and the opacity of IGM in the universe.  For each model, the fluxes in all bands
are pre-computed and stored in a grid database.  When fitting a galaxy, we scan
the database and calculate $\chi^2$ values for models in all grid points over
the whole parameter space.  The $\chi^2$ value is calculated as
\begin{equation}
  \chi^2 = \Sigma_i \frac{(F_{obs,i}-\alpha F_{model,i})^2}{\sigma_{i}^2},
\label{eq:chi2}
\end{equation}
where $F_{obs,i}$, $F_{model,i}$, and $\sigma_{i}$ are the observed flux, model
flux, and observational uncertainty in the $i$th band. $\alpha$ is a
normalization factor, which is equal to stellar mass if $F_{model,i}$ is
normalized to ${\rm 1 M_{\odot}}$ in our pre-computed database.  The model with
the least $\chi^2$ is considered the best-fit model and its parameters are used
as the measurements of the properties of the galaxy. During the SED-fitting,
the redshift of a galaxy is fixed as its photo-z or its spec-z, if the latter
is available.  

In addition to the SED fitting with bands from U-band all the way to IRAC 8.0
$\mu$m, we also estimate the E(B-V)s and SFRs of sVJL galaxies by using the
slopes and fluxes of their rest-frame UV continuum. Compared with SED-fitting,
this method is less model-dependent and requires no prior information on the
SFH of galaxies. In this method, we use the Calzetti law \citep{calzetti94,
calzetti00} to convert the rest-frame UV-slope of a galaxy into its dust
reddening,  and calculate the unobscured SFR from its dust-corrected rest-frame
UV continuum by using the formula in \citet{kennicutt98}. Since the SFHs of
high-redshift SFGs are controversial
\citep[e.g.,][]{joshualee10,maraston10,papovich11}, we prefer to use E(B-V)s
and SFRs estimated from UV continuum for SVJLs.

\begin{table}[h]
\caption{Parameter Space Used for SED-Fitting \label{tb:param}}
\begin{tabular}{ccc}
\hline\hline
Parameter & &  Range \\
\hline
Redshift & & 0.0 to 7.0 with a bin size of 0.01 \\
E(B-V)\footnote{E(B-V) runs up to 0.3 for models with $t/\tau>=4.0$.} & & 0.0 to 1.0, $\Delta E(B-V)=0.05$ \\
Metallicity & & 0.004, 0.02, 0.08 \\
Age (Gyr) & & (1, 2, 3, 5, 8) $\times$ $10^{-3}, 10^{-2}, 10^{-1}, 10^{0}, 10^{1}$, up to 13 \\
$\tau$ (Gyr) & & (1, 2, 3, 5, 8) $\times$ $10^{-3}, 10^{-2}, 10^{-1}, 10^{0}, 10^{1}$, and $\infty$ \\
\hline
\end{tabular}
\end{table}

\subsection{Redshift Distribution}
\label{svjl:redshift}

\begin{figure}[htbp]
\center{\includegraphics[scale=0.8, angle=0]{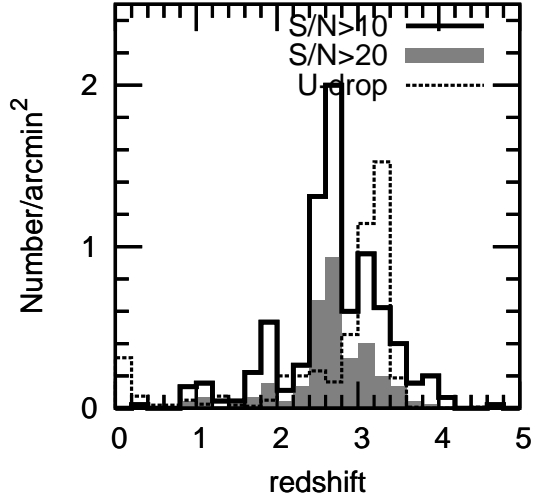}}
\caption[]{Redshift distributions of star-forming VJL galaxies.  Solid line
stands for the sample with S/N$>$10 in J- and L-bands, while filled gray
histogram for the sample with S/N$>$20 in the two bands.  For comparison, 
the distribution of U-band dropouts selected in GOODS-N is plotted 
with dotted line. \\
\label{fig:svjlzdist}}
\vspace{-0.2cm}
\end{figure}

The redshift distributions of our sVJL samples with S/N$>$10 and 20 are shown
in Figure \ref{fig:svjlzdist}. Both distributions highly peak around z$\sim$2.7
and extend to z$>$3.5, demonstrating that, as we expected, our sVJL criterion
is effective at selecting galaxies between $2.3 \leq z \leq 3.5$. The S/N$>$10
sample has a secondary peak around z=1.8, which is implying that the main
contamination of our sVJL selection is coming from galaxies at z$\sim$2.
Fortunately, this secondary peak is largely diminished in the S/N$>$20 sample.
The number ratio between galaxies at z$\sim$1.8 and at z$\sim$2.8 decreases
from 0.27 in the S/N$>$10 sample to 0.17 in the S/N$>$20 one. It suggests that
the low-z contamination in our sVJL sample is induced by photometric
uncertainty rather than the deficit of our method and hence can be removed by
increasing the S/N cuts in J- and L-band. In later study, to balance the
fraction of contamination and the number of statistics, we use the S/N$>$10
sample as our fiducial sample.

\subsection{Comparison with LBGs}
\label{svjl:lbgs}

Nowadays, high-redshift SFGs are commonly selected through the Lyman break
technique. In order to avoid the contamination zone of elliptical galaxies,
this technique compromises to only select galaxies with a bright and blue
rest-frame UV continuum, namely SFGs with low or no dust extinction.  Dusty
SFGs, whose rest-frame UV color mimics that of elliptical galaxies, are missed
by this technique. Because of this bias, the existence and contribution of
dusty SFGs to the cosmic SFRD at z$\sim$3 has been the topic of considerable
debate. To shed a light on the above question, we compare galaxies selected
through our sVJL method, which is designed to select both low-dust and dusty
SFGs, with LBGs at z$\sim$3. 

The U-band dropout method is used to select LBGs
at z$\sim$3, because the Lyman break of an SFG is redshifted to between the
U-band and the B-band. A sample of 1161 U-band dropouts is selected from GOODS
North (878) and South (283) fields with the following criteria:
\begin{eqnarray}
\label{eq:udrop}
U-B & \ge & 0.75 + 0.5 \times (B-z), \\ \nonumber
U-B & \ge & 0.9, \\ \nonumber
B-z & \le & 4.0, \\ \nonumber
S/N_{B} \ge 3 & {\rm and} & S/N_{z} \ge 3. \nonumber
\end{eqnarray}
We note that the number of U-band dropouts in GOODS-S is significantly less 
than that in GOODS-N, because the CTIO U-band image in GOODS-S is 1.5 mag
shallower than the KPNO U-band image in GOODS-N. The physical properties
of the U-band dropouts are measured in the same way used for the sVJL galaxies. 

In Figure \ref{fig:svjlzdist}, we overplot the redshift distribution of the
U-band dropout sample (dotted line) selected from GOODS-N. The distribution
peaks around z$\sim$3, being consistent with the expectation of LBGs, but
significantly deviates from the peak of our sVJL sample. Since the offset between
the peaks of the two samples is larger than 2$\sigma$ deviation of our photo-z
measurement ($\Delta z/(1+z)=0.037$), it is an intrinsic difference between the
two methods rather than due to photo-z uncertainty. However, since the cosmic
time interval between the two redshift peaks (250 Myr) is about 10 times less
than the age of the universe at z$\sim$3 ($\sim$2.2 Gyr), we assume that the
evolution of galaxies between the two redshifts is negligible. Under this
assumption, any difference between the two samples is considered due to the
fact that the two methods select galaxies with different physical properties
rather than select galaxies with different redshifts. Also, in order to
eliminate the effect of possible contamination, we only compare galaxies within
the range of $2.3<z<3.5$ in the two methods.

\begin{figure}[htbp]
\center{\includegraphics[scale=0.45, angle=0]{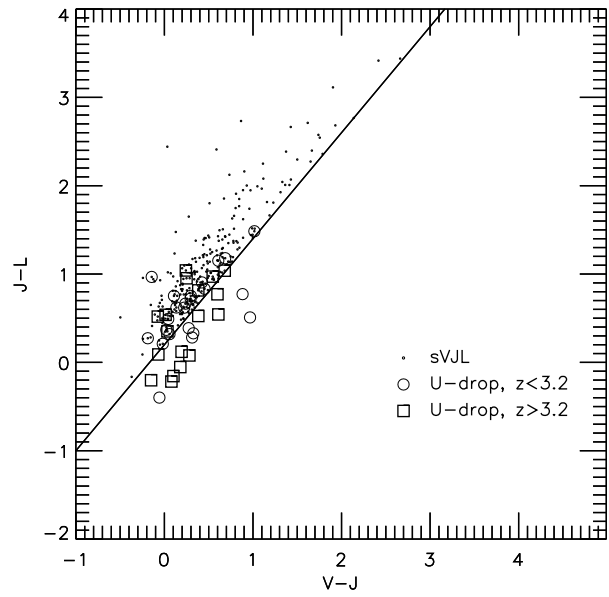}}
\caption[]{Star-forming VJLs (points) and U-band dropouts (circles: z$<$3.2;
squares: z$>$3.2) in the (J-L) versus (V-J) color-color diagram. Only U-band
dropouts that fall into the ERS field are plotted. Both sVJL and U-band dropout
samples have S/N$>$10 in J- and L-bands. \\
\label{fig:udropinvjl}}
\vspace{-0.2cm}
\end{figure}

A direct and illustrative way to compare both methods is to study the location
of the U-band dropouts in the (J-L) versus (V-J) plot. We match the U-band
dropouts that are selected from GOODS-S TFIT catalog to the ERS TFIT catalog to
measure their (J-L) and (V-J) colors.  Figure \ref{fig:udropinvjl} shows the
positions of 41 matched U-band dropouts (circles and squares) with S/N$>$10 in
J- and L-bands in the (J-L) versus (V-J) diagram, together with sVJLs (points). We
note that the U-band dropouts are scattered along the edge of our sVJL
selection window. Among 41 U-band dropouts, 16 fall outside our sVJL selection
window. Although photometric uncertainty could contribute to the scatter, we
suspect that the primary reason is due to the different redshift distribution
between sVJLs and U-band dropouts. As shown in Figure \ref{fig:svjlzdist}, the
U-band dropouts have systematically higher redshift than sVJLs and is hence
easier for them to be scattered out of the selection window. To examine our
suspicion, we divide the U-band dropouts sample into two sub-samples: z$<$3.2
(circles) and z$>$3.2 (squares). 10 out of 16 (63\%) U-band dropouts outside
the sVJL selection window have z$>$3.2, suggesting that redshift is the main
reason for these galaxies not being selected by our sVJL method.

The other feature of U-band dropouts is more prominent and physical: no U-band
dropout has J-L color redder than 2.0 mag. As shown in the {\it left} panel of
Figure \ref{fig:model}, CSF galaxies with E(B-V)$>$0.3 would have J-L color
redder than 2.0 mag. Therefore, (J-L)$>$2.0 mag can be treated as a rough
division for weakly and strongly obscured galaxies. The lack of red U-band
dropouts confirms conclusions of previous studies that LBGs miss highly
obscured galaxies \citep[e.g., ][]{bouwens09,ly11,riguccini11}. On the other
side, our sVJL method selects galaxies up to J-L around 3.0 mag, suggesting its
ability to select highly obscured SFGs.

\begin{figure*}[htbp]
\center{\includegraphics[scale=0.6, angle=0]{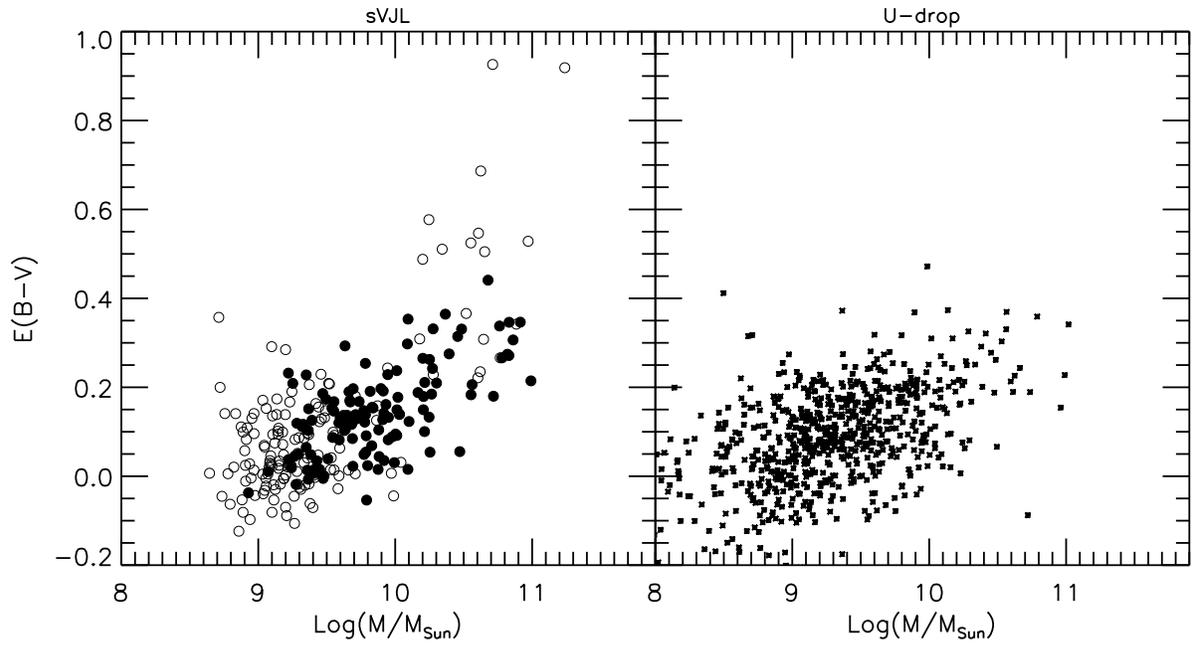}}
\caption[]{E(B-V) distribution as a function of stellar mass for star-forming
VJL galaxies ({\it left}) and U-band dropouts ({\it right}). Empty and solid
circles in the {\it left} panels show sVJLs in the S/N$>$10 and S/N$>$20
samples, respectively. \\
\label{fig:ebmvvsms}}
\vspace{-0.2cm}
\end{figure*}

The difference of the E(B-V) distributions of samples selected by the two
methods can be clearly seen from Figure \ref{fig:ebmvvsms}, where E(B-V) is
measured from the slope of rest-frame UV continuum and plotted as a function of
stellar mass of galaxies. Both samples have similar E(B-V) distribution in the
stellar mass range of ${\rm 9<log(M/M_\odot)<10}$. But in the range of ${\rm
10<log(M/M_\odot)<11}$, their E(B-V) distributions differ: the distribution of
U-band dropouts ends around E(B-V)=0.4, while that of sVJLs in the S/N$>$10
sample extends beyond E(B-V)=0.6. 

\begin{figure*}[htbp]
\center{\includegraphics[scale=0.4, angle=0]{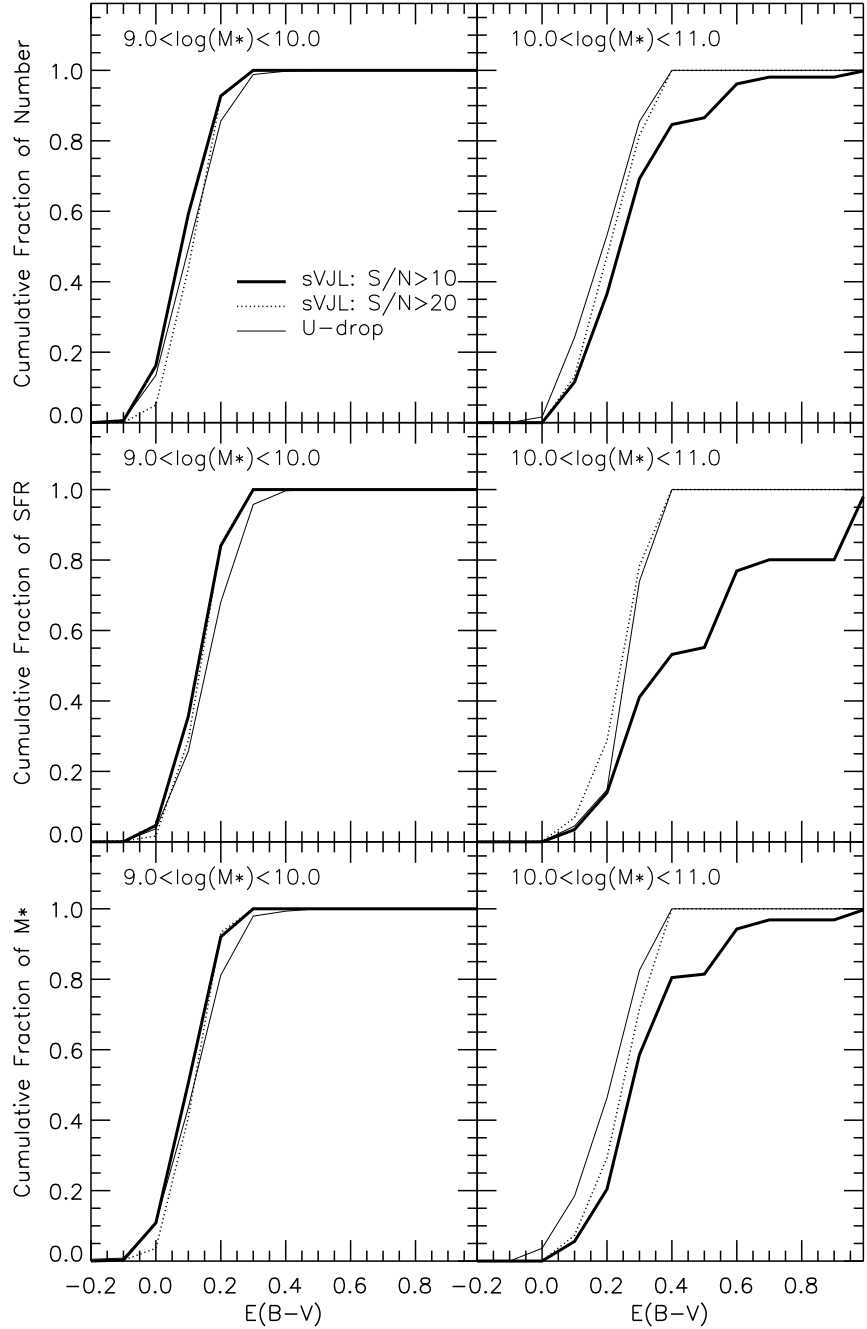}}
\caption[]{{\it Top}: cumulative fraction of number of galaxies as a function
of E(B-V) for sVJLs (thick solid for S/N$>$10 and dotted for S/N$>$20) and U-band
dropouts (thin solid) in two stellar mass ranges. {\it Middle}: cumulative
fraction of SFR for sVJLs and U-band dropouts. {\it Bottom}: cumulative
fraction of stellar mass for sVJLs and U-band dropouts. SVJLs and U-band 
dropouts in two stellar mass bins: ${\rm 9.0<log(M_{star}/M_\odot)<10.0}$ 
(left) and ${\rm 10.0<log(M_{star}/M_\odot)<11.0}$ (right) are plotted. \\
\label{fig:ebmvcurve}}
\vspace{-0.2cm}
\end{figure*}

The two upper panels of Figure \ref{fig:ebmvcurve} show the cumulative fraction
of number of galaxies as a function of E(B-V) in both stellar mass ranges for
sVJLs and U-band dropouts. In the range of ${\rm 9<log(M/M_\odot)<10}$, both
sVJL and U-band dropout samples have similar cumulative fraction curve and only
contain galaxies with E(B-V)$<$0.4.  In the range of ${\rm
10<log(M/M_\odot)<11}$, the U-band dropout sample still only contains
E(B-V)$<$0.4 galaxies, while about 20\% of sVJLs (in the S/N$>$10 sample) have
E(B-V)$>$0.4. The E(B-V) distribution of massive (${\rm 10<log(M/M_\odot)<11}$)
sVJLs drops quickly beyond E(B-V)=0.6 in the S/N$>$10 sample (only 5\% have
E(B-V)$>$0.6). This could be attributed to two factors: (1) the real lack of
very dusty SFGs at z$\sim$3 or (2) the sensitivity of the catalog detection
band image of ERS (H-band) is not deep enough to detect these galaxies. Either
way, we can still conclude that, compared to U-band dropout method, our sVJL
selection method can select moderate dusty (E(B-V)$<=$0.6) SFGs at $2.3<z<3.5$.

We also note that the distribution of sVJLs in the S/N$>$20 sample is similar
to that of U-band dropouts, even in the range of ${\rm 10<log(M/M_\odot)<11}$.
This reflects that an overcut on the J-band S/N would reduce our ability to
detect dusty SFGs at z$\sim$2.8. The S/N$>$20 sVJL sample also contains fewer
low-mass (around ${\rm 10^9M_\odot}$) galaxies than the S/N$>$10 sVJL sample.
This can also be attributed to the overcut on the L-band S/N in the latter.

\subsection{Dusty Star-forming Galaxies}
\label{svjl:dusty}

Although red (J-L$>$2.0) sVJLs are likely to be dusty SFGs at z$\sim$2.7, a
more careful census is needed to distinguish them from possible contamination.
Specifically, PEGs at similar redshift have similar red rest-frame UV colors
and hence can easily enter our sVJL sample due to photometric uncertainty. In
order to clean our dusty sVJL sample, we have to break the age--dust
degeneracy, which, however, cannot be broken by simply using rest-frame UV and
optical data. Fortunately, at z$\sim$2.7, the rest-frame 6 $\mu$m emission from
polycyclic aromatic hydrocarbons (PAHs), a feature of dusty SFGs, falls into
the MIPS 24 $\mu$m bandpass, and can help to separate dusty SFGs from PEGs.
Any 24 $\mu$m fluxes that are significantly brighter than the prediction of
pure stellar emission should be dominantly contributed by dust emission and
hence indicate a high amount of dust in the galaxies.

We match our sVJLs to GOODS-S MIPS 24 $\mu$m catalog (see the description in
Sec. \ref{data:cat}), with a matching radius of 1.0\arcsec\ . Galaxies without
MIPS 24 $\mu$m counterparts are assigned a flux upper limit of 3 $\mu$Jy, 
which is the upper envelope of the S/N--flux relation at S/N=1 in our MIPS 24 
$\mu$m catalog. (However, we note that we do not use 24$\mu$m sources with 
flux level of 3 $\mu$Jy for any scientific purpose. Sources with detection lower
than 3$\sigma$ should be treated with caution.)
A potential issue of measuring MIPS 24 $\mu$m flux of galaxies is the
uncertainty raised by confusion and crowding. Our 24 $\mu$m catalog over the
GOODS-S field contains about 22,000 sources, deducing an average number density
of 1.2 (4.8) sources in each circle with radius of 3\arcsec\ (6\arcsec\ ),
which is 0.5 (1.0) times the FWHM of MIPS 24 $\mu$m PSF. This implies that 60\%
of light of a source is overlapping with the light of other sources. The
PSF-fitting technique that we use to construct the catalog ideally reduces the
influence to the lowest level by fitting nearby sources simultaneously. In this
method, however, a slight oversubtraction (undersubtraction) of a bright
source would result in a significant underestimation (overestimation) of
fluxes of nearby sources.

\begin{figure}[htbp]
\center{\includegraphics[scale=0.4, angle=0]{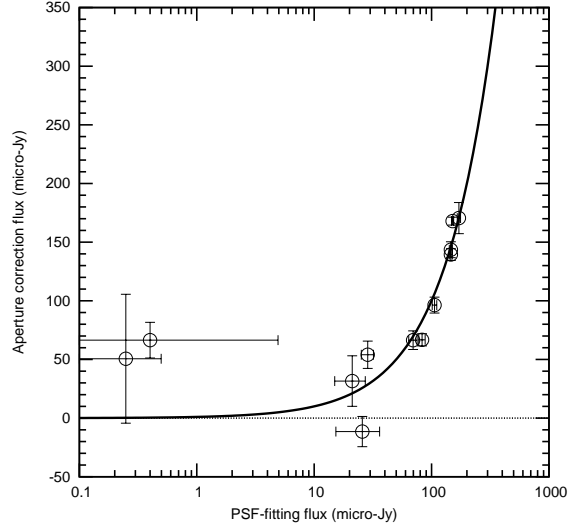}}
\caption[]{Comparison of MIPS 24$\mu$m fluxes derived through PSF-fitting and 
aperture correction for sVJLs with E(B-V)$\geq$0.4. The solid line shows 
one-to-one correspondence, while dotted line shows zero aperture corrected
fluxes. Since a few sources have negative aperture corrected fluxes, we only 
use logarithmic scale for PSF-fitting fluxes. \\
\label{fig:mipsconf}}
\vspace{-0.2cm}
\end{figure}

In order to evaluate whether MIPS 24 $\mu$m fluxes are correctly measured, we
compare our PSF-fitting fluxes to fluxes that are derived through aperture
correction. In a crowding environment, aperture correction on the flux measured
through the central region (e.g., within an aperture with size of 1 FWHM of
PSF) of a faint object tends to overestimate its flux, since the central region
of the object could be polluted by the light of its nearby sources. In this
case, the aperture corrected flux can be used as an upper limit. To obtain a
conservative estimation of the contribution of dusty SFGs to the cosmic SFRD,
we care more about sources whose fluxes are overestimated by PSF-fitting than
those whose fluxes are underestimated, as the former could be PEGs but
misclassified as dusty SFGs. Such misclassification would result in a severe
overestimation of their SFR and hence their contribution to the SFRD. For this
purpose, any sources whose PSF-fitting fluxes are significantly larger than
their aperture corrected fluxes are thought to have incorrect PSF-fitting
fluxes, and aperture corrected fluxes will be used for them.

Figure \ref{fig:mipsconf} shows the comparison between aperture corrected
fluxes and PSF-fitting fluxes for sVJLs with E(B-V)$\geq$0.4. For sources with
PSF-fitting fluxes larger than 40 $\mu$Jy, fluxes derived by both methods are
in good agreement. This is not surprising though, as both methods are robust
for bright sources. For source with PSF-fitting fluxes less than 10 $\mu$Jy,
aperture correction overestimates their fluxes due to the issue of confusion
and crowding, as these sources are faint sources around bright sources.  We
use the PSF-fitting fluxes for these sources, as they are the best
solution we can have for them. For sources with PSF-fitting fluxes between 10
$\mu$Jy and 40 $\mu$Jy, particularly of our attention is one source whose
aperture correction flux is less than zero but whose PSF-fitting flux is larger
than 10 $\mu$Jy. As we discuss above, the incorrectly high PSF-fitting flux of
this source is due to the under-subtraction of its nearby bright sources. We
will mark this problematic source in later analysis.

\begin{figure}[htbp]
\center{\includegraphics[scale=0.45, angle=0]{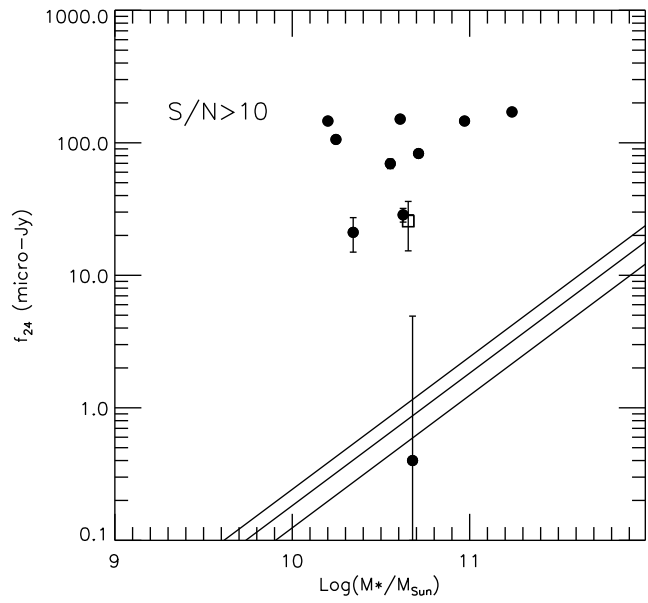}}
\caption[]{MIPS 24 $\mu$m fluxes of dusty sVJLs (S/N$>$10) as a function 
of stellar mass. Here we only show galaxies with E(B-V)$\geq$0.4 and 
$2.3<z<3.5$. Error bars show the photometric uncertainties. 
Three solid lines show the prediction of z$\sim$2.7 PEGs with age of
0.5, 1.0, and 2.0 Gyr. The square shows the source with possibly problematic
PSF-fitting flux. \\
\label{fig:mipsflx}}
\vspace{-0.2cm}
\end{figure}

Figure \ref{fig:mipsflx} shows the 24 $\mu$m fluxes of our sVJLs as a function
of stellar mass.  For simplicity, we only plot sVJLs with E(B-V)$>$0.4 and
$2.3<z<3.5$. Over-plotted (black lines) in the figure are the predictions of
the 24 $\mu$m flux--stellar mass relation for dust-free SSP models at 
z$\sim$2.7, with age of 0.5, 1.0 and 2.0 Gyr (from
top to bottom). Galaxies with 24 $\mu$m fluxes significantly brighter than the
prediction of SSP models are thought to be dusty SFGs, because their 24 $\mu$m
fluxes cannot be explained by pure stellar emission and hence should be
contributed by PAH emission. On the other hand, galaxies whose 24 $\mu$m fluxes
are consistent with the predictions of SSP models are thought as contamination.
We also mark the galaxy with problematic PSF-fitting flux with squares in
the figure. If the problematic galaxy is treated as a PEG, the fraction of
contamination is about 18\% (2 out of 11). This result is encouraging, as it
shows our sVJL method can select dusty SFGs with a low level of contamination.

Another possible source of contamination in our dusty SFGs is from AGN host
galaxies. The warm dust around AGN can absorb and reprocess the energetic
photons of AGN into IR emission that can be observed in the MIPS 24 $\mu$m
channel. We use the Chandra deep 4Ms X-ray image of
CDFS\footnote{http://cxc.harvard.edu/cda/Contrib/CDFS.html} to study the
possible AGN contamination. None of our nine dusty SFGs is individually detected
in the 4Ms Chandra catalog of \citet{xue11}. The stacked hard X-ray image of
them also reports a detection comparable to the noise level.  However, the
stacked soft X-ray image has a detection of 3.5$\sigma$. 
The soft detection may indicate that our dusty SFG sample is contaminated by
AGN host galaxies. However, using the stacked image, we measured a hardness
ratio of $\sim$-1, which is softer than the predicted hardness ratio of even
the least absorbed AGN model (column density ${\rm N_H=10^{21} cm^{-2}}$) at
z$\sim$3 in \citet{wangjx04}. The ultra-soft spectrum of the stacked image
implies that our dusty SFG sample is not heavily contaminated by AGN. We
also calculate an average luminosity from the stacked soft-band X-ray images,
using a mean redshift of 2.7. The mean luminosity is ${\rm 1.2 \times 10^{42}
erg/s}$, with the lower and upper limits from the Poisson uncertainty on net
counts of ${\rm 8.3 \times 10^{41} erg/s}$ and ${\rm 1.5 \times 10^{42}
erg/s}$.  If we use the SFR--X-ray relation of \citet{ranalli03}: Lx/SFR ${\rm
\sim 10^{40} erg/s/(M_\odot/yr)}$, we get an average SFR of about 100 ${\rm
M_\odot yr^{-1}}$. This value is consistent with the SFR measured through the
rest-frame UV continuum of these galaxies. These galaxies are heavily obscured
and occupy the high SFR end of the whole star-forming VJL sample. Therefore,
we conclude that they are compatible with being star forming.


\subsection{Contributions of Dusty Star-forming Galaxies}
\label{svjl:dustcontribution}

One of our motivations of selecting dusty SFGs around z$\sim$3 is to evaluate
their contribution to the number density, stellar mass density, and SFRD of
SFGs. A precise measurement of the absolute contributions of dusty SFGs relies
on the accurate correction of the incompleteness of the sample, which is a
function of the redshift, surface brightness, color, and spectral types of
galaxies. The best way to measure the incompleteness is simulating the
detection ability of galaxies with different physical properties and
multi-wavelength photometry. We leave such simulations to a future paper.
Instead, in this paper, we try to estimate the relative contributions (compared
with those of low-dust galaxies) of dusty SFGs to the above quantities to the
first-order accuracy. 

In our sVJL method, both low-dust and dusty galaxies are selected with the same
color criterion from the same catalog. They are also aiming to the same
redshift range. As a result, the two main factors that determine the selection
incompleteness, namely, redshift and surface brightness limit of the survey, are
roughly same for both low-dust and dusty sub-samples. We can assume that, to
the first order, incompleteness is roughly same for both sub-samples.
Therefore, the ratio of total numbers, SFRs, and stellar masses of both
sub-samples should be immune to the incompleteness and accurate to the first
order even no correction on incompleteness is applied. We acknowledge that the
redder color and fainter rest-frame UV photometry of dusty galaxies may vary
the selection incompleteness. However, both factors tend to increase the
incompleteness of dusty SFGs so that our derived ratio is a conservative
estimation of  the contributions of dusty SFGs.

Based on above discussion, a simple way to measure the relative contributions
of low-dust and dusty sub-samples is to study the cumulative number, SFR and
stellar mass as functions of dust extinction E(B-V).  Since E(B-V) has a loose
relation with stellar mass (see Figure \ref{fig:ebmvvsms}), we study sVJLs in
two stellar mass ranges separately: ${\rm 9<log(M/M_\odot)<10}$ and ${\rm
10<log(M/M_\odot)<11}$.  We plot the cumulative curves of number (top panel),
SFR (middle), and stellar mass (bottom) of our S/N$>$10 and S/N$>$20 sVJL
samples in Figure \ref{fig:ebmvcurve}, together with the curves of the U-band
dropout sample as a reference.
For galaxies with ${\rm 9<log(M/M_\odot)<10}$ (left column), sVJLs (in both
S/N$>$10 and S/N$>$20 samples) have similar cumulative curves with U-band
dropouts, simply because there are almost no dusty (E(B-V)$\geq$0.4) galaxies
detected in this mass range, as shown by the left panel of Figure
\ref{fig:ebmvvsms}. The situation is same for S/N$>$20 sVJLs in the ${\rm
10<log(M/M_\odot)<11}$ range (right column), as the overcut on the J-band S/N
reduces our ability to detect dusty galaxies. The significant difference comes
from the S/N$>$10 sVJLs, whose cumulative SFR curve obviously deviates from
that of other samples in the ${\rm 10<log(M/M_\odot)<11}$ range.  About 50\% of
SFR is contributed by galaxies with E(B-V)$\geq$0.4, although these dusty
galaxies only contribute about 20\% to number and 20\% to stellar mass of
galaxies in the mass range, as shown by the top right and bottom right panel of
this figure. 

\begin{figure*}[htbp]
\center{\includegraphics[scale=0.4, angle=0]{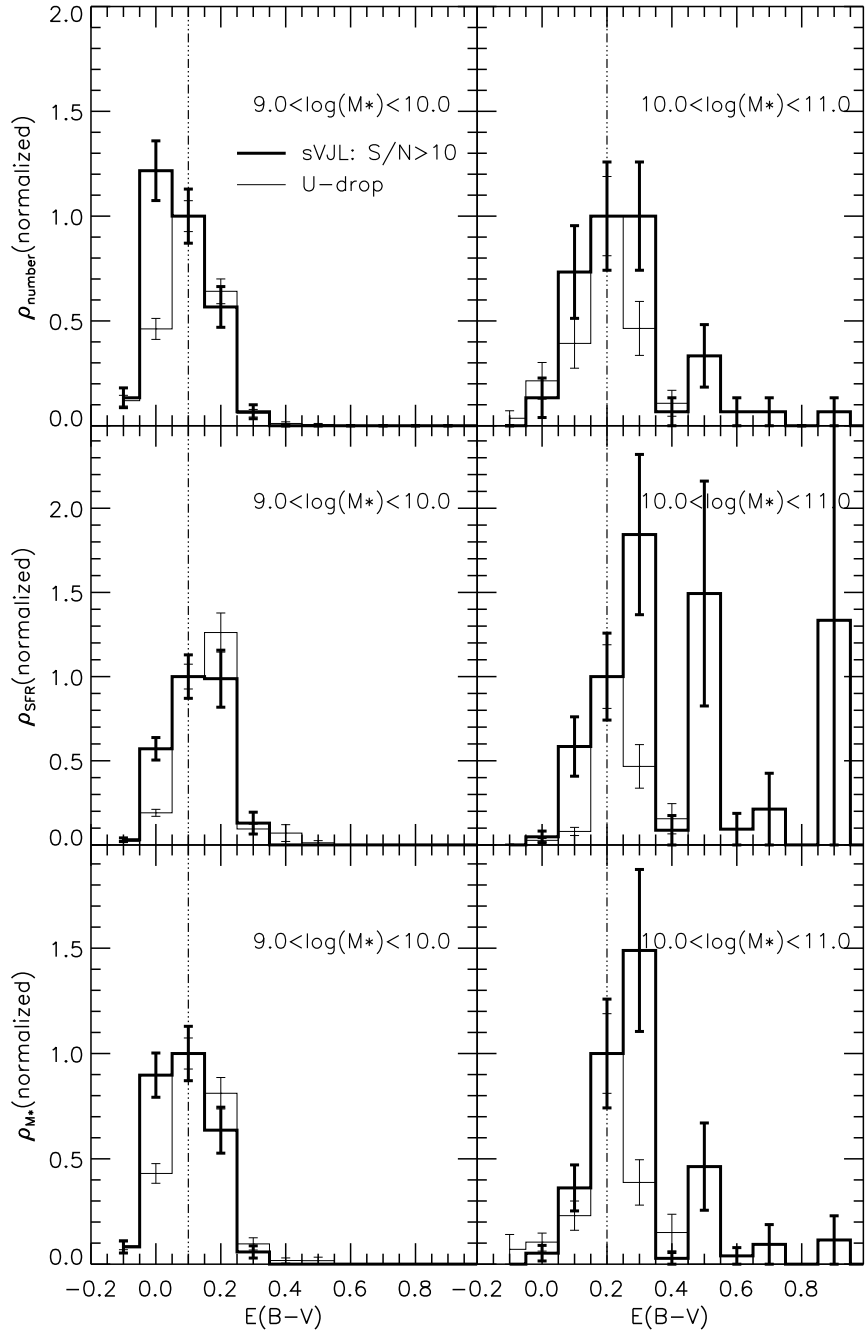}}
\caption[]{{\it Top}: comparison of normalized number density ({\it top}), SFRD
({\it middle}), and stellar mass density ({\it bottom}) of U-band dropouts (thin
lines) and S/N$>$10 sVJLs (thick lines), as a function of E(B-V). All densities
are normalized to E(B-V)=0.1 for the stellar mass bin of ${\rm
9.0<log(M_{star}/M_\odot)<10.0}$ and to E(B-V)=0.2 for ${\rm
10.0<log(M_{star}/M_\odot)<11.0}$ (dotted-dashed lines). Error bars in top
panels show the Poisson errors, while those in other panels show how the
Poisson error propagates into each quantity by assuming an average 
stellar mass and SFR for each galaxy.
\label{fig:ebmvcurve2}}
\vspace{-0.2cm}
\end{figure*}

An alternative way to evaluate the importance of dusty SFGs that are selected
by our sVJL method is to compare their contributions (on number, SFR, and
stellar mass densities) to those of U-band dropouts. The comparison again
relies on the accurate measurements of the incompleteness of the two selection
methods, but can be compromised through the following way. We choose a certain
population of galaxies that is highly completely selected by both methods so
that its three densities measured with both samples should be same even when no
correction on incompleteness is applied to this population. We then normalized
the densities of other populations in both samples to those of this population.
The normalized density distributions hence show the relative contributions of
each different populations in the two samples. We choose galaxies whose E(B-V)s
are within $\pm 0.05$ of the median E(B-V) of each sample as the "complete"
sub-sample.  In the stellar mass bin of ${\rm 9.0<log(M_{star}/M_\odot)<10.0}$,
this sub-sample consists of galaxies with ${\rm 0.05<E(B-V)<0.15}$ for both
U-band dropouts and sVJLs, while in the ${\rm 10.0<log(M_{star}/M_\odot)<11.0}$
bin, galaxies with ${\rm 0.15<E(B-V)<0.25}$.

The comparisons of normalized number density ({\it top}), SFRD ({\it middle}),
and stellar mass density ({\it bottom}) of U-band dropouts (thin lines) and
S/N$>$10 sVJLs (thick lines) as a function of E(B-V) are shown in Figure
\ref{fig:ebmvcurve2}. The same information of Figure \ref{fig:ebmvcurve}, that
about 20\% to 30\% of number density and about 50\% of stellar mass and SFR
densities in sVJLs at high mass end (${\rm 10.0<log(M_{star}/M_\odot)<11.0}$)
are contributed by galaxies with E(B-V)$>$0.3, can be inferred from this
figure. However, an important point of Figure \ref{fig:ebmvcurve2} is that the
densities of low-dust (E(B-V)$<$0.3) galaxies in the two samples are quite
similar in both stellar mass bins, with an only $\sim10$\% excess from the sVJL
sample, which demonstrates that although the cumulative distributions are
different in the high-mass end of the two samples, our sVJL method has the same 
ability to select low-dust galaxies as the U-band dropout method, in terms of 
the three densities. The $\sim$50\% of contributions to stellar mass and SFR
densities of dusty SFGs in our sVJL sample are "net" contributions, instead 
of due to the possibility that low-dust galaxies are largely missed in our sVJL 
sample. 

Our results, along with some recent studies, highlight the importance of
counting SFR from dusty galaxies, which occupy the high SFR (and massive) end
in the SFR--stellar mass plane, when calculating the cosmic SFRD.  These
galaxies are usually faint or even undetected in observed UV band at z$\sim$3
and could be missed by UV only selection (e.g., Lyman break technique).
\citet{ly11} carried out a census of SFGs at z = 1--3 in the Subaru Deep Field,
where good statistics and accurate measurements of photo-z and physical
properties are enabled by a large sample ($\sim$53000 galaxies) and 20 band
(1500{$\AA$} - 2.2{$\mu$}m) photometry. They compared the selection results of
BzK, LBG, and BX/BM, and found that among z=1-2.5 galaxies in their census,
81\%--90\% of them can be selected by combining the BzK selection with one of
the UV techniques (z$\sim$2 LBG or BX and BM). What is more important, they
found that for galaxies brighter than K$>$24 AB (roughly corresponding to ${\rm
log(M/M_\odot)>10}$ for SFGs at z$\sim$2), 65\% of the star formation in them
are contributed by galaxies with E(B-V)$>$0.25, even though they are only
one-fourth of the census by number. Their results are in very good agreement
with ours, although aiming to lower redshift. \citet{yun12}
studied the rest-frame UV and optical properties of sources detected by the
deep 1.1 mm wavelength imaging of the GOODS-S by AzTEC/ASTE \citep{scott10}.
They claimed that although not all sub-mm galaxies are faint and red in their
rest-frame UV and optical bands, the majority of the AzTEC GOODS sources, which
have a median redshift of 2.6\% and 80\% of which are at z$>$2.6, are too faint
and red to have been identified in previous surveys of SFGs and are likely be
entirely missed in the current measurements of the cosmic SFRD.

\section{Passively Evolving VJL Galaxies}
\label{pvjl}

In this section, we apply Eq. \ref{eq:pvjl} to the ERS field to select PEGs at
z$\sim$3. With a concern that a high S/N threshold in rest-frame optical band
would exclude real PEGs from our sample, we tune down the threshold to S/N$>$5
in both J- and L-bands. However, we still construct samples with S/N$>$10 and 20
to provide a reference on how photometric uncertainty affects our selection
results. We find 32, 27, and 13 galaxies falling into our pVJL selection window
for S/N$>$5, 10, and 20. However, as shown in Figure \ref{fig:model}, both low-z
and high-z dusty SFGs also enter our pVJL selection window so that a fraction
of our pVJL selected galaxies may not be real passive and old galaxies, but
rather dusty SFGs. We will estimate the fraction of contamination in our pVJL
sample and discuss how to clean the sample. 

\subsection{Clean Sample}
\label{pvjl:clean}

As similar as in \S\ref{svjl:dusty}, we use MIPS 24 $\mu$m flux to help
identify contamination in our pVJL sample. Galaxies whose observed 24 $\mu$m
fluxes are 3$\sigma$ higher than the prediction of a dust-free
passively evolving model (SSP with age of 2 Gyr) with the same redshift and
stellar mass are considered as contaminating dusty galaxies, because their 24
$\mu$m fluxes cannot be explained by pure stellar emission and hence are
dominated by dust emission.  The same issue we face here is again the confusion
and crowding of MIPS 24 $\mu$m image. We repeat the same test in Sec.
\ref{svjl:dusty} to compare PSF-fitting and aperture corrected fluxes. 
We use the aperture corrected fluxes for sources whose PSF-fitting fluxes are
larger than the 1$\sigma$ confidence level of their aperture corrected fluxes,
and use the PSF-fitting fluxes for other sources.

Comparing the observed 24 $\mu$m fluxes of our pVJLs with stellar models, we
find the contamination fraction of  59\%, 59\%, and 77\% for samples with S/N
cuts of 5, 10 and 20. The fraction does not decrease with the increase of S/N
thresholds, suggesting that simply increasing the S/N cuts cannot help clean
our pVJL sample. This is because such contamination is due to the intrinsic
deficit of our selection method (as shown by the left panel of Figure
\ref{fig:model}, where a few tracks of dusty SFGs also enter our pVJL selection
window) rather than due to photometric uncertainty.  Moreover, the fraction of
contamination is very high in all samples. This is not surprising though,
because the number density of PEGs is expected to be low at such high redshift
so that a small absolute number of contamination can occupy a relatively large
fraction of the sample.

An additional condition must be applied to remove contamination from our pVJL
sample. Although observations at longer wavelength, such as MIPS and Herschel
data, can readily help identify the contamination of dusty galaxies, we attempt
to restrict our selection criterion to using only V-, J- and L-band information
so that the method can be easily applied to large surveys where deep
observations at longer wavelengths may not be available. What is more important
is that using only the three-band information enables a relatively easy
multi-wavelength Monte Carlo simulation, which is essential to understand the
systematics and bias of our selections. In this study, longer wavelength
observations are only used to help calibrate and optimize our selection method.

A possible way to clean the sample is to examine the rest-frame optical size of
galaxies. \citet{cassata11} show that the fraction of compact galaxies in PEG
samples increases with redshift. At z$\sim$2, about 70\% of PEGs are compact.
Extrapolating their relation to z$\sim$3, we expect more than 90\% of PEGs to
have small size. If this expectation is true, galaxies with no 24 $\mu$m
detection is low should tend to have small radius and vice verse. 
 
\begin{figure}[htbp]
\center{\includegraphics[scale=0.45, angle=0]{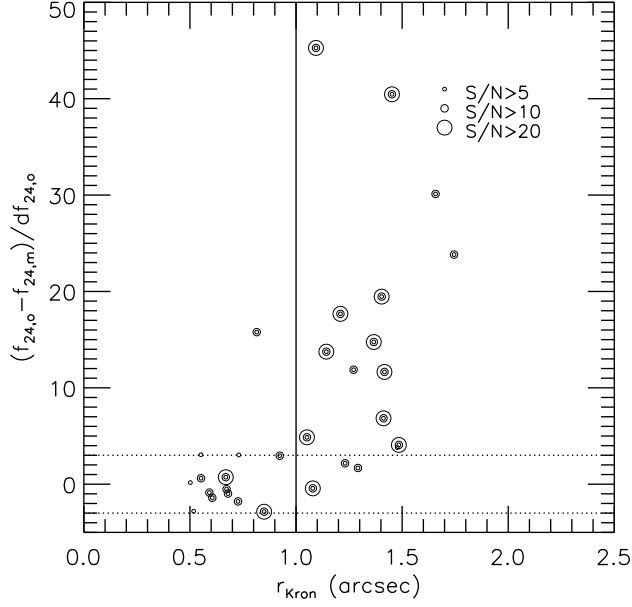}}
\caption[]{Deviation of MIPS 24 $\mu$m fluxes of pVJLs from the predication
of pure stellar emission as a function of Kron radius. Galaxies with different
S/N cuts are shown with different point sizes, as labels show. The dotted lines
show $\pm 3\sigma$ deviation from the prediction, and the solid line shows the
extra criterion (${\rm r_{Kron,J}<}$1\arcsec\ ) that we add to our pVJL 
selection method. \\
\label{fig:mipssize}}
\vspace{-0.2cm}
\end{figure}

Figure \ref{fig:mipssize} confirms our speculation by showing the relation
between the significance of 24 $\mu$m flux and J-band Kron radius. In the
S/N$>$5 sample, 85\% of galaxies whose MIPS 24 $\mu$m fluxes are within
3$\sigma$ deviation of a pure passive stellar emission have J-band Kron radius
less than 1\arcsec. On the other side, 84\% of galaxies with significant 24
$\mu$m fluxes, an indicator of dust emission, are larger than 1\arcsec\  in
terms of Kron radius. This interesting finding of the relation between MIPS 24
$\mu$m fluxes and galaxy sizes is a reflection of the size--star formation
relation of massive galaxies at $z\gtrsim2$ \citep[e.g.,][]{zirm07,toft09} and
suggests that using size can effectively distinguish real PEGs from dusty SFGs.
Moreover, in the small size (${\rm r_{Kron}<1}$\arcsec\ ) sample with S/N$>$5,
only 2 out of 14 galaxies have significant 24 $\mu$m fluxes. Therefore, based
on the high efficiency and low contamination of using small size to select
PEGs, we add the condition ${\rm r_{Kron, J}<}$1\arcsec\ to our pVJL criterion
(Eq.  \ref{eq:pvjl}).  After this extra condition being applied, our samples
now contain 14, 10, and 2 galaxies with S/N cuts of 5, 10, and 20. And the
contamination level is reduced to 14\%, 10\%, and 0\% in the three samples.

\begin{table*}[htbp]
\begin{minipage}[center]{\textwidth}\footnotesize
\caption{Passively Evolving Candidates at z$>$3 \label{tb:pvjlz3}}
\begin{tabular}{cccccccccc}
\hline\hline
ID & R.A. & Decl. & Photo-z & E(B-V) & Z & Age & $\tau$ & ${\rm M_{star}}$ & SFR \\
 & J2000 & J2000 & & & & Gyr & Gyr & Log(M*/M$_\odot$) & M$_\odot$/yr \\
\hline
2318 & 53.07387680 & -27.72217050 & 3.43 & 0.00 &   0.004 &   1.00  &  0.1  &  10.54 & 0.04 \\
2414 & 52.99881320 & -27.72097790 & 3.08 & 0.00 &   0.020 &   0.80  &  0.1  &  10.56 & 0.27 \\
2454 & 53.06628720 & -27.72043590 & 3.35 & 0.00 &   0.020 &   0.80  &  0.1  &  10.28 & 0.14 \\
3222 & 53.10302370 & -27.71234920 & 4.52 & 0.65 &   0.004 &   0.02  & 99.99  &  10.55 & 1.87E+03 \\
5218 & 53.17444360 & -27.69261340 & 4.56 & 0.05 &   0.050 &   0.50  &  0.1  &  10.46 & 4.33 \\
8124\footnote{The K-band image of this source is very faint so that TFIT likely has difficulty to measure reliable photometry for it. TFIT measures a negative flux with a large error bar. We carried out an aperture photometry with the aperture size of 1.0\arcsec\  and got a flux of 0.23$\pm$0.11 $\mu$Jy. This is broadly consistent with the prediction of the best-fit SED (solid line in Fig. \ref{fig:passsed}). We note that we did not include the K-band in the SED-fitting because of the negative TFIT flux. We also note that the marginal (1.3$\sigma$) detection of the source in K-band is somehow due to the lower sensitivity of the K-band image in this tile. The 5$\sigma$ limiting magnitude of this tile is 24.28 AB, while its 1$\sigma$ limiting magnitude is 26.03 AB. Our aperture photometry (0.23 $\mu$Jy, namely 25.50 AB) is broadly consistent with an about 2$\sigma$ detection.} & 53.14818030 & -27.71810980 & 4.81 & 0.00 &   0.050 &   0.50  &  0.1  &  10.21 & 2.44 \\
\hline
\end{tabular}
\end{minipage}
\end{table*}

\begin{figure*}[htbp]
\center{\includegraphics[scale=0.7, angle=0]{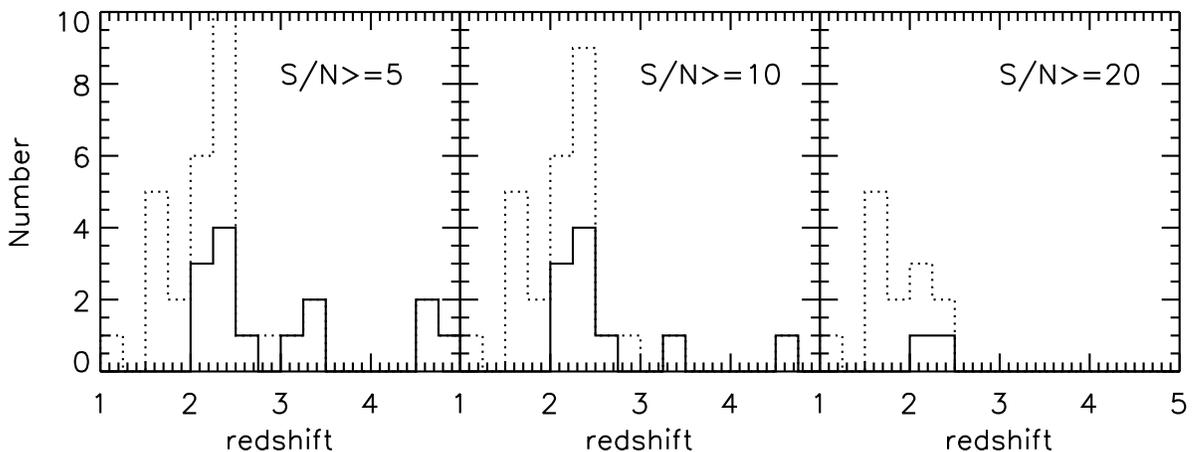}}
\caption[]{Redshift distributions of pVJL selected galaxies. Different panels
show cases with different S/N cuts, as labels show. Dotted lines show the
distributions of galaxies selected only through Eq. \ref{eq:pvjl}, while solid
lines show the distribution of galaxies that satisfy both Eq. \ref{eq:pvjl} and
the size criterion (${\rm r_{Kron}}<$1\arcsec\ ). \\
\label{fig:pvjlzdist}}
\vspace{-0.2cm}
\end{figure*}

The redshift distributions of galaxies in our final pVJL samples (solid lines)
are shown in Figure \ref{fig:pvjlzdist}. Comparison between samples with (solid
lines) or without (dotted lines) the size criterion shows the efficiency of
the additional size criterion on removing contamination from low redshift 
(z$<$2.0). The redshift distribution of our final sample peaks around
z$\sim$2.5 regardless the applied S/N cut. This distribution is a little lower
than our expectation (z$\sim$3.0) but consistent with our previous analysis
based on the color of stellar population synthetic models. As shown by the
dotted line with squares in the left panel of Figure \ref{fig:model}, the track
of SSP galaxy with t=1.0 Gyr begins to enter our pVJL selection window at
z$\sim$2.  Although the track stays in our pVJL selection window at higher
redshift, the number density of PEGs is expected to
decline with redshift.  As a result, it is not surprising that the redshift
distribution peaks at a point where the number density of galaxies is still
high and the photometric uncertainty cannot easily scatter galaxies out of the
selection window. 

\subsection{Passively Evolving Galaxies at z$>$3?}
\label{pvjl:highz} 

Recently, PEGs are occasionally found at z$>$3
\citep[e.g.,][]{mancini09,marchesini10}.  These galaxies contain important
information of when and how galaxies stopped their star formation activity.
Their number density, or even their existence itself, can set strong
constraints on current theories of galaxy formation and evolution. Six galaxies
in our S/N$>$5 sample are at z$>$3.  Although they do not enter our S/N$>$20
sample because of the low S/N of their rest-frame optical photometry, it is
still intriguing to study their physical properties and examine if they are
real PEGs at z$>$3.

Table \ref{tb:pvjlz3} summarizes the best-fit parameters of the six high-z PEG
candidates.  The ages of five galaxies are significantly (at least five times)
older than their characteristic star-formation time-scale (${\rm \tau}$),
suggesting that they have already passed their star-formation peaks and become
quiescent. Only one galaxy (ID 3222) is fitted as a dusty star-burst galaxy
with SFR $>$ 1000 M$_\odot$/yr.  Although the best-fit parameters support the
passive natures of the majority of our candidates, the SED-fitting procedure,
which only uses the rest-frame UV to NIR data, suffers from the age--dust
degeneracy and is hence not capable of perfectly distinguishing dusty
star-forming and old populations. If we assume that these galaxies are forming
stars and that their red rest-frame UV colors are caused by dust obscuration
rather than old stellar populations, their E(B-V)s and obscuration corrected
SFRs measured from their rest-frame UV continuum slopes would be much higher
than the SED-fitting derived values, with all E(B-V)s$>$0.3 and SFR on average
a few hundred times higher than the best SED-fitting values. Such high E(B-V)s
and SFRs together suggest that these galaxies should have
significant dust emission exists in longer wavelength (e.g., rest-frame IR and
sub-mm), where dust emission dominates the radiative spectrum, if their dusty
star-forming nature is true. 

\begin{figure*}[htbp]
\center{\includegraphics[scale=0.6, angle=0]{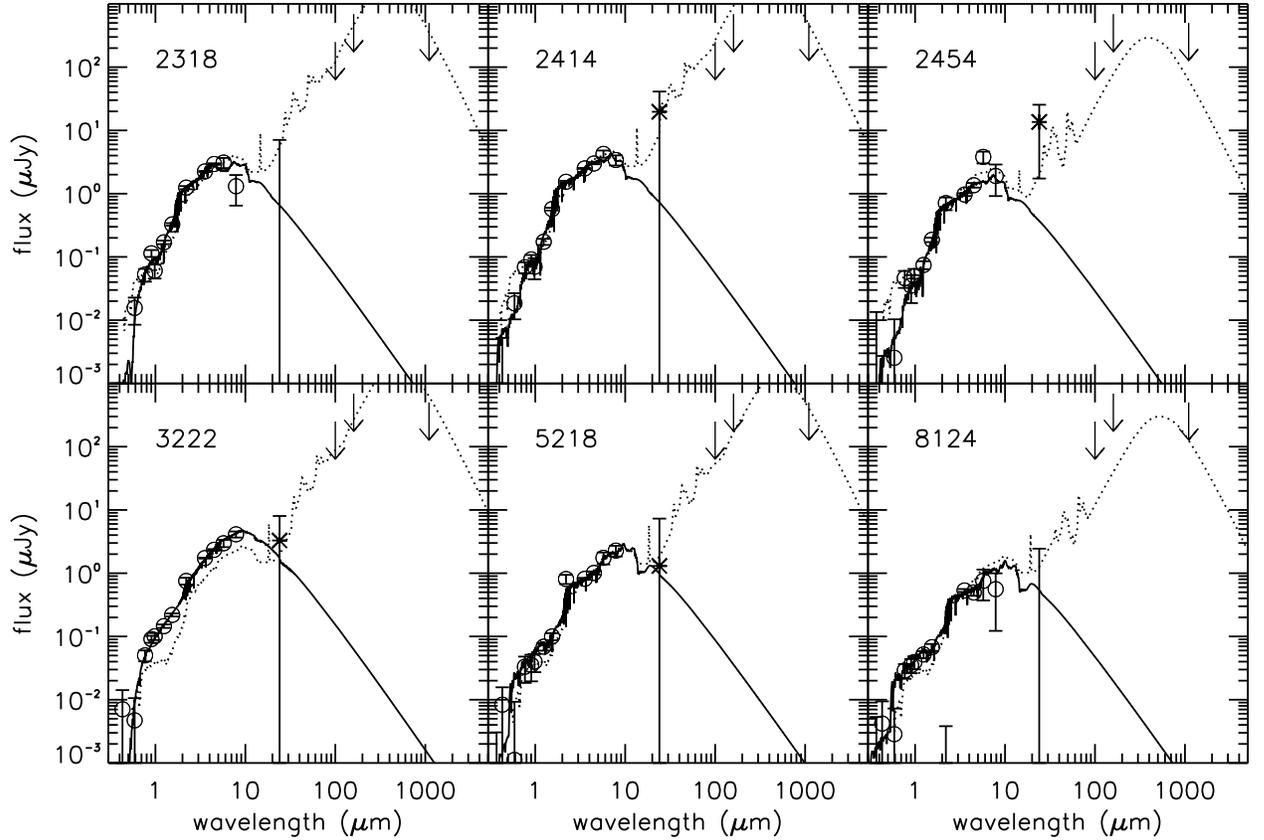}}
\caption[]{Observed and best-fit SEDs for six galaxies at z$>3$ in our
S/N$>$5 pVJL sample. Open circles with error bars are fluxes and their
uncertainties that are used for SED-fitting. Stars with error bars show the
MIPS 24 $\mu$m fluxes and 3$\sigma$ uncertainties. Arrows show the 1$\sigma$
detection limits of GOODS-{\it Herschel} (PI Elbaz) 100 and 160 $\mu$m and AzTEC 1.1 mm
images \citep{scott10}. MIPS, Hershel and AzTEC points are not used for
SED-fitting. The best-fit models are shown by solid lines.  Dotted lines are
reference model of SFGs of \citet{ce01}. \\
\label{fig:passsed}}
\vspace{-0.2cm}
\end{figure*}

Figure \ref{fig:passsed} shows the best-fit stellar population SEDs (solid
line) of the six galaxies. For comparison, we also plot templates of the SFGs
(dotted line) retrieved from the templates of \citet{ce01}. The star-forming
templates are not chosen by fitting to rest-frame UV and optical data to
models. Instead, we calculate the obscured SFR (total SFR minus unobscured SFR)
of these galaxies from their rest-frame UV continuum, assuming they are dusty
SFGs. Then, for each galaxy, we convert the obscured SFR to the bolometric IR
luminosity and choose the template whose bolometric IR luminosity best matches
the luminosity of the galaxy. It is interesting to find from the plot that
although we do not fit the templates to the rest-frame UV and optical data, the
templates match the data fairly well (except Galaxy 3222). The best chosen
template gives us an estimate of the fluxes from dust emission, which, if
existing, can be observed by our current MIPS 24 $\mu$m, GOODS-{\it Herschel}
(PI Elbaz) 100 and 160 $\mu$m, and AzTEC 1.1 mm observations \citep{scott10}.

At such high redshift, dust emission within the MIPS 24 $\mu$m bandpass is
still comparable to the stellar emission, as the stellar emission peak is just
a little blueward of the MIPS bandpass. As seen from the plot, in four out of
six galaxies, the MIPS 24 $\mu$m fluxes cannot help distinguish old and dusty
populations at all. In two galaxies (2414 and 2454), the observed 24 $\mu$m
fluxes lean toward the prediction of dust emission, however, the prediction of
pure stellar emission is still within the 3$\sigma$ level of the observation
and cannot be fully ruled out. 

In principle, GOODS-{\it Herschel} and AzTEC data can be effective at
distinguishing PEGs from dusty SFGs by sampling the blackbody radiation of cold
dust. Unfortunately, the detection thresholds of these surveys are so high that
the fluxes of dusty templates in Figure \ref{fig:passsed} are almost all under
their 1$\sigma$ detection limits.  Indeed, flux measurements of individual
galaxies in the AzTEC image suffer from a very low S/N, $\lesssim 1 \sigma$,
being comparable to the noise level. Due to the high detection thresholds, we
cannot conclude that if the non-detections in the AzTEC image provide a tight
constraint on the nature of our candidates. However, as shown in the figure,
the predicted dust emission from three or four dusty star-forming templates is
touching the 1$\sigma$ detection limit of these long-wavelength bands. We
expect an at least a 2$\sigma$ detection in the stacked images if {\it all} our
galaxies are dusty SFGs. In the stacked AzTEC image, we detect a signal with
S/N=1.1 in the central pixels (with size of 3\arcsec), still comparable to
noise. Such low S/N in the stacked AzTEC image suggests that at least some of
our candidates are not dusty SFGs but real PEGs at z$>$3. It also rule out our
suspicion that Galaxy 3233 has SFR over 1000 M$_\odot$/yr, as its best
SED-fitting shows in Table \ref{tb:pvjlz3}. Such huge SFR should have been
easily detected in the AzTEC image.

We also use the Chandra deep 4Ms X-ray image of CDFS to examine if 
AGN host galaxies contaminate our PEG candidates at z$>$3. None of our 
candidates is individually detected in the 4Ms Chandra catalog of \citet{xue11}.
The stacked images in both soft and hard bands show signals comparable 
to noises, with an S/N of 1.75 and 1.17, respectively. We conclude that our
PEG candidates at z$>$3 are not contaminated by AGN host galaxies.

We note that the two galaxies with the highest redshifts (5218 and 8124) have
the largest SFRs. Their best-fit SFRs are comparable to that of our Milky Way,
while their stellar masses are lower than that of Milky Way. The SSFR of these
galaxies are higher than ${\rm 10^{-11}/yr}$, the usual value used to
distinguish SFGs and PEGs. We suspect that it is possible that although these
galaxies have passed their peaks of star formation, their star formation
activity has not yet been fully ceased. They could be in a transition stage
from star forming to fully quiescent, since their rest-frame UV and optical
light is already dominated by old stellar populations. At lower redshift
(z$<$3.5), galaxies all have SSFR less than ${\rm 10^{-11}/yr}$, very well fit
to the usual criterion of PEGs. We speculate that galaxies in the universe
begin to transit from star forming to quiescent stages at z$\sim$4.5 and become
fully ceased PEGs at z$\sim$3.5. However, the fact of increasing SFR with
redshift could also be due to a selection effect of a flux-limited sample,
because SFR increases with luminosity so that galaxies with higher SFRs can be
observed out to higher redshift. Deep and large NIR band survey, such as
CANDELS, is required to observe galaxies down to a fainter luminosity (hence
lower SFR) level to provide a more accurate SED-fitting results to reveal the
secret of when galaxies began to cease their star formation.

\subsection{The Evolution of Integrated Stellar Mass Density of Passively Evolving Galaxies}
\label{pvjl:massdensity} 

The integrated stellar mass density (ISMD) of PEGs is a key parameter for
understanding the formation and evolution of the galaxies. It quantifies how
many stars have been locked in passive systems at a given cosmic epoch.
Currently, most studies on the evolution of stellar mass function and stellar
mass density focus on all (both star-forming and passively evolving) massive
galaxies at z$>$2 \citep[e.g.,][]{fontana06,marchesini09,marchesini10}.  Only
few works \citep[][]{mancini09,ilbert10,brammer11,cassata11} have been devoted to the study of
the evolution of PEGs (or quiescent galaxies) only, partly due to the
difficulty of identifying these galaxies at high redshift. However, the
evolution of the passive population only is as important as that of all
populations together, because it records when and how stars migrate from the
star-forming population to passive population, which are critical for us to
understand the physics that governs the ongoing and ceasing of star formation
activity in the universe.
 
In this section, we estimate the ISMD at 2$<$z$<$3 using our clean pVJL
samples. The precise measurement of the function should be obtained by
integrating the stellar mass function, either the analytic Schechter form or
the stepwise one. However, our small number samples (only 14 pVJLs even in the
S/N$>$5 sample) limit our ability to obtain an accurate measurement of the
stellar mass function at 2$<$z$<$3. We leave such an accurate measurement to a
forthcoming paper (Y. Guo et al.  in prep.) that employs the advantage of the
large survey area of the upcoming CANDELS. In this paper, instead, we simply
carry out a shortcut measurement of the ISMD to its first-order accuracy. 

We calculate the ISMD as follow: 
\begin{equation} 
\rho* = \frac{\int \int M N_{obs}(M,z) C(M,z) dz dM}{\int \frac{dV}{dz} dz}, 
\label{eq:massdense}
\end{equation} 
where $M$ is the stellar mass, $N_{obs}(M,z)$ is the observed
number of galaxies with stellar mass $M$ and redshift $z$, and 
$\frac{dV}{dz}$ is the
differential cosmic volume at $z$. The lower and upper limits of the integral
over $z$ are 2 and 3, while the lower limit of the integral over $M$ is ${\rm
10^{10}M_\odot}$.  $C(M,z)$ is a factor to correct the incompleteness caused by
observation and selection for galaxies with $M$ and $z$. As referred from
Figure \ref{fig:pvjlzdist}, the redshift distributions of our pVJL samples are
very well peaked around z$\sim$2.5 and have a narrow scatter. Therefore, it is
safe to assume that $C(M,z)$ is primarily dominated by $M$ and only has a weak
relation on $z$ in our sample. We choose z=2.5 for calculation $C(M,z)$ for all
pVJLs. The uncertainty induced by such an assumption is less than that induced
by the measurement of stellar mass of galaxies. Under this assumption, we place
an SSP model with age of 1 Gyr and stellar mass $M$ at z=2.5 and perturb its V-,
J- and L-band photometry using Gaussian random deviation with the variance set
equal to a photometric error that is randomly drawn from the distribution of
observed photometric uncertainties for a given magnitude of a given band in our
multi-wavelength catalog. The perturbation is repeated 1000 times and for each
time we justify whether the perturbed galaxy can be selected as a pVJL
according to our criterion, Equation \ref{eq:pvjl}, and different S/N cuts. The
factor $C(M,z)$ is defined the reciprocal of the rate of successful selections. 

\begin{figure}[htbp]
\center{\includegraphics[scale=0.45, angle=0]{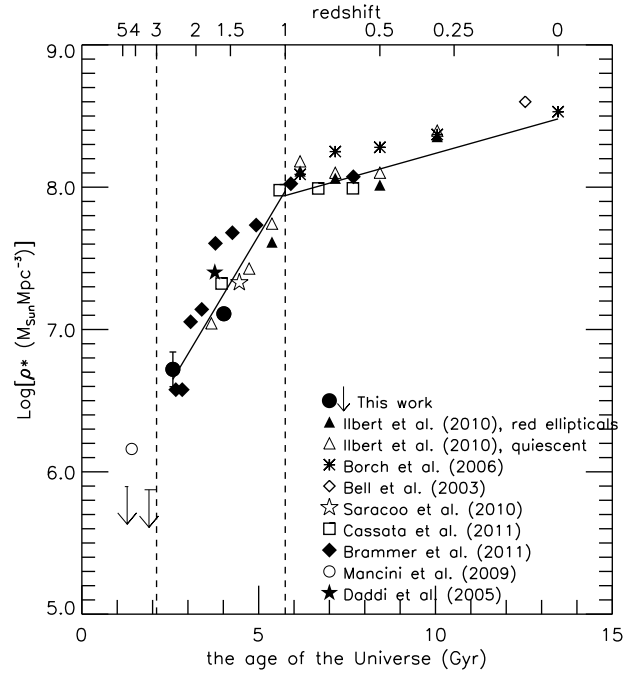}}
\caption[]{Evolution of ISMD for PEGs with 
${\rm M_{star} > 10^{10}M_\odot}$. Results of different works are shown by 
different symbols. The evolution can be schematically divided into three 
stages, as indicated by the two vertical dashed lines. The two solid lines
are the best fit to the evolution of ISMDs of 1$<$z$<$3 and z$<$1.
\label{fig:massdense}}
\vspace{-0.2cm}
\end{figure}

The ISMD of our pVJLs at 2$<$z$<$3 is shown in Figure \ref{fig:massdense},
together with measurements for lower redshift from other studies
\citep{bell03,borch06}. We measure the ISMD for each of our three samples with
different S/N cut and plot the mean and standard deviation of the three
samples. As shown by the filled point with error bars at z$\sim$2.5, the
1$\sigma$ deviation of the three samples is about 0.2 dex, comparable to the
typical stellar mass uncertainty obtained through SED-fitting at such redshift.
The small deviation also demonstrates that the incompleteness is fairly
accurately estimated for our samples so that the ISMDs of samples with
different mass limits that are induced by different S/N cuts are in very good
agreement. 

To further test the reliability of our measurement of the ISMD, we apply our
method to galaxies that are selected from GOODS-S using the passively evolving
criterion of the BzK method \citep[pBzK;][]{daddi04bzk}. The redshift
distribution of pBzKs peaks at z$\sim$1.5, where a number of measurements of
the ISMD \citep[][]{ilbert10,saracco10,brammer11,cassata11} can be used as references to test the accuracy
of our measurement. For pBzKs, we choose an SSP model with age of 2 Gyr at
z=1.5, perturb its B-, z- and K-band photometry according to photometric
uncertainties, and calculate the ISMD with the same formula as we use for
pVJLs. We also calculate the ISMD using three pBzK samples with different S/N
thresholds at z- and K-bands (S/N$>$5, 10, and 20). Thanks to the relatively
large number of galaxies in each sample, the ISMD of different pBzK samples
agree with each other better than that of different pVJL samples, with the
standard deviation less than 0.1 dex. 

Figure \ref{fig:massdense} illustrates the evolution of the ISMD of PEGs from
z$>$3 to z=0. We compile measurements of several previous studies and compare
them with our results.  In particular, we take the best-fit Schechter
parameters by \citet{bell03,borch06,ilbert10} and integrate their Schechter
functions down to mass limit ${\rm M_{*} > 10^{10} M_\odot}$. We also take the
ISMD listed in the tables of \citet{saracco10} and \citet{brammer11}. The ISMD
of PEGs at 1.3$<$z$<$2.0 in HUDF measured by \citet{daddi05} and the
measurement from one of our companion paper \citep{cassata11}
are also plotted. All adopted measurements are scaled to match our
Salpeter IMF with the following relations: ${\rm log(M_{Salpeter}) =
log(M_{Chabrier}) + 0.24}$ \citep{salimbeni09} and ${\rm log(M_{Salpeter}) =
log(M_{Kroupa}) + 0.20}$ \citep{marchesini09}. \citet{salimbeni09} also
compared stellar masses measured with different stellar synthesis libraries,
i.e., BC03, CB09, and \citet[][hereafter M05]{maraston05} and found the following
relations: ${\rm log(M_{CB09}) = log(M_{M05})}$ at all redshift; ${\rm
log(M_{CB09}) = log(M_{BC03}) + 0.20}$ at z$<$1.5; and \\
${\rm log(M_{CB09}) =
log(M_{BC03}) + 0.10}$ at 1.5$<$z$<$4. We use these relations to scale stellar
masses in other works to CB09.

Our ISMD at z$\sim$1.5 (pBzK) agrees well with that of quiescent galaxies of
\citet{ilbert10}, with difference less than 0.1 dex. However, our ISMD deviates
from other studies at z$\sim$1.5 by a few tenth dex. \citet{cassata11}
constructed a fairly complete and clean sample by using not only SSFR but also
morphology and MIPS 24 $\mu$m flux. Their ISMD should suffer the least from
incompleteness and contamination. However, their field, namely the ERS field,
is occupied by an over-dense large-scale structure at z$\sim$1.6
\citep{salimbeni09lss}, which might boost the ISMD upward.  \citet{daddi05}
used only a small sample (six galaxies) over the 12.2 arcmin$^2$ HUDF area so
that their result may suffer from both small number statistics and large cosmic
variance. The scheme of separating quiescent and dusty SFGs by
two rest-frame colors of \citet{brammer11} may induce into their quiescent
sample a fraction of dusty contamination, which could partly explain the
largest ISMD at z$\sim$1.5 measured by them. Despite the discrepancy, ISMDs at
z$\sim$1.5 measured by different authors scatter around the best fit of the
evolution of ISMD of PEGs at 1$<$z$<$3 (solid line in the plot) within
$\sim$0.3 dex, which is just slightly larger than the typical uncertainty of
deriving stellar mass through SED-fitting at this redshift ($\sim$0.2 dex).
This suggests that the uncertainty of stellar mass is the dominant source of
ISMD uncertainty, and that our simplified incompleteness correction is accurate
to the first order.

Only our work and \citet{brammer11} measure the ISMD of PEGs at z$\sim$2.5.
The ISMD of \citet{brammer11} is 0.2 dex lower than that of ours, again 
within the typical uncertainty of stellar mass. Besides the stellar mass
uncertainty, the discrepancy could also be due to the fact that
\citet{brammer11} only integrate their stellar mass function at z$>$2.0 down to
${\rm M_{*}> 10^{11} M_\odot}$, whereas the stellar mass function of PEGs is
dominated by galaxies around ${\rm M^{*}}$, typically ${\rm M_{*} = 10^{10.6}
M_\odot}$ \citep{ilbert10,peng10}.

We even extend our measurement to z$>$3, where we only have a few PEG
candidates though. We measure the ISMD for three candidates at 3$<$z$<$3.5
using an SSP model of 1 Gyr at z=3.3 and the ISMD for the other three candidates
at z$>$4 using an SSP model of 0.5 Gyr at z=4.5. Since there might be
contamination of dusty SFGs among our candidates (as discussed in
\S\ref{pvjl:highz}), the ISMDs at z$>$3 can be only treated as an upper limit.
The upper limit of ISMD at z$>$3 was also measured by \citet{mancini09}, who
found 21 z$>$3.5 quiescent candidates which are selected at IRAC 4.5 $\mu$m
channel but have no MIPS 24 $\mu$m detection in GOODS-N. As argued by them
as well as indicated by Figure \ref{fig:passsed}, the lack of 24 $\mu$m 
emission is a necessary but insufficient condition for determining a galaxy 
to be quiescent. Our upper limit of ISDM at z$>$3 is about 0.3 dex lower than 
that of theirs (also shown in Figure \ref{fig:massdense}), but still within 
the error bars of their upper limit. In this sense, the two measurements 
are not inconsistent.

\subsection{Stellar Mass Locked in Passively Evolving Galaxies}
\label{pvjl:massseg}

\begin{figure}[htbp]
\center{\includegraphics[scale=0.45, angle=0]{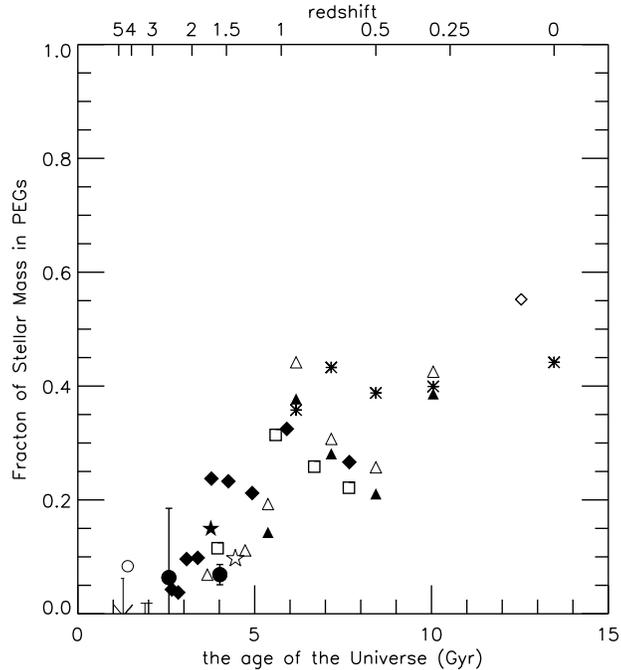}}
\caption[]{Fraction of stellar mass locked in PEGs as a function of redshift.
Results of different works are shown by different symbols. See labels in 
Figure \ref{fig:massdense} for the meanings of symbols.
\label{fig:massseg}}
\vspace{-0.2cm}
\end{figure}

The evolution of ISMD of PEGs can be easily converted into the evolution of
fraction of stellar mass in PEGs, if an underlying global
stellar mass density (GSMD) is measured for all types of galaxies. We obtain
such a measurement by fitting a linear relation to the evolution of GSMD
(log(GSMD) versus redshift) of Figure 12 of \citet{marchesini09}, which compiles
measurements of GSMD from several previous studies. We then divide the ISMD of
PEGs by the GSMD at a given redshift to obtain the fraction of stellar mass in
PEGs. 

The evolution of the mass fraction is shown in Figure \ref{fig:massseg}. At
z$>$3, the mass fraction of PEGs is less than 5\%. This fraction then increases
from 5\% to about 40\% from z=3 to z=1. However, there is large discrepancy
among measurements of the fraction at z$\sim$2, from 5\% of our study and
\citet{ilbert10} to 25\% of \citet{brammer11}. The reason of such large
discrepancy, as discussed above in the measurement of ISMD, is complicated,
possibly due to sample selection, stellar mass density measurement method,
and/or cosmic variance. A more accurate measure is needed in future to
constrain this fraction and hence the mechanisms that are responsible for
quenching the star formation activity during the peak of the cosmic SFRD.  It
should also be noted that there is about 0.2 dex deviation for the GSMD at
z$\sim$2 measured by different authors (see Figure 12 of \citet{marchesini09}).
Therefore the accurate measurement of stellar mass fraction in PEGs requires an
improvement on measuring the stellar mass densities of both PEGs and all types
of galaxies. Our work does not provide measurement on the mass fraction at
z$<$1, but we still plot the measurements of other authors for readers to
obtain a sight of the evolution trend in lower redshift.

\subsection{Discussion}
\label{pvjl:discussion}

The evolution of ISMD of PEGs can be schematically divided into three stages,
as indicated by the vertical dashed lines in Figure \ref{fig:massseg}. 
The physical mechanisms that govern
the formation and evolution of PEGs in each stage may be different.

The first stage (z$>$3) could be called as the formation (or present) stage.  The
existence of PEGs in the stage is still controversial \citep{mancini09,
marchesini10}. In our study, we find six candidates at z$>$3. Individual and
stack analysis of the sub-mm AzTEC images show that at least some
of them could be really passive. We cannot, however, draw a firm conclusion on
which one is real passive. If we treat the ISMD that we measured at z$>$3 as an
upper limit, the ISMD grows by a factor of 10, or even larger 
in 1 Gyr from z$\sim$4 to z$\sim$2.5. The existence of
PEGs of age of 1 Gyr at z$\sim$3.5 suggests that these galaxies begin to form
their stars at z$\gtrsim$5 or 6. Due to the small sample and limited
information, we cannot discuss the formation mechanism of these galaxies.
Future studies in the following two aspects would shed a light on this
question: (1) confirming or excluding the passive properties of our candidates
by using other facilities and (2) exploiting larger and deeper NIR survey
(e.g., CANDELS) to construct a large sample with a good statistics.

The second stage (1$<$z$<$3) is the rapid growth stage, during which the ISMD
of PEGs grows by a factor of 10 in 3.5 Gyr. Stars are extensively formed in or
migrated into passive systems in this period. This stage is coincident with the
broad peak of the cosmic SFRD \citep[e.g., ][]{hopkinsa04,
hopkinsa06,pg08,chary10}, suggesting that the formation of stars and the
migration of stars from star-forming systems to passive systems are happening
simultaneously during this epoch. \citet{cassata11} studied
the size distribution of PEGs during this stage and found $\sim$80\% of PEGs at
z$\sim$1.5 are compact. The mechanisms that are responsible for the rapid mass
growth of PEGs (e.g., gas-rich major merger, collapse of unstable disks and
monolithic collapse) also tend to produce passive remnants that are compact and
small with respect to local early-type galaxies.

The third stage (z$<$1) is the slow growth stage, during which the ISMD only
increases by a factor of $\sim$3 in $\sim$7 Gyr. This suggests that the
majority of PEGs has already been formed before this stage. This stage happens
when the cosmic SFRD begins to rapidly decline from its peak \citep[e.g.,
][]{hopkinsa04, hopkinsa06,pg08,chary10}, indicating that newly formed stars
may not be enough for explaining the steady growth of the passive systems from
z=1 to z=0. It requires stars that already formed in other systems to migrate
into the passive systems. A joint analysis of stellar mass density, number
density, and size distribution of PEGs in this stage by \citet{cassata11} found
that the number density increases by a factor of 1.5 from z=1 to z=0.5, while
the ISMD keeps almost constant at the same time. And the average size of PEGs
increases by a factor of about 2.5 in the same epoch.  These findings imply
that the mechanisms that increase PEGs' sizes during this redshift range would
not significantly increase their stellar masses, most likely being minor merges
and slow accretion \citep{hopkins08_2,vandokkum10}. Also, the newly formed PEGs
that increase the number density at this time would have small stellar masses
and larger sizes than those formed at z$<$1, indicating a different formation
mechanism.

\section{Summary and Conclusions}
\label{summary}

In this paper, we introduce a new method of selecting both SFGs and PEGs at
$2.3\lesssim z \lesssim 3.5$ using rest-frame UV-optical (V-J versus J-L)
colors. We apply our VJL criteria to select galaxies in the WFC3 ERS field and
study the physical properties of the selected galaxies. We also discuss the
implications of our selected galaxies on galaxy formation and evolution,
especially the contribution of dusty SFGs to the cosmic SFRD at z$\sim$3 and
the evolution of ISMD of PEGs. The paper is summarized below.

Our VJL criteria are thoroughly tested with theoretical stellar population
synthesis models and real galaxies with spectroscopic redshifts. The tests show
that our criteria for SFGs (Equation \ref{eq:svjl}, sVJL) is able to select
galaxies with constant or exponentially declining SFH
independently of their dust reddening. Our criteria for PEGs (Equation
\ref{eq:pvjl}, pVJL) can select galaxies with properties similar to SSP
models around z$\sim$2.5 and above. The tests also show that,
however, (1) the main source of contamination in our sVJL sample is the SFGs at
z$\sim$2 and z$\gtrsim$3.5 and (2) contamination in our pVJL sample is mainly
from dusty SFGs at z$\sim$2. 

We apply our sVJL criterion to the WFC3 ERS field to select 354 and 146
galaxies with J- and L-band S/N cuts greater than 10 and 20. The redshift
distribution of our sVJL sample peaks at z$\sim$2.7. However, it also has a
secondary peak around z$\sim$1.8. This secondary peak is induced by the color
uncertainty, as the power of the secondary peak decreases with the increase of
the S/N threshold.

We compare our sVJLs with Lyman break galaxies at z$\sim$3 (U-band dropouts),
assuming that the slight difference in the peak redshifts ($<z>\sim2.7$ for
sVJLs and $<z>\sim3.0$ for U-band dropouts) would not result in any significant
difference of properties of the two samples. In the ERS field, 39\% of U-band
dropouts are outside our sVJL selection window. Among the outsiders, 63\% of
them have redshift greater than 3.2, where our sVJL selection ability drops
sharply. 

Unlike the Lyman break technique, our sVJL method can select galaxies whose
(J-L) color redder than 2.0, which implies high dust extinction in these
galaxies. The measurement of E(B-V)s from the rest-frame UV continuum shows
that U-band dropouts all have E(B-V)$<$0.4, while the distribution of E(B-V)s
of sVJLs extended to E(B-V)$\sim$1.0. 

We evaluate the fraction of contamination from old galaxies in our sample of
dusty SFGs (E(B-V)$>$0.3) by comparing their observed 24 $\mu$m fluxes to that
predicted by an SSP model. We find that 18\% of our
galaxies have 24 $\mu$m fluxes that match the prediction of pure stellar
emission. The low fraction of contamination indicates that our sVJL method is
effective at selecting dusty galaxies around z$\sim$3.  

The dusty (E(B-V)$>$0.4) galaxies selected by sVJLs reside in the massive end
(${\rm M_{star} > 10^{10} M_\odot}$)of the mass distribution of sVJLs. Although
they only counts for $\sim$20\% of the number density in the mass bin ${\rm
10^{10} M_\odot < M_{star} < 10^{11} M_\odot}$, they contribute about half of
the star formation in this mass range.  In the low-mass end ${\rm 10^{9}
M_\odot < M_{star} < 10^{10} M_\odot}$, sVJLs and LBGs have no obvious
difference on their color, E(B-V), and SFR.

We also apply our criteria to the WFC3 ERS field to select PEGs at z$\sim$3.
Through a similar comparison between the observed and predicted MIPS 24 $\mu$m
fluxes, however, we find that our pVJL samples are heavily contaminated by
dusty SFGs. An additional condition is needed to clean the samples. Inspired by
the fact that the majority of PEGs at z$>$2 is compact, we require galaxies to
have a small radius (J-band Kron radius less than 1.0\arcsec) to enter our pVJL
sample. This extra criterion is proved to be able to effectively separate
passive and dusty galaxies in our samples.

The redshift distribution of our clean pVJL samples peaks at z$\sim$2.5 and
extends to z$\sim$3, and even to z$>$4 when low S/N cuts are employed. We carry
out case studies to examine the physical properties of our PEG candidates at
z$>$3. Most of these galaxies have very low SFRs derived through SED-fitting
but high SFRs derived from their rest-frame UV continuum. We try to use
observations at longer wavelengths (MIPS, Herschel and AzTEC) to break the
age--dust degeneracy and understand the nature of these galaxies. Unfortunately,
the detection limits of these long-wavelength observations are too high to help
achieve a firm conclusion. However, we find no significant detection even in
the stacked image of AzTEC, suggesting that some our candidates are real PEGs
at z$>$3. We speculate that galaxies with very low SFR, possibly a transition
stage from star-forming to passive, begin to exist at z$>$4 and PEGs begin to
exist at z$>$3. 

We estimate the ISMD of PEGs at z$\sim$2.5 by using our clean pVJL sample.  We
evaluate the incompleteness of observation and selection in a simplified way,
which is proved to be accurate to the first order by comparing our results with
other studies as well as by comparing results of samples with different S/N
cuts.  We also extend our measurement to z$>$3 and obtain a constraint on the
ISMD at z$>$3. Combining this with low-redshift observations from previous
studies, we find that the evolution of the ISMD can be divided into three
stages: (1) formation stage (z$>$3), when PEGs begin to form and their ISMD
grows by at least a factor of 10 in 1 Gyr; (2) rapid growth stage (1$<$z$<3$),
when the ISMD of PEGs grows by another factor of 10 in 3.5 Gyr; and (3) slow
growth stage (z$<$1), when the ISMD of PEGs grows by a factor of 3 in 7 Gyr. We
discuss the possible mechanisms that drive the growth in each stage.

We conclude that our new color selection criteria are effective at selecting
SFGs independent of dust reddening as well as PEGs at z$\sim$3.  This method is
less model-dependent and easier to reproduce than methods based on SED-fitting
so that it can be quickly applied to upcoming large optical and NIR surveys,
such as CANDELS, where large samples obtained through wide survey areas would
set stronger constraints and shed new light on our understanding of galaxy
formation and evolution.

\ \\

We thank Danilo Marchesini for useful discussions. We thank the anonymous
referee for constructive comments that improve this article.  Y.G., M.G., P.C.,
C.W., and S.S. acknowledge support from NASA grant HST-GO-12060, from STScI,
which is operated by AURA, Inc., under NASA contract NAS5-2655. ET acknowledges
support from contract ASI-INAF I/023/05/0 and from PD51 INFN grant. The work
presented here is partly based on observations obtained with WIRCam, a joint
project of Canada--France--Hawaii Telescope (CFHT), Taiwan, Korea, Canada,
France, at the CFHT which is operated by the National Research Council (NRC) of
Canada, the Institute National des Sciences de l’Univers of the Centre National
de la Recherche Scientifique of France, and the University of Hawaii.


\begin{thebibliography}{116}
\expandafter\ifx\csname natexlab\endcsname\relax\def\natexlab#1{#1}\fi

\bibitem[{{Adelberger} {et~al.}(2004){Adelberger}, {Steidel}, {Shapley},
  {Hunt}, {Erb}, {Reddy}, \& {Pettini}}]{adelberger04}
{Adelberger}, K.~L., {Steidel}, C.~C., {Shapley}, A.~E., {Hunt}, M.~P., {Erb},
  D.~K., {Reddy}, N.~A., \& {Pettini}, M. 2004, \apj, 607, 226

\bibitem[{{Bell} {et~al.}(2003){Bell}, {McIntosh}, {Katz}, \&
  {Weinberg}}]{bell03}
{Bell}, E.~F., {McIntosh}, D.~H., {Katz}, N., \& {Weinberg}, M.~D. 2003, \apjs,
  149, 289

\bibitem[{{Bell} {et~al.}(2004){Bell}, {Wolf}, {Meisenheimer}, {Rix}, {Borch},
  {Dye}, {Kleinheinrich}, {Wisotzki}, \& {McIntosh}}]{bell04}
{Bell}, E.~F., {Wolf}, C., {Meisenheimer}, K., {Rix}, H.-W., {Borch}, A.,
  {Dye}, S., {Kleinheinrich}, M., {Wisotzki}, L., \& {McIntosh}, D.~H. 2004,
  \apj, 608, 752

\bibitem[{{Benson} {et~al.}(2003){Benson}, {Bower}, {Frenk}, {Lacey}, {Baugh},
  \& {Cole}}]{benson03}
{Benson}, A.~J., {Bower}, R.~G., {Frenk}, C.~S., {Lacey}, C.~G., {Baugh},
  C.~M., \& {Cole}, S. 2003, \apj, 599, 38

\bibitem[{{Blain} {et~al.}(2002){Blain}, {Smail}, {Ivison}, {Kneib}, \&
  {Frayer}}]{blain02}
{Blain}, A.~W., {Smail}, I., {Ivison}, R.~J., {Kneib}, J.-P., \& {Frayer},
  D.~T. 2002, \physrep, 369, 111

\bibitem[{{Blanc} {et~al.}(2008){Blanc}, {Lira}, {Barrientos}, {Aguirre},
  {Francke}, {Taylor}, {Quadri}, {Marchesini}, {Infante}, {Gawiser}, {Hall},
  {Willis}, {Herrera}, \& {Maza}}]{blanc08}
{Blanc}, G.~A., {Lira}, P., {Barrientos}, L.~F., {Aguirre}, P., {Francke}, H.,
  {Taylor}, E.~N., {Quadri}, R., {Marchesini}, D., {Infante}, L., {Gawiser},
  E., {Hall}, P.~B., {Willis}, J.~P., {Herrera}, D., \& {Maza}, J. 2008, \apj,
  681, 1099

\bibitem[{{Blanton} {et~al.}(2001){Blanton}, {Dalcanton}, {Eisenstein},
  {Loveday}, {Strauss}, {SubbaRao}, {Weinberg}, {Anderson}, {Annis}, {Bahcall},
  {Bernardi}, {Brinkmann}, {Brunner}, {Burles}, {Carey}, {Castander},
  {Connolly}, {Csabai}, {Doi}, {Finkbeiner}, {Friedman}, {Frieman}, {Fukugita},
  {Gunn}, {Hennessy}, {Hindsley}, {Hogg}, {Ichikawa}, {Ivezi{\'c}}, {Kent},
  {Knapp}, {Lamb}, {Leger}, {Long}, {Lupton}, {McKay}, {Meiksin}, {Merelli},
  {Munn}, {Narayanan}, {Newcomb}, {Nichol}, {Okamura}, {Owen}, {Pier}, {Pope},
  {Postman}, {Quinn}, {Rockosi}, {Schlegel}, {Schneider}, {Shimasaku},
  {Siegmund}, {Smee}, {Snir}, {Stoughton}, {Stubbs}, {Szalay}, {Szokoly},
  {Thakar}, {Tremonti}, {Tucker}, {Uomoto}, {Vanden Berk}, {Vogeley},
  {Waddell}, {Yanny}, {Yasuda}, \& {York}}]{blanton01}
{Blanton}, M.~R., {Dalcanton}, J., {Eisenstein}, D., {Loveday}, J., {Strauss},
  M.~A., {SubbaRao}, M., {Weinberg}, D.~H., {Anderson}, Jr., J.~E., {Annis},
  J., {Bahcall}, N.~A., {Bernardi}, M., {Brinkmann}, J., {Brunner}, R.~J.,
  {Burles}, S., {Carey}, L., {Castander}, F.~J., {Connolly}, A.~J., {Csabai},
  I., {Doi}, M., {Finkbeiner}, D., {Friedman}, S., {Frieman}, J.~A.,
  {Fukugita}, M., {Gunn}, J.~E., {Hennessy}, G.~S., {Hindsley}, R.~B., {Hogg},
  D.~W., {Ichikawa}, T., {Ivezi{\'c}}, {\v Z}., {Kent}, S., {Knapp}, G.~R.,
  {Lamb}, D.~Q., {Leger}, R.~F., {Long}, D.~C., {Lupton}, R.~H., {McKay},
  T.~A., {Meiksin}, A., {Merelli}, A., {Munn}, J.~A., {Narayanan}, V.,
  {Newcomb}, M., {Nichol}, R.~C., {Okamura}, S., {Owen}, R., {Pier}, J.~R.,
  {Pope}, A., {Postman}, M., {Quinn}, T., {Rockosi}, C.~M., {Schlegel}, D.~J.,
  {Schneider}, D.~P., {Shimasaku}, K., {Siegmund}, W.~A., {Smee}, S., {Snir},
  Y., {Stoughton}, C., {Stubbs}, C., {Szalay}, A.~S., {Szokoly}, G.~P.,
  {Thakar}, A.~R., {Tremonti}, C., {Tucker}, D.~L., {Uomoto}, A., {Vanden
  Berk}, D., {Vogeley}, M.~S., {Waddell}, P., {Yanny}, B., {Yasuda}, N., \&
  {York}, D.~G. 2001, \aj, 121, 2358

\bibitem[{{Blanton} {et~al.}(2005){Blanton}, {Eisenstein}, {Hogg}, {Schlegel},
  \& {Brinkmann}}]{blanton05}
{Blanton}, M.~R., {Eisenstein}, D., {Hogg}, D.~W., {Schlegel}, D.~J., \&
  {Brinkmann}, J. 2005, \apj, 629, 143

\bibitem[{{Blanton} {et~al.}(2003){Blanton}, {Hogg}, {Bahcall}, {Brinkmann},
  {Britton}, {Connolly}, {Csabai}, {Fukugita}, {Loveday}, {Meiksin}, {Munn},
  {Nichol}, {Okamura}, {Quinn}, {Schneider}, {Shimasaku}, {Strauss}, {Tegmark},
  {Vogeley}, \& {Weinberg}}]{blanton03a}
{Blanton}, M.~R., {Hogg}, D.~W., {Bahcall}, N.~A., {Brinkmann}, J., {Britton},
  M., {Connolly}, A.~J., {Csabai}, I., {Fukugita}, M., {Loveday}, J.,
  {Meiksin}, A., {Munn}, J.~A., {Nichol}, R.~C., {Okamura}, S., {Quinn}, T.,
  {Schneider}, D.~P., {Shimasaku}, K., {Strauss}, M.~A., {Tegmark}, M.,
  {Vogeley}, M.~S., \& {Weinberg}, D.~H. 2003, \apj, 592, 819

\bibitem[{{Borch} {et~al.}(2006){Borch}, {Meisenheimer}, {Bell}, {Rix}, {Wolf},
  {Dye}, {Kleinheinrich}, {Kovacs}, \& {Wisotzki}}]{borch06}
{Borch}, A., {Meisenheimer}, K., {Bell}, E.~F., {Rix}, H.-W., {Wolf}, C.,
  {Dye}, S., {Kleinheinrich}, M., {Kovacs}, Z., \& {Wisotzki}, L. 2006, \aap,
  453, 869

\bibitem[{{Bouwens} {et~al.}(2009){Bouwens}, {Illingworth}, {Franx}, {Chary},
  {Meurer}, {Conselice}, {Ford}, {Giavalisco}, \& {van Dokkum}}]{bouwens09}
{Bouwens}, R.~J., {Illingworth}, G.~D., {Franx}, M., {Chary}, R.-R., {Meurer},
  G.~R., {Conselice}, C.~J., {Ford}, H., {Giavalisco}, M., \& {van Dokkum}, P.
  2009, \apj, 705, 936

\bibitem[{{Bower} {et~al.}(2006){Bower}, {Benson}, {Malbon}, {Helly}, {Frenk},
  {Baugh}, {Cole}, \& {Lacey}}]{bower06}
{Bower}, R.~G., {Benson}, A.~J., {Malbon}, R., {Helly}, J.~C., {Frenk}, C.~S.,
  {Baugh}, C.~M., {Cole}, S., \& {Lacey}, C.~G. 2006, \mnras, 370, 645

\bibitem[{{Brammer} {et~al.}(2011){Brammer}, {Whitaker}, {van Dokkum},
  {Marchesini}, {Franx}, {Kriek}, {Labb{\'e}}, {Lee}, {Muzzin}, {Quadri},
  {Rudnick}, \& {Williams}}]{brammer11}
{Brammer}, G.~B., {Whitaker}, K.~E., {van Dokkum}, P.~G., {Marchesini}, D.,
  {Franx}, M., {Kriek}, M., {Labb{\'e}}, I., {Lee}, K.-S., {Muzzin}, A.,
  {Quadri}, R.~F., {Rudnick}, G., \& {Williams}, R. 2011, \apj, 739, 24

\bibitem[{{Bruzual} \& {Charlot}(2003)}]{bc03}
{Bruzual}, G., \& {Charlot}, S. 2003, \mnras, 344, 1000

\bibitem[{{Calzetti} {et~al.}(2000){Calzetti}, {Armus}, {Bohlin}, {Kinney},
  {Koornneef}, \& {Storchi-Bergmann}}]{calzetti00}
{Calzetti}, D., {Armus}, L., {Bohlin}, R.~C., {Kinney}, A.~L., {Koornneef}, J.,
  \& {Storchi-Bergmann}, T. 2000, \apj, 533, 682

\bibitem[{{Calzetti} {et~al.}(1994){Calzetti}, {Kinney}, \&
  {Storchi-Bergmann}}]{calzetti94}
{Calzetti}, D., {Kinney}, A.~L., \& {Storchi-Bergmann}, T. 1994, \apj, 429, 582

\bibitem[{{Calzetti} {et~al.}(1997){Calzetti}, {Meurer}, {Bohlin}, {Garnett},
  {Kinney}, {Leitherer}, \& {Storchi-Bergmann}}]{calzetti97}
{Calzetti}, D., {Meurer}, G.~R., {Bohlin}, R.~C., {Garnett}, D.~R., {Kinney},
  A.~L., {Leitherer}, C., \& {Storchi-Bergmann}, T. 1997, \aj, 114, 1834

\bibitem[{{Cameron} {et~al.}(2011){Cameron}, {Carollo}, {Oesch}, {Bouwens},
  {Illingworth}, {Trenti}, {Labb{\'e}}, \& {Magee}}]{cameron11}
{Cameron}, E., {Carollo}, C.~M., {Oesch}, P.~A., {Bouwens}, R.~J.,
  {Illingworth}, G.~D., {Trenti}, M., {Labb{\'e}}, I., \& {Magee}, D. 2011,
  \apj, 743, 146

\bibitem[{{Cassata} {et~al.}(2011){Cassata}, {Giavalisco}, {Guo}, {Renzini},
  {Ferguson}, {Koekemoer}, {Salimbeni}, {Scarlata}, {Grogin}, {Conselice},
  {Dahlen}, {Lotz}, {Dickinson}, \& {Lin}}]{cassata11}
{Cassata}, P., {Giavalisco}, M., {Guo}, Y., {Renzini}, A., {Ferguson}, H.,
  {Koekemoer}, A.~M., {Salimbeni}, S., {Scarlata}, C., {Grogin}, N.~A.,
  {Conselice}, C.~J., {Dahlen}, T., {Lotz}, J.~M., {Dickinson}, M., \& {Lin},
  L. 2011, \apj, 743, 96

\bibitem[{{Chapman} {et~al.}(2003){Chapman}, {Blain}, {Ivison}, \&
  {Smail}}]{chapman03}
{Chapman}, S.~C., {Blain}, A.~W., {Ivison}, R.~J., \& {Smail}, I.~R. 2003,
  \nat, 422, 695

\bibitem[{{Chapman} {et~al.}(2005){Chapman}, {Blain}, {Smail}, \&
  {Ivison}}]{chapman05}
{Chapman}, S.~C., {Blain}, A.~W., {Smail}, I., \& {Ivison}, R.~J. 2005, \apj,
  622, 772

\bibitem[{{Chary} \& {Elbaz}(2001)}]{ce01}
{Chary}, R., \& {Elbaz}, D. 2001, \apj, 556, 562

\bibitem[{{Chary} \& {Pope}(2010)}]{chary10}
{Chary}, R.-R., \& {Pope}, A. 2010, ArXiv e-prints, 1003.1731

\bibitem[{{Cimatti} {et~al.}(2003){Cimatti}, {Daddi}, {Cassata}, {Pignatelli},
  {Fasano}, {Vernet}, {Fomalont}, {Kellermann}, {Zamorani}, {Mignoli},
  {Pozzetti}, {Renzini}, {di Serego Alighieri}, {Franceschini}, {Giallongo}, \&
  {Fontana}}]{cimatti03}
{Cimatti}, A., {Daddi}, E., {Cassata}, P., {Pignatelli}, E., {Fasano}, G.,
  {Vernet}, J., {Fomalont}, E., {Kellermann}, K., {Zamorani}, G., {Mignoli},
  M., {Pozzetti}, L., {Renzini}, A., {di Serego Alighieri}, S., {Franceschini},
  A., {Giallongo}, E., \& {Fontana}, A. 2003, \aap, 412, L1

\bibitem[{{Cimatti} {et~al.}(2002){Cimatti}, {Daddi}, {Mignoli}, {Pozzetti},
  {Renzini}, {Zamorani}, {Broadhurst}, {Fontana}, {Saracco}, {Poli},
  {Cristiani}, {D'Odorico}, {Giallongo}, {Gilmozzi}, \& {Menci}}]{cimatti02}
{Cimatti}, A., {Daddi}, E., {Mignoli}, M., {Pozzetti}, L., {Renzini}, A.,
  {Zamorani}, G., {Broadhurst}, T., {Fontana}, A., {Saracco}, P., {Poli}, F.,
  {Cristiani}, S., {D'Odorico}, S., {Giallongo}, E., {Gilmozzi}, R., \&
  {Menci}, N. 2002, \aap, 381, L68

\bibitem[{{Civano} {et~al.}(2011){Civano}, {Brusa}, {Comastri}, {Elvis},
  {Salvato}, {Zamorani}, {Capak}, {Fiore}, {Gilli}, {Hao}, {Ikeda}, {Kakazu},
  {Kartaltepe}, {Masters}, {Miyaji}, {Mignoli}, {Puccetti}, {Shankar},
  {Silverman}, {Vignali}, {Zezas}, \& {Koekemoer}}]{civano11}
{Civano}, F., {Brusa}, M., {Comastri}, A., {Elvis}, M., {Salvato}, M.,
  {Zamorani}, G., {Capak}, P., {Fiore}, F., {Gilli}, R., {Hao}, H., {Ikeda},
  H., {Kakazu}, Y., {Kartaltepe}, J.~S., {Masters}, D., {Miyaji}, T.,
  {Mignoli}, M., {Puccetti}, S., {Shankar}, F., {Silverman}, J., {Vignali}, C.,
  {Zezas}, A., \& {Koekemoer}, A.~M. 2011, \apj, 741, 91

\bibitem[{{Conselice} {et~al.}(2004){Conselice}, {Grogin}, {Jogee}, {Lucas},
  {Dahlen}, {de Mello}, {Gardner}, {Mobasher}, \& {Ravindranath}}]{conselice04}
{Conselice}, C.~J., {Grogin}, N.~A., {Jogee}, S., {Lucas}, R.~A., {Dahlen}, T.,
  {de Mello}, D., {Gardner}, J.~P., {Mobasher}, B., \& {Ravindranath}, S. 2004,
  \apjl, 600, L139

\bibitem[{{Croton} {et~al.}(2006){Croton}, {Springel}, {White}, {De Lucia},
  {Frenk}, {Gao}, {Jenkins}, {Kauffmann}, {Navarro}, \& {Yoshida}}]{croton06}
{Croton}, D.~J., {Springel}, V., {White}, S.~D.~M., {De Lucia}, G., {Frenk},
  C.~S., {Gao}, L., {Jenkins}, A., {Kauffmann}, G., {Navarro}, J.~F., \&
  {Yoshida}, N. 2006, \mnras, 365, 11

\bibitem[{{Daddi} {et~al.}(2007{\natexlab{a}}){Daddi}, {Alexander},
  {Dickinson}, {Gilli}, {Renzini}, {Elbaz}, {Cimatti}, {Chary}, {Frayer},
  {Bauer}, {Brandt}, {Giavalisco}, {Grogin}, {Huynh}, {Kurk}, {Mignoli},
  {Morrison}, {Pope}, \& {Ravindranath}}]{daddi07b}
{Daddi}, E., {Alexander}, D.~M., {Dickinson}, M., {Gilli}, R., {Renzini}, A.,
  {Elbaz}, D., {Cimatti}, A., {Chary}, R., {Frayer}, D., {Bauer}, F.~E.,
  {Brandt}, W.~N., {Giavalisco}, M., {Grogin}, N.~A., {Huynh}, M., {Kurk}, J.,
  {Mignoli}, M., {Morrison}, G., {Pope}, A., \& {Ravindranath}, S.
  2007{\natexlab{a}}, \apj, 670, 173

\bibitem[{{Daddi} {et~al.}(2000){Daddi}, {Cimatti}, {Pozzetti}, {Hoekstra},
  {R{\"o}ttgering}, {Renzini}, {Zamorani}, \& {Mannucci}}]{daddi00}
{Daddi}, E., {Cimatti}, A., {Pozzetti}, L., {Hoekstra}, H., {R{\"o}ttgering},
  H.~J.~A., {Renzini}, A., {Zamorani}, G., \& {Mannucci}, F. 2000, \aap, 361,
  535

\bibitem[{{Daddi} {et~al.}(2004{\natexlab{a}}){Daddi}, {Cimatti}, {Renzini},
  {Fontana}, {Mignoli}, {Pozzetti}, {Tozzi}, \& {Zamorani}}]{daddi04bzk}
{Daddi}, E., {Cimatti}, A., {Renzini}, A., {Fontana}, A., {Mignoli}, M.,
  {Pozzetti}, L., {Tozzi}, P., \& {Zamorani}, G. 2004{\natexlab{a}}, \apj, 617,
  746

\bibitem[{{Daddi} {et~al.}(2004{\natexlab{b}}){Daddi}, {Cimatti}, {Renzini},
  {Vernet}, {Conselice}, {Pozzetti}, {Mignoli}, {Tozzi}, {Broadhurst}, {di
  Serego Alighieri}, {Fontana}, {Nonino}, {Rosati}, \& {Zamorani}}]{daddi04a}
{Daddi}, E., {Cimatti}, A., {Renzini}, A., {Vernet}, J., {Conselice}, C.,
  {Pozzetti}, L., {Mignoli}, M., {Tozzi}, P., {Broadhurst}, T., {di Serego
  Alighieri}, S., {Fontana}, A., {Nonino}, M., {Rosati}, P., \& {Zamorani}, G.
  2004{\natexlab{b}}, \apjl, 600, L127

\bibitem[{{Daddi} {et~al.}(2007{\natexlab{b}}){Daddi}, {Dickinson}, {Morrison},
  {Chary}, {Cimatti}, {Elbaz}, {Frayer}, {Renzini}, {Pope}, {Alexander},
  {Bauer}, {Giavalisco}, {Huynh}, {Kurk}, \& {Mignoli}}]{daddi07a}
{Daddi}, E., {Dickinson}, M., {Morrison}, G., {Chary}, R., {Cimatti}, A.,
  {Elbaz}, D., {Frayer}, D., {Renzini}, A., {Pope}, A., {Alexander}, D.~M.,
  {Bauer}, F.~E., {Giavalisco}, M., {Huynh}, M., {Kurk}, J., \& {Mignoli}, M.
  2007{\natexlab{b}}, \apj, 670, 156

\bibitem[{{Daddi} {et~al.}(2005){Daddi}, {Renzini}, {Pirzkal}, {Cimatti},
  {Malhotra}, {Stiavelli}, {Xu}, {Pasquali}, {Rhoads}, {Brusa}, {di Serego
  Alighieri}, {Ferguson}, {Koekemoer}, {Moustakas}, {Panagia}, \&
  {Windhorst}}]{daddi05}
{Daddi}, E., {Renzini}, A., {Pirzkal}, N., {Cimatti}, A., {Malhotra}, S.,
  {Stiavelli}, M., {Xu}, C., {Pasquali}, A., {Rhoads}, J.~E., {Brusa}, M., {di
  Serego Alighieri}, S., {Ferguson}, H.~C., {Koekemoer}, A.~M., {Moustakas},
  L.~A., {Panagia}, N., \& {Windhorst}, R.~A. 2005, \apj, 626, 680

\bibitem[{{Dahlen} {et~al.}(2010){Dahlen}, {Mobasher}, {Dickinson}, {Ferguson},
  {Giavalisco}, {Grogin}, {Guo}, {Koekemoer}, {Lee}, {Lee}, {Nonino}, {Riess},
  \& {Salimbeni}}]{dahlen10}
{Dahlen}, T., {Mobasher}, B., {Dickinson}, M., {Ferguson}, H.~C., {Giavalisco},
  M., {Grogin}, N.~A., {Guo}, Y., {Koekemoer}, A., {Lee}, K.-S., {Lee}, S.-K.,
  {Nonino}, M., {Riess}, A.~G., \& {Salimbeni}, S. 2010, \apj, 724, 425

\bibitem[{{De Lucia} {et~al.}(2006){De Lucia}, {Springel}, {White}, {Croton},
  \& {Kauffmann}}]{delucia06}
{De Lucia}, G., {Springel}, V., {White}, S.~D.~M., {Croton}, D., \&
  {Kauffmann}, G. 2006, \mnras, 366, 499

\bibitem[{{Dunlop} {et~al.}(2007){Dunlop}, {Cirasuolo}, \& {McLure}}]{dunlop07}
{Dunlop}, J.~S., {Cirasuolo}, M., \& {McLure}, R.~J. 2007, \mnras, 376, 1054

\bibitem[{{Fioc} \& {Rocca-Volmerange}(1997)}]{pegase}
{Fioc}, M., \& {Rocca-Volmerange}, B. 1997, \aap, 326, 950

\bibitem[{{Fontana} {et~al.}(2004){Fontana}, {Pozzetti}, {Donnarumma},
  {Renzini}, {Cimatti}, {Zamorani}, {Menci}, {Daddi}, {Giallongo}, {Mignoli},
  {Perna}, {Salimbeni}, {Saracco}, {Broadhurst}, {Cristiani}, {D'Odorico}, \&
  {Gilmozzi}}]{fontana04}
{Fontana}, A., {Pozzetti}, L., {Donnarumma}, I., {Renzini}, A., {Cimatti}, A.,
  {Zamorani}, G., {Menci}, N., {Daddi}, E., {Giallongo}, E., {Mignoli}, M.,
  {Perna}, C., {Salimbeni}, S., {Saracco}, P., {Broadhurst}, T., {Cristiani},
  S., {D'Odorico}, S., \& {Gilmozzi}, R. 2004, \aap, 424, 23

\bibitem[{{Fontana} {et~al.}(2006){Fontana}, {Salimbeni}, {Grazian},
  {Giallongo}, {Pentericci}, {Nonino}, {Fontanot}, {Menci}, {Monaco},
  {Cristiani}, {Vanzella}, {de Santis}, \& {Gallozzi}}]{fontana06}
{Fontana}, A., {Salimbeni}, S., {Grazian}, A., {Giallongo}, E., {Pentericci},
  L., {Nonino}, M., {Fontanot}, F., {Menci}, N., {Monaco}, P., {Cristiani}, S.,
  {Vanzella}, E., {de Santis}, C., \& {Gallozzi}, S. 2006, \aap, 459, 745

\bibitem[{{F{\"o}rster Schreiber} {et~al.}(2004){F{\"o}rster Schreiber}, {van
  Dokkum}, {Franx}, {Labb{\'e}}, {Rudnick}, {Daddi}, {Illingworth}, {Kriek},
  {Moorwood}, {Rix}, {R{\"o}ttgering}, {Trujillo}, {van der Werf}, {van
  Starkenburg}, \& {Wuyts}}]{fs04}
{F{\"o}rster Schreiber}, N.~M., {van Dokkum}, P.~G., {Franx}, M., {Labb{\'e}},
  I., {Rudnick}, G., {Daddi}, E., {Illingworth}, G.~D., {Kriek}, M.,
  {Moorwood}, A.~F.~M., {Rix}, H.-W., {R{\"o}ttgering}, H., {Trujillo}, I.,
  {van der Werf}, P., {van Starkenburg}, L., \& {Wuyts}, S. 2004, \apj, 616, 40

\bibitem[{{Franx} {et~al.}(2003){Franx}, {Labb{\'e}}, {Rudnick}, {van Dokkum},
  {Daddi}, {F{\"o}rster Schreiber}, {Moorwood}, {Rix}, {R{\"o}ttgering}, {van
  de Wel}, {van der Werf}, \& {van Starkenburg}}]{franx03}
{Franx}, M., {Labb{\'e}}, I., {Rudnick}, G., {van Dokkum}, P.~G., {Daddi}, E.,
  {F{\"o}rster Schreiber}, N.~M., {Moorwood}, A., {Rix}, H.-W.,
  {R{\"o}ttgering}, H., {van de Wel}, A., {van der Werf}, P., \& {van
  Starkenburg}, L. 2003, \apjl, 587, L79

\bibitem[{{Giavalisco}(2002)}]{giavalisco02}
{Giavalisco}, M. 2002, \araa, 40, 579

\bibitem[{{Giavalisco} {et~al.}(2004){Giavalisco}, {Dickinson}, {Ferguson},
  {Ravindranath}, {Kretchmer}, {Moustakas}, {Madau}, {Fall}, {Gardner},
  {Livio}, {Papovich}, {Renzini}, {Spinrad}, {Stern}, \&
  {Riess}}]{giavalisco04}
{Giavalisco}, M., {Dickinson}, M., {Ferguson}, H.~C., {Ravindranath}, S.,
  {Kretchmer}, C., {Moustakas}, L.~A., {Madau}, P., {Fall}, S.~M., {Gardner},
  J.~P., {Livio}, M., {Papovich}, C., {Renzini}, A., {Spinrad}, H., {Stern},
  D., \& {Riess}, A. 2004, \apjl, 600, L103

\bibitem[{{Giavalisco} {et~al.}(1996){Giavalisco}, {Steidel}, \&
  {Macchetto}}]{giavalisco96}
{Giavalisco}, M., {Steidel}, C.~C., \& {Macchetto}, F.~D. 1996, \apj, 470, 189

\bibitem[{{Glazebrook} {et~al.}(2004){Glazebrook}, {Abraham}, {McCarthy},
  {Savaglio}, {Chen}, {Crampton}, {Murowinski}, {J{\o}rgensen}, {Roth}, {Hook},
  {Marzke}, \& {Carlberg}}]{glazebrook04}
{Glazebrook}, K., {Abraham}, R.~G., {McCarthy}, P.~J., {Savaglio}, S., {Chen},
  H.-W., {Crampton}, D., {Murowinski}, R., {J{\o}rgensen}, I., {Roth}, K.,
  {Hook}, I., {Marzke}, R.~O., \& {Carlberg}, R.~G. 2004, \nat, 430, 181

\bibitem[{{Grazian} {et~al.}(2007){Grazian}, {Salimbeni}, {Pentericci},
  {Fontana}, {Nonino}, {Vanzella}, {Cristiani}, {de Santis}, {Gallozzi},
  {Giallongo}, \& {Santini}}]{grazian07}
{Grazian}, A., {Salimbeni}, S., {Pentericci}, L., {Fontana}, A., {Nonino}, M.,
  {Vanzella}, E., {Cristiani}, S., {de Santis}, C., {Gallozzi}, S.,
  {Giallongo}, E., \& {Santini}, P. 2007, \aap, 465, 393

\bibitem[{{Grogin} {et~al.}(2011){Grogin}, {Kocevski}, {Faber}, {Ferguson},
  {Koekemoer}, {Riess}, {Acquaviva}, {Alexander}, {Almaini}, {Ashby}, {Barden},
  {Bell}, {Bournaud}, {Brown}, {Caputi}, {Casertano}, {Cassata}, {Castellano},
  {Challis}, {Chary}, {Cheung}, {Cirasuolo}, {Conselice}, {Roshan Cooray},
  {Croton}, {Daddi}, {Dahlen}, {Dav{\'e}}, {de Mello}, {Dekel}, {Dickinson},
  {Dolch}, {Donley}, {Dunlop}, {Dutton}, {Elbaz}, {Fazio}, {Filippenko},
  {Finkelstein}, {Fontana}, {Gardner}, {Garnavich}, {Gawiser}, {Giavalisco},
  {Grazian}, {Guo}, {Hathi}, {H{\"a}ussler}, {Hopkins}, {Huang}, {Huang},
  {Jha}, {Kartaltepe}, {Kirshner}, {Koo}, {Lai}, {Lee}, {Li}, {Lotz}, {Lucas},
  {Madau}, {McCarthy}, {McGrath}, {McIntosh}, {McLure}, {Mobasher},
  {Moustakas}, {Mozena}, {Nandra}, {Newman}, {Niemi}, {Noeske}, {Papovich},
  {Pentericci}, {Pope}, {Primack}, {Rajan}, {Ravindranath}, {Reddy}, {Renzini},
  {Rix}, {Robaina}, {Rodney}, {Rosario}, {Rosati}, {Salimbeni}, {Scarlata},
  {Siana}, {Simard}, {Smidt}, {Somerville}, {Spinrad}, {Straughn}, {Strolger},
  {Telford}, {Teplitz}, {Trump}, {van der Wel}, {Villforth}, {Wechsler},
  {Weiner}, {Wiklind}, {Wild}, {Wilson}, {Wuyts}, {Yan}, \&
  {Yun}}]{candelsoverview}
{Grogin}, N.~A., {Kocevski}, D.~D., {Faber}, S.~M., {Ferguson}, H.~C.,
  {Koekemoer}, A.~M., {Riess}, A.~G., {Acquaviva}, V., {Alexander}, D.~M.,
  {Almaini}, O., {Ashby}, M.~L.~N., {Barden}, M., {Bell}, E.~F., {Bournaud},
  F., {Brown}, T.~M., {Caputi}, K.~I., {Casertano}, S., {Cassata}, P.,
  {Castellano}, M., {Challis}, P., {Chary}, R.-R., {Cheung}, E., {Cirasuolo},
  M., {Conselice}, C.~J., {Roshan Cooray}, A., {Croton}, D.~J., {Daddi}, E.,
  {Dahlen}, T., {Dav{\'e}}, R., {de Mello}, D.~F., {Dekel}, A., {Dickinson},
  M., {Dolch}, T., {Donley}, J.~L., {Dunlop}, J.~S., {Dutton}, A.~A., {Elbaz},
  D., {Fazio}, G.~G., {Filippenko}, A.~V., {Finkelstein}, S.~L., {Fontana}, A.,
  {Gardner}, J.~P., {Garnavich}, P.~M., {Gawiser}, E., {Giavalisco}, M.,
  {Grazian}, A., {Guo}, Y., {Hathi}, N.~P., {H{\"a}ussler}, B., {Hopkins},
  P.~F., {Huang}, J.-S., {Huang}, K.-H., {Jha}, S.~W., {Kartaltepe}, J.~S.,
  {Kirshner}, R.~P., {Koo}, D.~C., {Lai}, K., {Lee}, K.-S., {Li}, W., {Lotz},
  J.~M., {Lucas}, R.~A., {Madau}, P., {McCarthy}, P.~J., {McGrath}, E.~J.,
  {McIntosh}, D.~H., {McLure}, R.~J., {Mobasher}, B., {Moustakas}, L.~A.,
  {Mozena}, M., {Nandra}, K., {Newman}, J.~A., {Niemi}, S.-M., {Noeske}, K.~G.,
  {Papovich}, C.~J., {Pentericci}, L., {Pope}, A., {Primack}, J.~R., {Rajan},
  A., {Ravindranath}, S., {Reddy}, N.~A., {Renzini}, A., {Rix}, H.-W.,
  {Robaina}, A.~R., {Rodney}, S.~A., {Rosario}, D.~J., {Rosati}, P.,
  {Salimbeni}, S., {Scarlata}, C., {Siana}, B., {Simard}, L., {Smidt}, J.,
  {Somerville}, R.~S., {Spinrad}, H., {Straughn}, A.~N., {Strolger}, L.-G.,
  {Telford}, O., {Teplitz}, H.~I., {Trump}, J.~R., {van der Wel}, A.,
  {Villforth}, C., {Wechsler}, R.~H., {Weiner}, B.~J., {Wiklind}, T., {Wild},
  V., {Wilson}, G., {Wuyts}, S., {Yan}, H.-J., \& {Yun}, M.~S. 2011, \apjs,
  197, 35

\bibitem[{{Hopkins}(2004)}]{hopkinsa04}
{Hopkins}, A.~M. 2004, \apj, 615, 209

\bibitem[{{Hopkins} \& {Beacom}(2006)}]{hopkinsa06}
{Hopkins}, A.~M., \& {Beacom}, J.~F. 2006, \apj, 651, 142

\bibitem[{{Hopkins} {et~al.}(2008){Hopkins}, {Cox}, {Kere{\v s}}, \&
  {Hernquist}}]{hopkins08_2}
{Hopkins}, P.~F., {Cox}, T.~J., {Kere{\v s}}, D., \& {Hernquist}, L. 2008,
  \apjs, 175, 390

\bibitem[{{Ilbert} {et~al.}(2009){Ilbert}, {Capak}, {Salvato}, {Aussel},
  {McCracken}, {Sanders}, {Scoville}, {Kartaltepe}, {Arnouts}, {Le Floc'h},
  {Mobasher}, {Taniguchi}, {Lamareille}, {Leauthaud}, {Sasaki}, {Thompson},
  {Zamojski}, {Zamorani}, {Bardelli}, {Bolzonella}, {Bongiorno}, {Brusa},
  {Caputi}, {Carollo}, {Contini}, {Cook}, {Coppa}, {Cucciati}, {de la Torre},
  {de Ravel}, {Franzetti}, {Garilli}, {Hasinger}, {Iovino}, {Kampczyk},
  {Kneib}, {Knobel}, {Kovac}, {Le Borgne}, {Le Brun}, {F{\`e}vre}, {Lilly},
  {Looper}, {Maier}, {Mainieri}, {Mellier}, {Mignoli}, {Murayama}, {Pell{\`o}},
  {Peng}, {P{\'e}rez-Montero}, {Renzini}, {Ricciardelli}, {Schiminovich},
  {Scodeggio}, {Shioya}, {Silverman}, {Surace}, {Tanaka}, {Tasca}, {Tresse},
  {Vergani}, \& {Zucca}}]{ilbert09}
{Ilbert}, O., {Capak}, P., {Salvato}, M., {Aussel}, H., {McCracken}, H.~J.,
  {Sanders}, D.~B., {Scoville}, N., {Kartaltepe}, J., {Arnouts}, S., {Le
  Floc'h}, E., {Mobasher}, B., {Taniguchi}, Y., {Lamareille}, F., {Leauthaud},
  A., {Sasaki}, S., {Thompson}, D., {Zamojski}, M., {Zamorani}, G., {Bardelli},
  S., {Bolzonella}, M., {Bongiorno}, A., {Brusa}, M., {Caputi}, K.~I.,
  {Carollo}, C.~M., {Contini}, T., {Cook}, R., {Coppa}, G., {Cucciati}, O., {de
  la Torre}, S., {de Ravel}, L., {Franzetti}, P., {Garilli}, B., {Hasinger},
  G., {Iovino}, A., {Kampczyk}, P., {Kneib}, J.-P., {Knobel}, C., {Kovac}, K.,
  {Le Borgne}, J.~F., {Le Brun}, V., {F{\`e}vre}, O.~L., {Lilly}, S., {Looper},
  D., {Maier}, C., {Mainieri}, V., {Mellier}, Y., {Mignoli}, M., {Murayama},
  T., {Pell{\`o}}, R., {Peng}, Y., {P{\'e}rez-Montero}, E., {Renzini}, A.,
  {Ricciardelli}, E., {Schiminovich}, D., {Scodeggio}, M., {Shioya}, Y.,
  {Silverman}, J., {Surace}, J., {Tanaka}, M., {Tasca}, L., {Tresse}, L.,
  {Vergani}, D., \& {Zucca}, E. 2009, \apj, 690, 1236

\bibitem[{{Ilbert} {et~al.}(2010){Ilbert}, {Salvato}, {Le Floc'h}, {Aussel},
  {Capak}, {McCracken}, {Mobasher}, {Kartaltepe}, {Scoville}, {Sanders},
  {Arnouts}, {Bundy}, {Cassata}, {Kneib}, {Koekemoer}, {Le F{\`e}vre}, {Lilly},
  {Surace}, {Taniguchi}, {Tasca}, {Thompson}, {Tresse}, {Zamojski}, {Zamorani},
  \& {Zucca}}]{ilbert10}
{Ilbert}, O., {Salvato}, M., {Le Floc'h}, E., {Aussel}, H., {Capak}, P.,
  {McCracken}, H.~J., {Mobasher}, B., {Kartaltepe}, J., {Scoville}, N.,
  {Sanders}, D.~B., {Arnouts}, S., {Bundy}, K., {Cassata}, P., {Kneib}, J.-P.,
  {Koekemoer}, A., {Le F{\`e}vre}, O., {Lilly}, S., {Surace}, J., {Taniguchi},
  Y., {Tasca}, L., {Thompson}, D., {Tresse}, L., {Zamojski}, M., {Zamorani},
  G., \& {Zucca}, E. 2010, \apj, 709, 644

\bibitem[{{Kauffmann} {et~al.}(2003){Kauffmann}, {Heckman}, {White}, {Charlot},
  {Tremonti}, {Peng}, {Seibert}, {Brinkmann}, {Nichol}, {SubbaRao}, \&
  {York}}]{kauffmann03}
{Kauffmann}, G., {Heckman}, T.~M., {White}, S.~D.~M., {Charlot}, S.,
  {Tremonti}, C., {Peng}, E.~W., {Seibert}, M., {Brinkmann}, J., {Nichol},
  R.~C., {SubbaRao}, M., \& {York}, D. 2003, \mnras, 341, 54

\bibitem[{{Kennicutt}(1998)}]{kennicutt98}
{Kennicutt}, Jr., R.~C. 1998, \araa, 36, 189

\bibitem[{{Koekemoer} {et~al.}(2011){Koekemoer}, {Faber}, {Ferguson}, {Grogin},
  {Kocevski}, {Koo}, {Lai}, {Lotz}, {Lucas}, {McGrath}, {Ogaz}, {Rajan},
  {Riess}, {Rodney}, {Strolger}, {Casertano}, {Castellano}, {Dahlen},
  {Dickinson}, {Dolch}, {Fontana}, {Giavalisco}, {Grazian}, {Guo}, {Hathi},
  {Huang}, {van der Wel}, {Yan}, {Acquaviva}, {Alexander}, {Almaini}, {Ashby},
  {Barden}, {Bell}, {Bournaud}, {Brown}, {Caputi}, {Cassata}, {Challis},
  {Chary}, {Cheung}, {Cirasuolo}, {Conselice}, {Roshan Cooray}, {Croton},
  {Daddi}, {Dav{\'e}}, {de Mello}, {de Ravel}, {Dekel}, {Donley}, {Dunlop},
  {Dutton}, {Elbaz}, {Fazio}, {Filippenko}, {Finkelstein}, {Frazer}, {Gardner},
  {Garnavich}, {Gawiser}, {Gruetzbauch}, {Hartley}, {H{\"a}ussler},
  {Herrington}, {Hopkins}, {Huang}, {Jha}, {Johnson}, {Kartaltepe},
  {Khostovan}, {Kirshner}, {Lani}, {Lee}, {Li}, {Madau}, {McCarthy},
  {McIntosh}, {McLure}, {McPartland}, {Mobasher}, {Moreira}, {Mortlock},
  {Moustakas}, {Mozena}, {Nandra}, {Newman}, {Nielsen}, {Niemi}, {Noeske},
  {Papovich}, {Pentericci}, {Pope}, {Primack}, {Ravindranath}, {Reddy},
  {Renzini}, {Rix}, {Robaina}, {Rosario}, {Rosati}, {Salimbeni}, {Scarlata},
  {Siana}, {Simard}, {Smidt}, {Snyder}, {Somerville}, {Spinrad}, {Straughn},
  {Telford}, {Teplitz}, {Trump}, {Vargas}, {Villforth}, {Wagner}, {Wandro},
  {Wechsler}, {Weiner}, {Wiklind}, {Wild}, {Wilson}, {Wuyts}, \&
  {Yun}}]{candelshst}
{Koekemoer}, A.~M., {Faber}, S.~M., {Ferguson}, H.~C., {Grogin}, N.~A.,
  {Kocevski}, D.~D., {Koo}, D.~C., {Lai}, K., {Lotz}, J.~M., {Lucas}, R.~A.,
  {McGrath}, E.~J., {Ogaz}, S., {Rajan}, A., {Riess}, A.~G., {Rodney}, S.~A.,
  {Strolger}, L., {Casertano}, S., {Castellano}, M., {Dahlen}, T., {Dickinson},
  M., {Dolch}, T., {Fontana}, A., {Giavalisco}, M., {Grazian}, A., {Guo}, Y.,
  {Hathi}, N.~P., {Huang}, K.-H., {van der Wel}, A., {Yan}, H.-J., {Acquaviva},
  V., {Alexander}, D.~M., {Almaini}, O., {Ashby}, M.~L.~N., {Barden}, M.,
  {Bell}, E.~F., {Bournaud}, F., {Brown}, T.~M., {Caputi}, K.~I., {Cassata},
  P., {Challis}, P.~J., {Chary}, R.-R., {Cheung}, E., {Cirasuolo}, M.,
  {Conselice}, C.~J., {Roshan Cooray}, A., {Croton}, D.~J., {Daddi}, E.,
  {Dav{\'e}}, R., {de Mello}, D.~F., {de Ravel}, L., {Dekel}, A., {Donley},
  J.~L., {Dunlop}, J.~S., {Dutton}, A.~A., {Elbaz}, D., {Fazio}, G.~G.,
  {Filippenko}, A.~V., {Finkelstein}, S.~L., {Frazer}, C., {Gardner}, J.~P.,
  {Garnavich}, P.~M., {Gawiser}, E., {Gruetzbauch}, R., {Hartley}, W.~G.,
  {H{\"a}ussler}, B., {Herrington}, J., {Hopkins}, P.~F., {Huang}, J.-S.,
  {Jha}, S.~W., {Johnson}, A., {Kartaltepe}, J.~S., {Khostovan}, A.~A.,
  {Kirshner}, R.~P., {Lani}, C., {Lee}, K.-S., {Li}, W., {Madau}, P.,
  {McCarthy}, P.~J., {McIntosh}, D.~H., {McLure}, R.~J., {McPartland}, C.,
  {Mobasher}, B., {Moreira}, H., {Mortlock}, A., {Moustakas}, L.~A., {Mozena},
  M., {Nandra}, K., {Newman}, J.~A., {Nielsen}, J.~L., {Niemi}, S., {Noeske},
  K.~G., {Papovich}, C.~J., {Pentericci}, L., {Pope}, A., {Primack}, J.~R.,
  {Ravindranath}, S., {Reddy}, N.~A., {Renzini}, A., {Rix}, H.-W., {Robaina},
  A.~R., {Rosario}, D.~J., {Rosati}, P., {Salimbeni}, S., {Scarlata}, C.,
  {Siana}, B., {Simard}, L., {Smidt}, J., {Snyder}, D., {Somerville}, R.~S.,
  {Spinrad}, H., {Straughn}, A.~N., {Telford}, O., {Teplitz}, H.~I., {Trump},
  J.~R., {Vargas}, C., {Villforth}, C., {Wagner}, C.~R., {Wandro}, P.,
  {Wechsler}, R.~H., {Weiner}, B.~J., {Wiklind}, T., {Wild}, V., {Wilson}, G.,
  {Wuyts}, S., \& {Yun}, M.~S. 2011, \apjs, 197, 36

\bibitem[{{Kong} {et~al.}(2006){Kong}, {Daddi}, {Arimoto}, {Renzini},
  {Broadhurst}, {Cimatti}, {Ikuta}, {Ohta}, {da Costa}, {Olsen}, {Onodera}, \&
  {Tamura}}]{kong06}
{Kong}, X., {Daddi}, E., {Arimoto}, N., {Renzini}, A., {Broadhurst}, T.,
  {Cimatti}, A., {Ikuta}, C., {Ohta}, K., {da Costa}, L., {Olsen}, L.~F.,
  {Onodera}, M., \& {Tamura}, N. 2006, \apj, 638, 72

\bibitem[{{Laidler} {et~al.}(2007){Laidler}, {Papovich}, {Grogin}, {Idzi},
  {Dickinson}, {Ferguson}, {Hilbert}, {Clubb}, \& {Ravindranath}}]{laidler07}
{Laidler}, V.~G., {Papovich}, C., {Grogin}, N.~A., {Idzi}, R., {Dickinson}, M.,
  {Ferguson}, H.~C., {Hilbert}, B., {Clubb}, K., \& {Ravindranath}, S. 2007,
  \pasp, 119, 1325

\bibitem[{{Lane} {et~al.}(2007){Lane}, {Almaini}, {Foucaud}, {Simpson},
  {Smail}, {McLure}, {Conselice}, {Cirasuolo}, {Page}, {Dunlop}, {Hirst},
  {Watson}, \& {Sekiguchi}}]{lane07}
{Lane}, K.~P., {Almaini}, O., {Foucaud}, S., {Simpson}, C., {Smail}, I.,
  {McLure}, R.~J., {Conselice}, C.~J., {Cirasuolo}, M., {Page}, M.~J.,
  {Dunlop}, J.~S., {Hirst}, P., {Watson}, M.~G., \& {Sekiguchi}, K. 2007,
  \mnras, 379, L25

\bibitem[{{Lee} {et~al.}(2010){Lee}, {Ferguson}, {Somerville}, {Wiklind}, \&
  {Giavalisco}}]{joshualee10}
{Lee}, S., {Ferguson}, H.~C., {Somerville}, R.~S., {Wiklind}, T., \&
  {Giavalisco}, M. 2010, \apj, 725, 1644

\bibitem[{{Lejeune} {et~al.}(1997){Lejeune}, {Cuisinier}, \&
  {Buser}}]{lejeune97}
{Lejeune}, T., {Cuisinier}, F., \& {Buser}, R. 1997, \aaps, 125, 229

\bibitem[{{Lin} {et~al.}(2011){Lin}, {Dickinson}, {Jian}, {Merson}, {Baugh},
  {Scott}, {Foucaud}, {Wang}, {Yan}, {Yan}, {Cheng}, {Guo}, {Helly}, {Kirsten},
  {Koo}, {Lagos}, {Meger}, {Pope}, {Simard}, {Grogin}, {Messias}, \&
  {Wang}}]{linlihwai11}
{Lin}, L., {Dickinson}, M., {Jian}, H.-Y., {Merson}, A.~I., {Baugh}, C.~M.,
  {Scott}, D., {Foucaud}, S., {Wang}, W.-H., {Yan}, C.-H., {Yan}, H.-J.,
  {Cheng}, Y.-W., {Guo}, Y., {Helly}, J., {Kirsten}, F., {Koo}, D.~C., {Lagos},
  C.~d.~P., {Meger}, N., {Pope}, A., {Simard}, L., {Grogin}, N.~A., {Messias},
  H., \& {Wang}, S.-Y. 2011, ArXiv e-prints, 1111.2135

\bibitem[{{Ly} {et~al.}(2011){Ly}, {Malkan}, {Hayashi}, {Motohara},
  {Kashikawa}, {Shimasaku}, {Nagao}, \& {Grady}}]{ly11}
{Ly}, C., {Malkan}, M.~A., {Hayashi}, M., {Motohara}, K., {Kashikawa}, N.,
  {Shimasaku}, K., {Nagao}, T., \& {Grady}, C. 2011, \apj, 735, 91

\bibitem[{{Madau}(1995)}]{madau95}
{Madau}, P. 1995, \apj, 441, 18

\bibitem[{{Magnelli} {et~al.}(2011){Magnelli}, {Elbaz}, {Chary}, {Dickinson},
  {Le Borgne}, {Frayer}, \& {Willmer}}]{magnelli11}
{Magnelli}, B., {Elbaz}, D., {Chary}, R.~R., {Dickinson}, M., {Le Borgne}, D.,
  {Frayer}, D.~T., \& {Willmer}, C.~N.~A. 2011, \aap, 528, A35+

\bibitem[{{Mancini} {et~al.}(2009){Mancini}, {Matute}, {Cimatti}, {Daddi},
  {Dickinson}, {Rodighiero}, {Bolzonella}, \& {Pozzetti}}]{mancini09}
{Mancini}, C., {Matute}, I., {Cimatti}, A., {Daddi}, E., {Dickinson}, M.,
  {Rodighiero}, G., {Bolzonella}, M., \& {Pozzetti}, L. 2009, \aap, 500, 705

\bibitem[{{Maraston}(2005)}]{maraston05}
{Maraston}, C. 2005, \mnras, 362, 799

\bibitem[{{Maraston} {et~al.}(2010){Maraston}, {Pforr}, {Renzini}, {Daddi},
  {Dickinson}, {Cimatti}, \& {Tonini}}]{maraston10}
{Maraston}, C., {Pforr}, J., {Renzini}, A., {Daddi}, E., {Dickinson}, M.,
  {Cimatti}, A., \& {Tonini}, C. 2010, \mnras, 407, 830

\bibitem[{{Marchesini} {et~al.}(2009){Marchesini}, {van Dokkum}, {F{\"o}rster
  Schreiber}, {Franx}, {Labb{\'e}}, \& {Wuyts}}]{marchesini09}
{Marchesini}, D., {van Dokkum}, P.~G., {F{\"o}rster Schreiber}, N.~M., {Franx},
  M., {Labb{\'e}}, I., \& {Wuyts}, S. 2009, \apj, 701, 1765

\bibitem[{{Marchesini} {et~al.}(2010){Marchesini}, {Whitaker}, {Brammer}, {van
  Dokkum}, {Labb{\'e}}, {Muzzin}, {Quadri}, {Kriek}, {Lee}, {Rudnick}, {Franx},
  {Illingworth}, \& {Wake}}]{marchesini10}
{Marchesini}, D., {Whitaker}, K.~E., {Brammer}, G., {van Dokkum}, P.~G.,
  {Labb{\'e}}, I., {Muzzin}, A., {Quadri}, R.~F., {Kriek}, M., {Lee}, K.-S.,
  {Rudnick}, G., {Franx}, M., {Illingworth}, G.~D., \& {Wake}, D. 2010, \apj,
  725, 1277

\bibitem[{{McCarthy}(2004)}]{mccarthy04}
{McCarthy}, P.~J. 2004, \araa, 42, 477

\bibitem[{{Mobasher} {et~al.}(2005){Mobasher}, {Dickinson}, {Ferguson},
  {Giavalisco}, {Wiklind}, {Stark}, {Ellis}, {Fall}, {Grogin}, {Moustakas},
  {Panagia}, {Sosey}, {Stiavelli}, {Bergeron}, {Casertano}, {Ingraham},
  {Koekemoer}, {Labb{\'e}}, {Livio}, {Rodgers}, {Scarlata}, {Vernet},
  {Renzini}, {Rosati}, {Kuntschner}, {K{\"u}mmel}, {Walsh}, {Chary},
  {Eisenhardt}, {Pirzkal}, \& {Stern}}]{mobasher05}
{Mobasher}, B., {Dickinson}, M., {Ferguson}, H.~C., {Giavalisco}, M.,
  {Wiklind}, T., {Stark}, D., {Ellis}, R.~S., {Fall}, S.~M., {Grogin}, N.~A.,
  {Moustakas}, L.~A., {Panagia}, N., {Sosey}, M., {Stiavelli}, M., {Bergeron},
  E., {Casertano}, S., {Ingraham}, P., {Koekemoer}, A., {Labb{\'e}}, I.,
  {Livio}, M., {Rodgers}, B., {Scarlata}, C., {Vernet}, J., {Renzini}, A.,
  {Rosati}, P., {Kuntschner}, H., {K{\"u}mmel}, M., {Walsh}, J.~R., {Chary},
  R., {Eisenhardt}, P., {Pirzkal}, N., \& {Stern}, D. 2005, \apj, 635, 832

\bibitem[{{Nonino} {et~al.}(2009){Nonino}, {Dickinson}, {Rosati}, {Grazian},
  {Reddy}, {Cristiani}, {Giavalisco}, {Kuntschner}, {Vanzella}, {Daddi},
  {Fosbury}, \& {Cesarsky}}]{nonino09}
{Nonino}, M., {Dickinson}, M., {Rosati}, P., {Grazian}, A., {Reddy}, N.,
  {Cristiani}, S., {Giavalisco}, M., {Kuntschner}, H., {Vanzella}, E., {Daddi},
  E., {Fosbury}, R.~A.~E., \& {Cesarsky}, C. 2009, \apjs, 183, 244

\bibitem[{{Norberg} {et~al.}(2002){Norberg}, {Cole}, {Baugh}, {Frenk},
  {Baldry}, {Bland-Hawthorn}, {Bridges}, {Cannon}, {Colless}, {Collins},
  {Couch}, {Cross}, {Dalton}, {De Propris}, {Driver}, {Efstathiou}, {Ellis},
  {Glazebrook}, {Jackson}, {Lahav}, {Lewis}, {Lumsden}, {Maddox}, {Madgwick},
  {Peacock}, {Peterson}, {Sutherland}, \& {Taylor}}]{norberg02}
{Norberg}, P., {Cole}, S., {Baugh}, C.~M., {Frenk}, C.~S., {Baldry}, I.,
  {Bland-Hawthorn}, J., {Bridges}, T., {Cannon}, R., {Colless}, M., {Collins},
  C., {Couch}, W., {Cross}, N.~J.~G., {Dalton}, G., {De Propris}, R., {Driver},
  S.~P., {Efstathiou}, G., {Ellis}, R.~S., {Glazebrook}, K., {Jackson}, C.,
  {Lahav}, O., {Lewis}, I., {Lumsden}, S., {Maddox}, S., {Madgwick}, D.,
  {Peacock}, J.~A., {Peterson}, B.~A., {Sutherland}, W., \& {Taylor}, K. 2002,
  \mnras, 336, 907

\bibitem[{{Oke}(1974)}]{oke74}
{Oke}, J.~B. 1974, \apjs, 27, 21

\bibitem[{{Papovich} {et~al.}(2011){Papovich}, {Finkelstein}, {Ferguson},
  {Lotz}, \& {Giavalisco}}]{papovich11}
{Papovich}, C., {Finkelstein}, S.~L., {Ferguson}, H.~C., {Lotz}, J.~M., \&
  {Giavalisco}, M. 2011, \mnras, 412, 1123

\bibitem[{{Papovich} {et~al.}(2006){Papovich}, {Moustakas}, {Dickinson}, {Le
  Floc'h}, {Rieke}, {Daddi}, {Alexander}, {Bauer}, {Brandt}, {Dahlen}, {Egami},
  {Eisenhardt}, {Elbaz}, {Ferguson}, {Giavalisco}, {Lucas}, {Mobasher},
  {P{\'e}rez-Gonz{\'a}lez}, {Stutz}, {Rieke}, \& {Yan}}]{papovich06}
{Papovich}, C., {Moustakas}, L.~A., {Dickinson}, M., {Le Floc'h}, E., {Rieke},
  G.~H., {Daddi}, E., {Alexander}, D.~M., {Bauer}, F., {Brandt}, W.~N.,
  {Dahlen}, T., {Egami}, E., {Eisenhardt}, P., {Elbaz}, D., {Ferguson}, H.~C.,
  {Giavalisco}, M., {Lucas}, R.~A., {Mobasher}, B., {P{\'e}rez-Gonz{\'a}lez},
  P.~G., {Stutz}, A., {Rieke}, M.~J., \& {Yan}, H. 2006, \apj, 640, 92

\bibitem[{{Peng} {et~al.}(2010){Peng}, {Lilly}, {Kova{\v c}}, {Bolzonella},
  {Pozzetti}, {Renzini}, {Zamorani}, {Ilbert}, {Knobel}, {Iovino}, {Maier},
  {Cucciati}, {Tasca}, {Carollo}, {Silverman}, {Kampczyk}, {de Ravel},
  {Sanders}, {Scoville}, {Contini}, {Mainieri}, {Scodeggio}, {Kneib}, {Le
  F{\`e}vre}, {Bardelli}, {Bongiorno}, {Caputi}, {Coppa}, {de la Torre},
  {Franzetti}, {Garilli}, {Lamareille}, {Le Borgne}, {Le Brun}, {Mignoli},
  {Perez Montero}, {Pello}, {Ricciardelli}, {Tanaka}, {Tresse}, {Vergani},
  {Welikala}, {Zucca}, {Oesch}, {Abbas}, {Barnes}, {Bordoloi}, {Bottini},
  {Cappi}, {Cassata}, {Cimatti}, {Fumana}, {Hasinger}, {Koekemoer},
  {Leauthaud}, {Maccagni}, {Marinoni}, {McCracken}, {Memeo}, {Meneux}, {Nair},
  {Porciani}, {Presotto}, \& {Scaramella}}]{peng10}
{Peng}, Y., {Lilly}, S.~J., {Kova{\v c}}, K., {Bolzonella}, M., {Pozzetti}, L.,
  {Renzini}, A., {Zamorani}, G., {Ilbert}, O., {Knobel}, C., {Iovino}, A.,
  {Maier}, C., {Cucciati}, O., {Tasca}, L., {Carollo}, C.~M., {Silverman}, J.,
  {Kampczyk}, P., {de Ravel}, L., {Sanders}, D., {Scoville}, N., {Contini}, T.,
  {Mainieri}, V., {Scodeggio}, M., {Kneib}, J., {Le F{\`e}vre}, O., {Bardelli},
  S., {Bongiorno}, A., {Caputi}, K., {Coppa}, G., {de la Torre}, S.,
  {Franzetti}, P., {Garilli}, B., {Lamareille}, F., {Le Borgne}, J., {Le Brun},
  V., {Mignoli}, M., {Perez Montero}, E., {Pello}, R., {Ricciardelli}, E.,
  {Tanaka}, M., {Tresse}, L., {Vergani}, D., {Welikala}, N., {Zucca}, E.,
  {Oesch}, P., {Abbas}, U., {Barnes}, L., {Bordoloi}, R., {Bottini}, D.,
  {Cappi}, A., {Cassata}, P., {Cimatti}, A., {Fumana}, M., {Hasinger}, G.,
  {Koekemoer}, A., {Leauthaud}, A., {Maccagni}, D., {Marinoni}, C.,
  {McCracken}, H., {Memeo}, P., {Meneux}, B., {Nair}, P., {Porciani}, C.,
  {Presotto}, V., \& {Scaramella}, R. 2010, \apj, 721, 193

\bibitem[{{P{\'e}rez-Gonz{\'a}lez} {et~al.}(2008){P{\'e}rez-Gonz{\'a}lez},
  {Rieke}, {Villar}, {Barro}, {Blaylock}, {Egami}, {Gallego}, {Gil de Paz},
  {Pascual}, {Zamorano}, \& {Donley}}]{pg08}
{P{\'e}rez-Gonz{\'a}lez}, P.~G., {Rieke}, G.~H., {Villar}, V., {Barro}, G.,
  {Blaylock}, M., {Egami}, E., {Gallego}, J., {Gil de Paz}, A., {Pascual}, S.,
  {Zamorano}, J., \& {Donley}, J.~L. 2008, \apj, 675, 234

\bibitem[{{Polletta} {et~al.}(2007){Polletta}, {Tajer}, {Maraschi},
  {Trinchieri}, {Lonsdale}, {Chiappetti}, {Andreon}, {Pierre}, {Le F{\`e}vre},
  {Zamorani}, {Maccagni}, {Garcet}, {Surdej}, {Franceschini}, {Alloin},
  {Shupe}, {Surace}, {Fang}, {Rowan-Robinson}, {Smith}, \&
  {Tresse}}]{polletta07}
{Polletta}, M., {Tajer}, M., {Maraschi}, L., {Trinchieri}, G., {Lonsdale},
  C.~J., {Chiappetti}, L., {Andreon}, S., {Pierre}, M., {Le F{\`e}vre}, O.,
  {Zamorani}, G., {Maccagni}, D., {Garcet}, O., {Surdej}, J., {Franceschini},
  A., {Alloin}, D., {Shupe}, D.~L., {Surace}, J.~A., {Fang}, F.,
  {Rowan-Robinson}, M., {Smith}, H.~E., \& {Tresse}, L. 2007, \apj, 663, 81

\bibitem[{{Ranalli} {et~al.}(2003){Ranalli}, {Comastri}, \&
  {Setti}}]{ranalli03}
{Ranalli}, P., {Comastri}, A., \& {Setti}, G. 2003, \aap, 399, 39

\bibitem[{{Ravindranath} {et~al.}(2006){Ravindranath}, {Giavalisco},
  {Ferguson}, {Conselice}, {Katz}, {Weinberg}, {Lotz}, {Dickinson}, {Fall},
  {Mobasher}, \& {Papovich}}]{ravindranath06}
{Ravindranath}, S., {Giavalisco}, M., {Ferguson}, H.~C., {Conselice}, C.,
  {Katz}, N., {Weinberg}, M., {Lotz}, J., {Dickinson}, M., {Fall}, S.~M.,
  {Mobasher}, B., \& {Papovich}, C. 2006, \apj, 652, 963

\bibitem[{{Reddy} {et~al.}(2005){Reddy}, {Erb}, {Steidel}, {Shapley},
  {Adelberger}, \& {Pettini}}]{reddy05}
{Reddy}, N.~A., {Erb}, D.~K., {Steidel}, C.~C., {Shapley}, A.~E., {Adelberger},
  K.~L., \& {Pettini}, M. 2005, \apj, 633, 748

\bibitem[{{Retzlaff} {et~al.}(2010){Retzlaff}, {Rosati}, {Dickinson},
  {Vandame}, {Rit{\'e}}, {Nonino}, {Cesarsky}, \& {GOODS Team}}]{retzlaff10}
{Retzlaff}, J., {Rosati}, P., {Dickinson}, M., {Vandame}, B., {Rit{\'e}}, C.,
  {Nonino}, M., {Cesarsky}, C., \& {GOODS Team}. 2010, \aap, 511, A50+

\bibitem[{{Riguccini} {et~al.}(2011){Riguccini}, {Le Floc'h}, {Ilbert},
  {Aussel}, {Salvato}, {Capak}, {McCracken}, {Kartaltepe}, {Sanders}, \&
  {Scoville}}]{riguccini11}
{Riguccini}, L., {Le Floc'h}, E., {Ilbert}, O., {Aussel}, H., {Salvato}, M.,
  {Capak}, P., {McCracken}, H., {Kartaltepe}, J., {Sanders}, D., \& {Scoville},
  N. 2011, \aap, 534, A81

\bibitem[{{Roche} {et~al.}(2002){Roche}, {Almaini}, {Dunlop}, {Ivison}, \&
  {Willott}}]{roche02}
{Roche}, N.~D., {Almaini}, O., {Dunlop}, J., {Ivison}, R.~J., \& {Willott},
  C.~J. 2002, \mnras, 337, 1282

\bibitem[{{Roche} {et~al.}(2003){Roche}, {Dunlop}, \& {Almaini}}]{roche03}
{Roche}, N.~D., {Dunlop}, J., \& {Almaini}, O. 2003, \mnras, 346, 803

\bibitem[{{Rodighiero} {et~al.}(2007){Rodighiero}, {Cimatti}, {Franceschini},
  {Brusa}, {Fritz}, \& {Bolzonella}}]{rodighiero07}
{Rodighiero}, G., {Cimatti}, A., {Franceschini}, A., {Brusa}, M., {Fritz}, J.,
  \& {Bolzonella}, M. 2007, \aap, 470, 21

\bibitem[{{Salimbeni} {et~al.}(2009{\natexlab{a}}){Salimbeni}, {Castellano},
  {Pentericci}, {Trevese}, {Fiore}, {Grazian}, {Fontana}, {Giallongo},
  {Boutsia}, {Cristiani}, {de Santis}, {Gallozzi}, {Menci}, {Nonino}, {Paris},
  {Santini}, \& {Vanzella}}]{salimbeni09lss}
{Salimbeni}, S., {Castellano}, M., {Pentericci}, L., {Trevese}, D., {Fiore},
  F., {Grazian}, A., {Fontana}, A., {Giallongo}, E., {Boutsia}, K.,
  {Cristiani}, S., {de Santis}, C., {Gallozzi}, S., {Menci}, N., {Nonino}, M.,
  {Paris}, D., {Santini}, P., \& {Vanzella}, E. 2009{\natexlab{a}}, \aap, 501,
  865

\bibitem[{{Salimbeni} {et~al.}(2009{\natexlab{b}}){Salimbeni}, {Fontana},
  {Giallongo}, {Grazian}, {Menci}, {Pentericci}, \& {Santini}}]{salimbeni09}
{Salimbeni}, S., {Fontana}, A., {Giallongo}, E., {Grazian}, A., {Menci}, N.,
  {Pentericci}, L., \& {Santini}, P. 2009{\natexlab{b}}, in American Institute
  of Physics Conference Series, Vol. 1111, American Institute of Physics
  Conference Series, ed. {G.~Giobbi, A.~Tornambe, G.~Raimondo, M.~Limongi,
  L.~A.~Antonelli, N.~Menci, \& E.~Brocato}, 207--211

\bibitem[{{Salpeter}(1955)}]{salpeter55}
{Salpeter}, E.~E. 1955, \apj, 121, 161

\bibitem[{{Saracco} {et~al.}(2010){Saracco}, {Longhetti}, \&
  {Gargiulo}}]{saracco10}
{Saracco}, P., {Longhetti}, M., \& {Gargiulo}, A. 2010, \mnras, 408, L21

\bibitem[{{Saracco} {et~al.}(2005){Saracco}, {Longhetti}, {Severgnini}, {Della
  Ceca}, {Braito}, {Mannucci}, {Bender}, {Drory}, {Feulner}, {Hopp}, \&
  {Maraston}}]{saracco05}
{Saracco}, P., {Longhetti}, M., {Severgnini}, P., {Della Ceca}, R., {Braito},
  V., {Mannucci}, F., {Bender}, R., {Drory}, N., {Feulner}, G., {Hopp}, U., \&
  {Maraston}, C. 2005, \mnras, 357, L40

\bibitem[{{Scott} {et~al.}(2010){Scott}, {Yun}, {Wilson}, {Austermann},
  {Aguilar}, {Aretxaga}, {Ezawa}, {Ferrusca}, {Hatsukade}, {Hughes}, {Iono},
  {Giavalisco}, {Kawabe}, {Kohno}, {Mauskopf}, {Oshima}, {Perera}, {Rand},
  {Tamura}, {Tosaki}, {Velazquez}, {Williams}, \& {Zeballos}}]{scott10}
{Scott}, K.~S., {Yun}, M.~S., {Wilson}, G.~W., {Austermann}, J.~E., {Aguilar},
  E., {Aretxaga}, I., {Ezawa}, H., {Ferrusca}, D., {Hatsukade}, B., {Hughes},
  D.~H., {Iono}, D., {Giavalisco}, M., {Kawabe}, R., {Kohno}, K., {Mauskopf},
  P.~D., {Oshima}, T., {Perera}, T.~A., {Rand}, J., {Tamura}, Y., {Tosaki}, T.,
  {Velazquez}, M., {Williams}, C.~C., \& {Zeballos}, M. 2010, \mnras, 405, 2260

\bibitem[{{Steidel} {et~al.}(1999){Steidel}, {Adelberger}, {Giavalisco},
  {Dickinson}, \& {Pettini}}]{steidel99}
{Steidel}, C.~C., {Adelberger}, K.~L., {Giavalisco}, M., {Dickinson}, M., \&
  {Pettini}, M. 1999, \apj, 519, 1

\bibitem[{{Steidel} {et~al.}(2003){Steidel}, {Adelberger}, {Shapley},
  {Pettini}, {Dickinson}, \& {Giavalisco}}]{steidel03}
{Steidel}, C.~C., {Adelberger}, K.~L., {Shapley}, A.~E., {Pettini}, M.,
  {Dickinson}, M., \& {Giavalisco}, M. 2003, \apj, 592, 728

\bibitem[{{Steidel} {et~al.}(1996{\natexlab{a}}){Steidel}, {Giavalisco},
  {Dickinson}, \& {Adelberger}}]{steidel96a}
{Steidel}, C.~C., {Giavalisco}, M., {Dickinson}, M., \& {Adelberger}, K.~L.
  1996{\natexlab{a}}, \aj, 112, 352

\bibitem[{{Steidel} {et~al.}(1996{\natexlab{b}}){Steidel}, {Giavalisco},
  {Pettini}, {Dickinson}, \& {Adelberger}}]{steidel96b}
{Steidel}, C.~C., {Giavalisco}, M., {Pettini}, M., {Dickinson}, M., \&
  {Adelberger}, K.~L. 1996{\natexlab{b}}, \apjl, 462, L17+

\bibitem[{{Steidel} {et~al.}(2004){Steidel}, {Shapley}, {Pettini},
  {Adelberger}, {Erb}, {Reddy}, \& {Hunt}}]{steidel04}
{Steidel}, C.~C., {Shapley}, A.~E., {Pettini}, M., {Adelberger}, K.~L., {Erb},
  D.~K., {Reddy}, N.~A., \& {Hunt}, M.~P. 2004, \apj, 604, 534

\bibitem[{{Swinbank} {et~al.}(2006){Swinbank}, {Chapman}, {Smail}, {Lindner},
  {Borys}, {Blain}, {Ivison}, \& {Lewis}}]{swinbank06}
{Swinbank}, A.~M., {Chapman}, S.~C., {Smail}, I., {Lindner}, C., {Borys}, C.,
  {Blain}, A.~W., {Ivison}, R.~J., \& {Lewis}, G.~F. 2006, \mnras, 371, 465

\bibitem[{{Thompson} {et~al.}(1999){Thompson}, {Beckwith}, {Fockenbrock},
  {Fried}, {Hippelein}, {Huang}, {von Kuhlmann}, {Leinert}, {Meisenheimer},
  {Phleps}, {R{\"o}ser}, {Thommes}, \& {Wolf}}]{thompson99}
{Thompson}, D., {Beckwith}, S.~V.~W., {Fockenbrock}, R., {Fried}, J.,
  {Hippelein}, H., {Huang}, J.-S., {von Kuhlmann}, B., {Leinert}, C.,
  {Meisenheimer}, K., {Phleps}, S., {R{\"o}ser}, H.-J., {Thommes}, E., \&
  {Wolf}, C. 1999, \apj, 523, 100

\bibitem[{{Toft} {et~al.}(2009){Toft}, {Franx}, {van Dokkum}, {F{\"o}rster
  Schreiber}, {Labbe}, {Wuyts}, \& {Marchesini}}]{toft09}
{Toft}, S., {Franx}, M., {van Dokkum}, P., {F{\"o}rster Schreiber}, N.~M.,
  {Labbe}, I., {Wuyts}, S., \& {Marchesini}, D. 2009, \apj, 705, 255

\bibitem[{{van den Bergh} {et~al.}(2000){van den Bergh}, {Cohen}, {Hogg}, \&
  {Blandford}}]{vandenbergh00}
{van den Bergh}, S., {Cohen}, J.~G., {Hogg}, D.~W., \& {Blandford}, R. 2000,
  \aj, 120, 2190

\bibitem[{{van Dokkum} {et~al.}(2003){van Dokkum}, {F{\"o}rster Schreiber},
  {Franx}, {Daddi}, {Illingworth}, {Labb{\'e}}, {Moorwood}, {Rix},
  {R{\"o}ttgering}, {Rudnick}, {van der Wel}, {van der Werf}, \& {van
  Starkenburg}}]{vandokkum03}
{van Dokkum}, P.~G., {F{\"o}rster Schreiber}, N.~M., {Franx}, M., {Daddi}, E.,
  {Illingworth}, G.~D., {Labb{\'e}}, I., {Moorwood}, A., {Rix}, H.-W.,
  {R{\"o}ttgering}, H., {Rudnick}, G., {van der Wel}, A., {van der Werf}, P.,
  \& {van Starkenburg}, L. 2003, \apjl, 587, L83

\bibitem[{{van Dokkum} {et~al.}(2004){van Dokkum}, {Franx}, {F{\"o}rster
  Schreiber}, {Illingworth}, {Daddi}, {Knudsen}, {Labb{\'e}}, {Moorwood},
  {Rix}, {R{\"o}ttgering}, {Rudnick}, {Trujillo}, {van der Werf}, {van der
  Wel}, {van Starkenburg}, \& {Wuyts}}]{vandokkum04}
{van Dokkum}, P.~G., {Franx}, M., {F{\"o}rster Schreiber}, N.~M.,
  {Illingworth}, G.~D., {Daddi}, E., {Knudsen}, K.~K., {Labb{\'e}}, I.,
  {Moorwood}, A., {Rix}, H.-W., {R{\"o}ttgering}, H., {Rudnick}, G.,
  {Trujillo}, I., {van der Werf}, P., {van der Wel}, A., {van Starkenburg}, L.,
  \& {Wuyts}, S. 2004, \apj, 611, 703

\bibitem[{{van Dokkum} {et~al.}(2006){van Dokkum}, {Quadri}, {Marchesini},
  {Rudnick}, {Franx}, {Gawiser}, {Herrera}, {Wuyts}, {Lira}, {Labb{\'e}},
  {Maza}, {Illingworth}, {F{\"o}rster Schreiber}, {Kriek}, {Rix}, {Taylor},
  {Toft}, {Webb}, \& {Yi}}]{vandokkum06}
{van Dokkum}, P.~G., {Quadri}, R., {Marchesini}, D., {Rudnick}, G., {Franx},
  M., {Gawiser}, E., {Herrera}, D., {Wuyts}, S., {Lira}, P., {Labb{\'e}}, I.,
  {Maza}, J., {Illingworth}, G.~D., {F{\"o}rster Schreiber}, N.~M., {Kriek},
  M., {Rix}, H.-W., {Taylor}, E.~N., {Toft}, S., {Webb}, T., \& {Yi}, S.~K.
  2006, \apjl, 638, L59

\bibitem[{{van Dokkum} {et~al.}(2010){van Dokkum}, {Whitaker}, {Brammer},
  {Franx}, {Kriek}, {Labb{\'e}}, {Marchesini}, {Quadri}, {Bezanson},
  {Illingworth}, {Muzzin}, {Rudnick}, {Tal}, \& {Wake}}]{vandokkum10}
{van Dokkum}, P.~G., {Whitaker}, K.~E., {Brammer}, G., {Franx}, M., {Kriek},
  M., {Labb{\'e}}, I., {Marchesini}, D., {Quadri}, R., {Bezanson}, R.,
  {Illingworth}, G.~D., {Muzzin}, A., {Rudnick}, G., {Tal}, T., \& {Wake}, D.
  2010, \apj, 709, 1018

\bibitem[{{Wang} {et~al.}(2004){Wang}, {Malhotra}, {Rhoads}, \&
  {Norman}}]{wangjx04}
{Wang}, J.~X., {Malhotra}, S., {Rhoads}, J.~E., \& {Norman}, C.~A. 2004, \apjl,
  612, L109

\bibitem[{{Wang} {et~al.}(2010){Wang}, {Cowie}, {Barger}, {Keenan}, \&
  {Ting}}]{wangweihao10}
{Wang}, W.-H., {Cowie}, L.~L., {Barger}, A.~J., {Keenan}, R.~C., \& {Ting},
  H.-C. 2010, \apjs, 187, 251

\bibitem[{{White} \& {Rees}(1978)}]{white78}
{White}, S.~D.~M., \& {Rees}, M.~J. 1978, \mnras, 183, 341

\bibitem[{{Wiklind} {et~al.}(2008){Wiklind}, {Dickinson}, {Ferguson},
  {Giavalisco}, {Mobasher}, {Grogin}, \& {Panagia}}]{wiklind08}
{Wiklind}, T., {Dickinson}, M., {Ferguson}, H.~C., {Giavalisco}, M.,
  {Mobasher}, B., {Grogin}, N.~A., \& {Panagia}, N. 2008, \apj, 676, 781

\bibitem[{{Windhorst} {et~al.}(2011){Windhorst}, {Cohen}, {Hathi}, {McCarthy},
  {Ryan}, {Yan}, {Baldry}, {Driver}, {Frogel}, {Hill}, {Kelvin}, {Koekemoer},
  {Mechtley}, {O'Connell}, {Robotham}, {Rutkowski}, {Seibert}, {Straughn},
  {Tuffs}, {Balick}, {Bond}, {Bushouse}, {Calzetti}, {Crockett}, {Disney},
  {Dopita}, {Hall}, {Holtzman}, {Kaviraj}, {Kimble}, {MacKenty}, {Mutchler},
  {Paresce}, {Saha}, {Silk}, {Trauger}, {Walker}, {Whitmore}, \&
  {Young}}]{windhorst11ers}
{Windhorst}, R.~A., {Cohen}, S.~H., {Hathi}, N.~P., {McCarthy}, P.~J., {Ryan},
  Jr., R.~E., {Yan}, H., {Baldry}, I.~K., {Driver}, S.~P., {Frogel}, J.~A.,
  {Hill}, D.~T., {Kelvin}, L.~S., {Koekemoer}, A.~M., {Mechtley}, M.,
  {O'Connell}, R.~W., {Robotham}, A.~S.~G., {Rutkowski}, M.~J., {Seibert}, M.,
  {Straughn}, A.~N., {Tuffs}, R.~J., {Balick}, B., {Bond}, H.~E., {Bushouse},
  H., {Calzetti}, D., {Crockett}, M., {Disney}, M.~J., {Dopita}, M.~A., {Hall},
  D.~N.~B., {Holtzman}, J.~A., {Kaviraj}, S., {Kimble}, R.~A., {MacKenty},
  J.~W., {Mutchler}, M., {Paresce}, F., {Saha}, A., {Silk}, J.~I., {Trauger},
  J.~T., {Walker}, A.~R., {Whitmore}, B.~C., \& {Young}, E.~T. 2011, \apjs,
  193, 27

\bibitem[{{Xue} {et~al.}(2011){Xue}, {Luo}, {Brandt}, {Bauer}, {Lehmer},
  {Broos}, {Schneider}, {Alexander}, {Brusa}, {Comastri}, {Fabian}, {Gilli},
  {Hasinger}, {Hornschemeier}, {Koekemoer}, {Liu}, {Mainieri}, {Paolillo},
  {Rafferty}, {Rosati}, {Shemmer}, {Silverman}, {Smail}, {Tozzi}, \&
  {Vignali}}]{xue11}
{Xue}, Y.~Q., {Luo}, B., {Brandt}, W.~N., {Bauer}, F.~E., {Lehmer}, B.~D.,
  {Broos}, P.~S., {Schneider}, D.~P., {Alexander}, D.~M., {Brusa}, M.,
  {Comastri}, A., {Fabian}, A.~C., {Gilli}, R., {Hasinger}, G.,
  {Hornschemeier}, A.~E., {Koekemoer}, A., {Liu}, T., {Mainieri}, V.,
  {Paolillo}, M., {Rafferty}, D.~A., {Rosati}, P., {Shemmer}, O., {Silverman},
  J.~D., {Smail}, I., {Tozzi}, P., \& {Vignali}, C. 2011, \apjs, 195, 10

\bibitem[{{Yan} {et~al.}(2004){Yan}, {Thompson}, \& {Soifer}}]{yanl04}
{Yan}, L., {Thompson}, D., \& {Soifer}, B.~T. 2004, \aj, 127, 1274

\bibitem[{{Yun} {et~al.}(2012){Yun}, {Scott}, {Guo}, {Aretxaga}, {Giavalisco},
  {Austermann}, {Capak}, {Chen}, {Ezawa}, {Hatsukade}, {Hughes}, {Iono},
  {Johnson}, {Kawabe}, {Kohno}, {Lowenthal}, {Miller}, {Morrison}, {Oshima},
  {Perera}, {Salvato}, {Silverman}, {Tamura}, {Williams}, \& {Wilson}}]{yun12}
{Yun}, M.~S., {Scott}, K.~S., {Guo}, Y., {Aretxaga}, I., {Giavalisco}, M.,
  {Austermann}, J.~E., {Capak}, P., {Chen}, Y., {Ezawa}, H., {Hatsukade}, B.,
  {Hughes}, D.~H., {Iono}, D., {Johnson}, S., {Kawabe}, R., {Kohno}, K.,
  {Lowenthal}, J., {Miller}, N., {Morrison}, G., {Oshima}, T., {Perera}, T.~A.,
  {Salvato}, M., {Silverman}, J., {Tamura}, Y., {Williams}, C.~C., \& {Wilson},
  G.~W. 2012, \mnras, 420, 957

\bibitem[{{Zirm} {et~al.}(2007){Zirm}, {van der Wel}, {Franx}, {Labb{\'e}},
  {Trujillo}, {van Dokkum}, {Toft}, {Daddi}, {Rudnick}, {Rix},
  {R{\"o}ttgering}, \& {van der Werf}}]{zirm07}
{Zirm}, A.~W., {van der Wel}, A., {Franx}, M., {Labb{\'e}}, I., {Trujillo}, I.,
  {van Dokkum}, P., {Toft}, S., {Daddi}, E., {Rudnick}, G., {Rix}, H.-W.,
  {R{\"o}ttgering}, H.~J.~A., \& {van der Werf}, P. 2007, \apj, 656, 66

\end{thebibliography}

\end{document}